\documentclass[a4paper,10pt]{article}

\pdfoutput=1
\usepackage[utf8]{inputenc}

\usepackage{authblk}

\usepackage{ifthen,ifpdf}

\ifpdf
  \usepackage{hyperref}
  \hypersetup{colorlinks=false,allcolors=blue}
  \usepackage{hypcap}
  \pdfpagewidth=\paperwidth
  \pdfpageheight=\paperheight
\fi
\usepackage{multirow}
\usepackage[utf8]{inputenc}
\usepackage[T1]{fontenc}
\usepackage{verbatim}
\usepackage{amsmath, amsthm, amssymb}
\usepackage{mathtools}
\usepackage{enumerate}
\usepackage{afterpage}
\usepackage{tabularx}
\usepackage{dsfont}
\usepackage{graphicx}

\usepackage{algorithm}
\usepackage{algpseudocode}
\usepackage{caption}
\usepackage[labelformat=simple]{subcaption}

\usepackage{color}
\usepackage{units}
\usepackage{array, multirow}
\usepackage{booktabs}
\usepackage{MnSymbol}
\newcolumntype{M}[1]{>{\vspace{3pt}\raggedleft\arraybackslash}m{#1}}

\usepackage{siunitx}
\usepackage{tikz}
\usepackage{pgfplots}
\usepackage{pgfplotstable}
\usepgfplotslibrary{units}
\pgfplotsset{compat=1.9}
\pgfplotstableset{col sep=comma}
\usepackage{adjustbox}

\usepackage{cite}

\usepackage{anysize}
\marginsize{2.5cm}{2.5cm}{2cm}{2cm}

\usepackage{wrapfig}
\newcommand{\fsigma}{\mbox{\boldmath $\sigma$}}
\newcommand{\fxi}{\mbox{\boldmath $\xi$}}

\theoremstyle{plain}

\theoremstyle{remark}

\numberwithin{equation}{section}

\newcommand{\R}{\mathds{R}}

\newcommand{\dd}{\text{d}}

\newcommand{\n}{\mathbf{n}}
\newcommand{\g}{\mathbf{g}}
\newcommand{\p}{\mathbf{p}}
\newcommand{\rr}{\mathbf{x}}
\newcommand{\setB}{\Omega_1}
\newcommand{\setA}{\Omega_0}
\newcommand{\e}{\mathbf{e}}
\newcommand{\lb}{\left(}
\newcommand{\rb}{\right)}
\newcommand{\sv}{\lb\begin{array}{cccccccccc}}
\newcommand{\ev}{\end{array}\rb}

\newcommand{\mean}{V_1}
\newcommand{\gauss}{V_0}

\newcommand{\gaussFilter}{\mathcal{G}_\sigma}
\newcommand{\ballFilter}{\mathcal{B}_\sigma}

\newcommand{\depth}{p}
\newcommand{\grayColor}{\mathcal{C}^\depth}

\DeclareMathOperator{\tr}{\textrm{tr}}
\DeclareMathOperator{\Id}{\text{Id}}

\newcommand{\Y}{Y}
\newcommand{\chara}{\chi}
\newcommand{\conv}{\mathcal{F}}
\newcommand{\smoothIm}{\mathcal{I}}

\newcommand{\normedW}{\text{{QNT}}}
\newcommand{\tensorError}{E}
\newcommand{\tensorDirError}{\overline{E}}

\title{Characterizing digital microstructures by\\ the Minkowski-based {quadratic} normal tensor}

\author[1]{Felix Ernesti}
\author[1,*]{Matti Schneider}
\author[2]{Steffen Winter}
\author[2]{Daniel Hug}
\author[2]{Günter Last}
\author[1]{Thomas Böhlke}

\affil[1]{Karlsruhe Institute of Technology (KIT), Institute of Engineering Mechanics}
\affil[2]{Karlsruhe Institute of Technology (KIT), Institute of Stochastics}
\affil[*]{correspondence to: \texttt{matti.schneider@kit.edu}}

\date{\today}

\begin{document}
\maketitle
\begin{abstract}
\noindent
For material modeling of microstructured media, an accurate characterization of the underlying microstructure is {indispensable}. Mathematically speaking, the overall goal of microstructure characterization is to find simple functionals which describe the geometric shape as well as the composition of the microstructures under consideration, and enable distinguishing microstructures with distinct effective material behavior. For this purpose, we propose using Minkowski tensors, in general, and the {quadratic} normal tensor, in particular, and introduce a computational algorithm applicable to voxel-based microstructure representations.\\
Rooted in the mathematical field of integral geometry, Minkowski tensors associate a tensor to rather general geometric shapes, which make them suitable for a wide range of microstructured material classes. Furthermore, they satisfy additivity and continuity properties, which makes them suitable and robust for large-scale applications. We present a modular algorithm for computing the {quadratic} normal tensor of digital microstructures. We demonstrate multigrid convergence for selected numerical examples and apply our approach to a variety of microstructures. Strikingly, the presented algorithm remains unaffected by inaccurate computation of the interface area.\\
The {quadratic} normal tensor may be used for engineering purposes, such as mean-field homogenization or as target value for generating synthetic microstructures.
\end{abstract}

\section{Introduction}
\label{sec:Intro}
 \subsection{State of the art}
 The effective mechanical and thermal behavior of heterogeneous materials is strongly affected by their microstructure and the local material properties. In particular, macroscopic material models need to account for the microstructure of these materials. Since resolving the microstructure for simulations on component scale is computationally expensive, homogenization-based multiscale approaches are popular, see Matou\v{s} et al.~\cite{MatousSummary} for a recent overview. These homogenization techniques compute the effective response of the heterogeneous material, taking the material behavior of the constituents and the microstructure into account. Therefore, the microstructure has to be quantified in terms of suitable data, which is where microstructure characterization comes into play.\\
 A common image-based microstructure-characterization method is scanning a material sample via micro-computed tomography ($\mu$-CT)~\cite{Schladitz2017,Cnudde2009}. After tomographic reconstruction, the {local mass density} of a material is determined and {stored} as 3D voxel data. In case of a two-phase material and after some processing, this voxel data may be interpreted as the characteristic function of the microstructure, i.e., the function which attains the value $1$ for one phase, and the value $0$ for the complementary phase. Correctly segmenting $\mu$-CT scans requires a certain contrast in the absorption rates of the constituents to be applicable, for instance for porous media or for a variety of composite materials.\\
 The mechanical behavior of composite and porous materials is strongly influenced by the volume fractions of the phases. If the characteristic function is accurately resolved by the $\mu$CT-scan, the volume fraction can be computed accurately by numerical integration. With the volume fraction at hand, bounds that predict the possible range of effective elastic and thermal material properties may be established, see Voigt\cite{Voigt} and Reuss~\cite{Reuss}. However, for a high material contrast, these bounds span a wide range and hence provide limited information.\\
 For higher accuracy, additional information is required, see Torquato~\cite{Torquato2002} for an overview. For instance, $n$-point correlation functions~\cite{Brown1955,Torquato1982} provide suitable additional information. Their applicability for anisotropic materials, however, is limited due to the high associated computational effort, see Eriksen et al.~\cite{Correlation}.\\
 For the class of fiber-reinforced composites, specific microstructure-characterization techniques have been established. In addition to the fiber volume-fraction, common characteristics include the fiber aspect-ratio and fiber-orientation tensors of second and fourth order~\cite{Kanatani1984,AdvaniTucker}, see for instance M\"uller and B\"ohlke~\cite{MuellerBoehlke}. A variety of methods for computing fiber-orientation tensors based on volumetric images has been established~\cite{FiberGauss,FiberHessian}. A common approach is based on the so-called structure tensor~\cite{StructureTensor}.\\
 For porous structures, different microstructure characteristics are of interest. For instance, the tortuosity~\cite{Neumann2019} and chord-length distribution~\cite{Matheron1975,Torquato2002}, as well as the pore-size distribution~\cite{poresize} are investigated. These measures are primarily responsible for the effective (isotropic) permeability of the porous medium in question.\\
 Polycrystalline materials require a different approach. Typically, the grains differ only in their crystalline orientation, but have identical absorption rates. Hence, $\mu$-CT scans are of limited use. Instead, for reconstructing the $3$D-microstructure of polycrystalline materials, focused ion beam - scanning electron microscopy (FIB-SEM)~\cite{FIBSEM1, FIBSEM2, FIBSEM3} or electron back-scattering diffraction (EBSD)~\cite{EBSD1, EBSD2, EBSD3} are preferred. Primary microstructure {characteristics of polycrystalline} materials are the grain-size distribution (morphological texture)~\cite{grainsize} and the orientation distribution (crystallographic texture)~\cite{Bunge1982,Texture1,Texture2,Texture3,Texture4}.\\
 {From a theoretical point of view, most materials undergoing a manufacturing process are influenced by stochastic factors, for instance due to slight variations in the composition or the seemingly chaotic behavior of the processing condition as a result of a high sensitivity to initial and boundary conditions. Still, the experimentally determined effective properties of such composites are often surprisingly deterministic. These observations may be formalized by the theory of stochastic homogenization~\cite{Koz1979,PapanicolaouVaradhan79}. From this stochastic point of view, any finite volume element represents only a fraction of a specific realization of a random material~\cite{Torquato2002,Kanit2003}. In particular, any quantity associated to such a volume element may also be considered as a \emph{random variable}. For this work, we assume the volume element to be given and fixed, and regard the associated quantities as \emph{deterministic}.}\\
 Minkowski functionals~\cite{SchneiderWeil,Hadwiger}, also known as intrinsic volumes, are a basic tool in stochastic geometry. They are defined for wide classes of shapes, including all convex sets and their finite unions as well as all bounded sets with smooth boundary. A Minkowski functional associates to any such shape a scalar quantity. If one requires such a functional to be invariant with respect to Euclidean motions, additive and to satisfy a certain continuity property, then it can be shown, see Hadwiger\cite{Hadwiger}, that in $3$D it can be written as linear combination of only four \emph{basic} functionals, the Minkowski functionals. Among them are the total volume, the total surface area, the Euler characteristic and one further functional, which for convex shapes may be interpreted as the mean width, or in the context of smooth boundaries as the integral of mean curvature. 
 {Approaches for computing Minkowski functionals are based, for instance, on marching squares\cite{MarchingSquare} or on Steiner's formula~\cite{Klenk2006,Gunderlei2007}.}\\
 Being scalar-valued and rotation invariant, Minkowski-functionals are intrinsically insensitive to anisotropic features of the shape in question. Therefore, tensor-valued analogs of Minkowski functionals, the so-called Minkowski tensors~\cite{Alesker,Hug2008a,Hug2008b,JK2017book}, were introduced and studied. In addition to additivity and continuity, Minkowski tensors are required to be equivariant w.r.t.\ Euclidean transformations. This means, for instance, that rotating a shape first and computing its Minkowski tensor afterwards leads to the same result as computing the Minkowski tensor first and rotating the tensor afterwards. A direct consequence of this property is that Minkowski tensors preserve axes of symmetry of structures, i.e., if a shape is rotationally invariant w.r.t.\ an axis $\p$, the Minkowski tensor will be rotationally invariant w.r.t.\ $\p$ as well.\\
 Minkowski tensors may be computed for general microstructures with distinct interfaces, such as porous media, foams, bones or granular structures~\cite{SchroederTurk2011}. For porous media, Klatt et al.~\cite{Klatt17} conducted a comparison between the common chord-length analysis and a Minkowski-tensor based approach. Schroeder-Turk et al.\cite{SchroederTurk2011,SchroederTurk2013} evaluated Minkowski tensors for a given triangulation of the interface via explicitly known expressions for polytopes. Their ansatz was successfully used for characterizing the anisotropy of granular matter and metal foams, as well as identifying defects in molecular dynamics simulations of metal phases.
 For $3$D gray-value images, Svane~\cite{Svane2014,Svane2015} introduced approximation formulae for Minkowski functionals and tensors, also establishing convergence upon mesh refinement, called multigrid convergence in this context. Unfortunately, the cited works~\cite{Svane2014,Svane2015} did not include numerical examples.\\
 For finite point samples, Voronoi-based estimators~\cite{MinkowskiVoronoi} may be used for approximating Minkowski tensors.

 \subsection{Contributions}
 We present an applied approach for characterizing digital microstructures of industrial complexity in terms of the {quadratic} normal tensor, a tensor-valued quantity based on Minkowski tensors, bringing these concepts to the attention of the engineering community.\\
  {For phenomenological continuum theories, which use microstructure information as state or microstructure variables to model the influence of microstructure on macroscopic material behavior, the Minkowski tensors are promising quantities, because they are in principle observable and can be effectively calculated from three-dimensional image data. The Minkowski tensors complement, e.g., the already widely used fiber-orientation tensors~\cite{AdvaniTucker,Kanatani1984}, which approximate the tangent distribution of the fiber centerline, and tensorial texture coefficients~\cite{Texture1,Texture4}, which describe the distribution of crystal orientations.}
 \\
 We introduce the relevant Minkowski functionals and tensors in Section \ref{sec:Theory} and isolate among them those suitable for microstructure characterization. In Section \ref{sec:Implementation}, we present a novel algorithm for computing the {quadratic} normal tensor. For large microstructures with complex geometry, finding triangulations of the interface may be a challenging task, in particular if the microstructure is described by voxel data. Therefore, we present here an alternative to triangulation-based algorithms~\cite{SchroederTurk2011}, that works directly with gray-value images as input. The outward-pointing unit normals on the materials interface are approximated by finite-difference gradients of the discretized characteristic function.\\
 We investigate multigrid convergence of our approach by numerical studies in Section \ref{sec:Numerics}. For fiber-reinforced composites, we compare the {quadratic} normal tensor to the more conventional fiber-orientation tensor of second order~\cite{Kanatani1984,AdvaniTucker}. We compare the accuracy of our approach to the commonly used {structure-tensor based algorithm}~\cite{StructureTensor} for computing fiber-orientation tensors. Last but not least, we study the anisotropy of sand grains and porous sand-binder aggregates based on the {quadratic} normal tensor.

\section{Using Minkowski tensors for describing microstructures}
\label{sec:Theory}

 \subsection{Minkowski tensors}
We briefly introduce Minkowski functionals and Minkowski tensors in a form suitable for our purposes and restrict to the $3$D case. We refer to Schr\"oder-Turk et al.~\cite{SchroederTurk2013, SchroederTurk2011} {or the lecture notes \cite{JK2017book}} for the general case.\\
Consider a solid body, by which we mean a bounded, not necessarily connected set $K$ in $\R^3$ with sufficiently regular boundary $\partial K$. Here regularity can mean smoothness or convexity of some form. For our purposes it will be completely sufficient to assume that $K$ is polyconvex, i.e.,\ $K$ can be represented as a finite union of (not necessarily disjoint) convex sets.

To gain insight into the morphology of $K$, a shape index $\varphi$ associates to any such set $K$ a scalar value. If one requires the shape index $\varphi$ to satisfy some natural basic properties, namely invariance with respect to rigid motions, additivity (meaning that $\varphi(K\cup L)=\varphi(K)+\varphi(L)-\varphi(K\cap L)$ for solid bodies $K,L$) and a certain continuity (for convex sets, and w.r.t.\ the Hausdorff distance, see e.g.\ Schneider-Weil~\cite[\S 12.3]{SchneiderWeil}), then it is a well-known fact due to Hadwiger~\cite{Hadwiger}, that $\varphi$ may be represented as a linear combination of only four basic functionals $V_0,\ldots, V_3$, known as Minkowski functionals or intrinsic volumes. The Minkowski functionals encompass the volume $V=V_3$, the surface area $S=2V_2$, and two further functionals, $\mean$ and $\gauss$, which in special situations can be interpreted as the total mean curvature and the total Gaussian curvature of the body $K$. The latter is proportional to the Euler characteristic of $K$, i.e., the genus of the surface $\partial K$, which is a topological invariant. 
Volume and surface area are computed by 
 \begin{align}
 V(K) = \int_K dV\quad \text{and}\qquad S(K) = \int_{\partial K}dS .
 \end{align}
If the boundary $\partial K$ is sufficiently smooth, then the local mean curvature $H$ and the Gaussian curvature $G$ (i.e.\ the average and the product of the principal curvatures) are well-defined at each boundary point and the total curvatures may be computed via
 \begin{align}
 {\mean(K) =\frac 1{\pi} \int_{\partial K}H \, dS\quad \text{and}\quad
 \gauss(K) = \frac 1{4\pi}\int_{\partial K} G \, dS. }
 \end{align}
 \begin{figure}
 \begin{center}
 \includegraphics[width = 0.5\textwidth, draft = false]{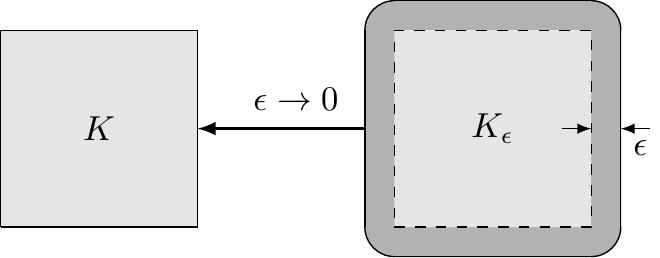}
 \caption{Illustration of the $\varepsilon$-parallel expansion $K_\varepsilon$ of the shape $K\subseteq \R^3$.}\label{fig:K_epsilon}
 \end{center}
 \end{figure}
 Such integral representations are also available for non-smooth bodies when one replaces $\partial K$ by an integration over the normal bundle of $K$~\cite{Z86}. 
 For practical computations, the additivity property is essential, allowing to decompose complex structures into simple convex pieces and to treat these pieces individually. For convex shapes, the Steiner formula provides another way to characterize the Minkowski functionals and another idea how to compute them.
 Consider, for a convex body $K$ and $\varepsilon>0$, the $\varepsilon$-approximation 
 \[
 	K_\varepsilon = \left\{\mathbf{x}\in\R^3: ||\mathbf{x}-\mathbf{y}||\leq \varepsilon \text{ for some } \mathbf{y}\in K\right\}.
 \]
The Steiner formula~\cite{Schneider14} states that the volume of $K_\varepsilon$ is a polynomial in $\varepsilon$, whose coefficients are (up to some normalization constants) the Minkowski functionals of $K$:
 \begin{align}\label{eq:Steiner}
 V(K_\varepsilon) = V(K) + \varepsilon S(K) +\pi \varepsilon^2 \mean(K) + \frac{4\pi}{3}\varepsilon^3\gauss(K).
\end{align}
This allows to recover the Minkowski functionals of $K$ by computing volumes of a number of {$\varepsilon$-ap\-proximations} and inverting the {above} formula, see {Klenk-Schmidt-Spodarev}~\cite{KSS06}.
One can also use the fact that the $\varepsilon$-ap\-proximations $K_\varepsilon$ are smooth even if $K$ is not, allowing to determine the Minkowski functionals $\mean$ and $\gauss$ by means of the limit procedure
 \[
  \mean(K) = \lim_{\varepsilon \rightarrow 0} \mean(K_\varepsilon) \quad \text{and} \quad \gauss(K) = \lim_{\varepsilon \rightarrow 0} \gauss(K_\varepsilon).
 \]
 {While these approximation results follow from the continuity of the Minkowski functionals, $\varepsilon$-approxima\-tion properties of more general classes of sets are discussed in Rataj~\cite{Rataj06}}. For more background to our informal discussion, we refer to Schr\"oder-Turk et al.~\cite{SchroederTurk2013} and the references therein.

Since Minkowski functionals are, by definition, invariant w.r.t.\ Euclidean motions {or change of frame}, they are insensitive to directional and positional information. Hence, they are inappropriate for detecting anisotropies in a shape $K$.
For this latter purpose and other applications, a more general theory of tensor-valued shape indices has been developed, which are covariant w.r.t.\ Euclidean motions, see Schr\"oder-Turk et al.\cite{SchroederTurk2013}. In analogy to Hadwiger's theorem~\cite{Hadwiger} and restricting  to $\R^3 \otimes_{\text{sym}} \R^3 \cong \R^{3\times 3}_{\text{sym}}$ tensors, there are only six linearly independent shape indices (in addition to the Minkowski functionals multiplied by the identity), see \cite{Alesker} and in particular \cite[\S4]{Hug2008a}. For a (convex) body $K$ with sufficiently smooth boundary, these may be expressed as
 \begin{align}\label{eq:MinkowskiTensors}
 \begin{aligned}
 W^{2,0}_0(K) &= \int_K \rr\otimes \rr \, dV,		
 & W^{2,0}_1(K) &= \frac{1}{3}\int_{\partial K} \rr \otimes \rr \, dS, 	
 & W^{2,0}_2(K) &= \frac 13\int_{\partial K} H(\rr)\rr \otimes \rr \, dS,
 \\
 W^{2,0}_3(K) &= \frac 13\int_{\partial K} G(\rr)\rr \otimes \rr \, dS,		
 & W^{0,2}_1(K) &= \frac 13\int_{\partial K} \n \otimes \n \, dS, 		
 & W^{0,2}_2(K) &= \frac 13\int_{\partial K} H(\rr) \n \otimes \n \, dS.
 \end{aligned}
 \end{align}
 Here, $\rr$ denotes the position vector of a point in $K$ (or $\partial K$) and $\n$ stands for the field of outward-pointing unit-normal vectors on $\partial K$. For Minkowski tensors, Steiner-type formulae based on support measures have been established, see Schneider~\cite{SteinerTensor}. Note that some Minkowski functionals can be recovered from Minkowski tensors. For instance, the surface area is given by the formula $S(K) = 3\tr(W^{0,2}_1(K))$.\\
Based on these Minkowski tensors, Schr\"oder-Turk et.\ al.~\cite{SchroederTurk2011} introduce the eigenvalue ratios
\begin{align} \label{eq:eigenval_ratio}
\beta(W) = \frac{ \min_{\lambda \in \text{E}(W)} |\lambda| }{\max_{\lambda \in \text{E}(W)} |\lambda|}, 
\end{align}
as scalar measures of anisotropy. Here $W$ stands for any of the six Minkowski tensors defined in \eqref{eq:MinkowskiTensors} and $\text{E}(W)$ is the set of eigenvalues of the symmetric matrix $W$. Clearly, $\beta(W) \in [0,1]$. For $W=W^{2,0}_0$, $W^{2,0}_1$ and $W^{0,2}_1$, the matrix $W(K)$ is positive semi-definite in general (and this is also true for the other Minkowski tensors if $K$ is a convex body), implying that all eigenvalues of $W$ are nonnegative. In this case, $\beta(W) = 1$ if and only if all eigenvalues are equal, i.e., if the tensor is a multiple of the identity. Note that smaller values of $\beta(W)$ correspond to a higher degree of anisotropy.

 \subsection{Minkowski-tensor based microstructure characterization}

Heterogeneous materials often exhibit random variations in their microstructure, but a resulting deterministic material behavior~\cite{Torquato2002}. For characterizing microstructures, we are interested in singling out a small number of tensor-valued descriptors, that may, in turn, be used as input for homogenization schemes, see Klusemann and Svendsen~\cite{Klusemann2010} for an overview.
These microstructure identifiers should preferably exhibit certain natural properties:

\begin{enumerate}
\item Respect for symmetries:
 We seek microstructure identifiers that preserve symmetry information. If a microstructure possesses some symmetry, then this is typically reflected in the macroscopic material behavior. Therefore, identifiers should capture such symmetry.
 \item
Robustness: To be of practical use, small changes in the microstructure should only result in small changes in the descriptor.
\item
Translation invariance: For homogenization, statistical homogeneity is essential~\cite{Torquato2002}. Thus, our identifiers should be invariant with respect to translations of the shape $K$. Furthermore, we want to explicitly include periodic structures, as periodic homogenization is often used for studying random microstructures~\cite{Kanit2003}.
\item
Universal applicability: Minimal assumptions on the geometry of the structure allow for general application on a variety of different microstructures.
\end{enumerate}
In the light of these criteria, Minkowski tensors are promising candidates for microstructure characteristics.
\begin{enumerate}
\item
Their covariant tensorial nature reflects the anisotropy and direction dependence of the structure in question.
\item They are robust due to their continuity properties w.r.t.\ the Hausdorff distance. For example, if a sequence of convex bodies $K_n$ converges to a convex body $K$, as $n\to\infty$, then all their Minkowski functionals and Minkowski tensors converge as well. Similar results hold e.g.\ if a polyconvex set $K$ is approximated by its parallel sets $K_\varepsilon$, see \cite{Schneider14,Rataj06}. There are also stability results showing Hölder continuity with exponent at least $1/2$, \cite{Hug2015}.
\item If translation invariance is required, then beside the Minkowski functionals among the above mentioned Minkowski tensors precisely $W^{0,2}_1$ and $W^{0,2}_2$ are suitable. As computing the curvature of interfaces of $3$D voxel images is not straightforward~\cite{Curvature1,Curvature2}, we restrict in this article to the volume $V$, the surface area $S$ and the Minkowski tensor $W^{0,2}_1$.
\item The Minkowski tensors are not restricted to specific shape assumptions on $K$. Indeed, only minimal assumptions on $K$ are required~\cite{SchroederTurk2011}. For any practical application it is probably sufficient to note, that any set (however complex) can be approximated arbitrarily well by a polyconvex set on which Minkowski tensors are defined. The tensor $W^{0,2}_1$ under consideration can in fact be defined under much weaker regularity assumptions, e.g.\ for sets with piecewise smooth boundaries. This flexibility distinguishes them from other approaches, where geometric priors are required for characterizing microstructures. For instance, for fiber-reinforced composites, fibers are often assumed to be (locally) cylindrical. Such geometrical priors run into problems for fibrous microstructure where the fibers deviate from their original cylindrical shape. For instance, during injection molding, fibers may be bent or twisted~\cite{FiberBending}. As they are independent of priors, Minkowski tensors may be suitable for characterizing fibers with distinct curvature.
\end{enumerate}
\begin{figure}
\begin{center}
\includegraphics[width=0.3\textwidth, draft = false]{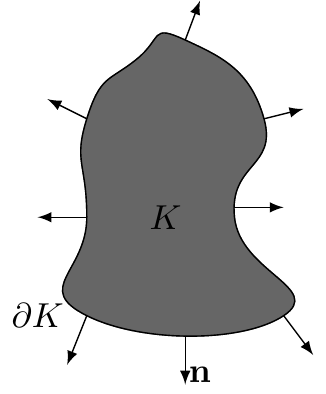}
\caption{Body $K$ with outward-pointing unit-normal field $\n$.}\label{fig:hedgehog}
\end{center}
\end{figure}
In the field of microstructure characterization, $K$ is often the set union of a multitude of bodies, for instance inclusions within a surrounding matrix material. In this context, we are interested in a {tensorial anisotropy-measure}, which is stable w.r.t.\ an infinite-volume limit, where the number of inclusions tends to infinity. Thus, we normalize $W^{0,2}_1$ to obtain the {quadratic} normal tensor ({QNT})
\begin{align}\label{eq:def_DNT}
	\normedW(K) = W^{0,2}_1(K)/\tr(W^{0,2}_1(K)),
\end{align}
which, for a {single body or microstructure} $K$ with sufficiently smooth boundary, may be written in the form
\begin{align*}
\normedW(K) = \frac{1}{S(K)}\int_{\partial K}\n \otimes \n \,\dd S,
\end{align*}
where again $\n = \n(\rr)$ is the field of normal vectors on $\partial K$, cf.\ Fig.~\ref{fig:hedgehog}. For a geometric interpretation of the $\normedW$, {observe that for any vector $\fxi\in\R^3$ the expression
\[
(\n\otimes \n) \fxi=\n\,(\n \cdot \fxi)
\]
describes the orthogonal projection of $\fxi$ onto the line spanned by the normal direction $\n$ at $\rr$.}
In this sense, $\normedW(K)$ may be interpreted as an average over the normal projections computed w.r.t.\ the uniform probability measure concentrated on the surface $\partial K$. (Note that the resulting average matrix is still symmetric and positive definite but does not represent a projection anymore.)\\
The $\normedW(K)$ admits an additional interpretation from a mechanical point of view. Suppose the {structure} $K$ deforms with a homogeneous stress $\fsigma$. Then, contracting the stress tensor with the $\normedW$
\begin{align*}
{
\normedW(K):\fsigma=\frac{1}{S(K)}\int_{\partial K} \n \cdot  (\fsigma\n) \, dS}
\end{align*}
computes the mean normal stress on the surface $\partial K$.\\
By construction, the $\normedW$ is symmetric, positive semi-definite and has trace $1$. In particular, $\normedW(K)$ admits an eigenvalue decomposition with real-valued, non-negative eigenvalues $\lambda_1$, $\lambda_2$ and $\lambda_3$, which sum to 1. In case of a convex $K$, certain eigenvalue combinations can directly be interpreted in terms of the resulting shape of $K$: $\lambda_1\gg\lambda_2=\lambda_3$, for instance, indicates a rather flat shape within the plane perpendicular to the eigenvector corresponding to $\lambda_1$. For $\lambda_1=\lambda_2\gg\lambda_3$ we expect $K$ to be a needle expanded in the direction of the eigenvector associated with $\lambda_3$, {see also Appendix~\ref{appendix:cylinder}, where the $\normedW$ is computed for a cylinder, and the sand grain experiments in Section~\ref{sec:4.4}.} 
\\
Another advantage of Minkowski tensors is that they are locally defined and therefore locally computable. Complex polyconvex shapes can be cut into simple pieces and each piece can be treated separately. Then the additivity allows to recover the Minkowski tensor of the whole body from the Minkowski tensors of the pieces, allowing for efficient computation and parallelization.

\section{Efficient implementation for 3D image data}
\label{sec:Implementation}
\subsection{Algorithmic overview}
\begin{figure}
 \begin{center}
 \begin{subfigure}{0.25\textwidth}
 \includegraphics[width=\textwidth]{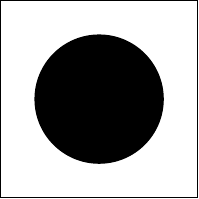}
     \vspace{-0.7cm}
    \caption{Characteristic function $\chi$}
    \label{fig:inclusionY}
    \end{subfigure}
    \hspace{0.2cm}
  \begin{subfigure}{0.25\textwidth}
 \includegraphics[width=\textwidth]{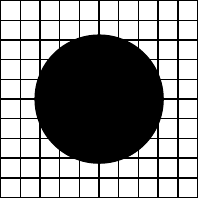}
 \vspace{-0.7cm}
    \caption{$\chi$ with background grid}
    \label{fig:inclusionY_H}
 \end{subfigure}
 \hspace{0.2cm}
  \begin{subfigure}{0.25\textwidth}
 \includegraphics[width=\textwidth]{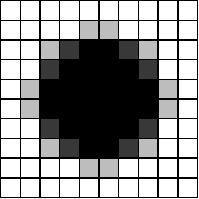}
  \vspace{-0.7cm}
    \caption{gray-value image $\chi_h$}
    \label{fig:chiH}
 \end{subfigure}
 \caption{Characteristic function of a ball and its discrete representation by a gray-value image on a regular voxel grid.}
 \label{fig:chi}
 \end{center}
 \end{figure}

Consider a (periodic) heterogeneous two-phase material on the domain $\Y = [0,L_x]\times [0,L_y]\times [0,L_z]$. The microstructure of the material is described by its characteristic function $\chara:\Y\rightarrow\{0,1\}$,
defining the two phases $\setA$ and $\setB$ via $\setA = \{ \rr \in \Y: \chara(\rr)=0\}$ and $\setB = \Y\backslash \setA$, respectively. Our aim is to describe phase $\setB$ using the Minkowski functionals and tensors $V(\setB)$, $S(\setB)$, $W^{0,2}_1(\setB)$ and $\normedW(\setB)$.\\
Note that $\setB$ is unknown in practice, only CT images of $\setB$ can be observed. $\mu$-CT data is typically stored as $3$D gray-value voxel data. 
We interpret the voxel data as a mapping $\chi_h:\Y_h\to [0,1]$ from the discrete set $\Y_h$, comprising the centers of a regular voxel grid with voxel length $h$, to the
unit interval representing gray values. The gray value $\chi_h(\mathbf{y})$ associated to a point $\mathbf{y}\in Y_h$ stands for the volume fraction of $\setB$ in the voxel centered at $\mathbf{y}$. 
The relation between $\chi$ and its discretization $\chi_h$ is demonstrated in Fig.~\ref{fig:chi}. Fig.~\ref{fig:inclusionY} shows the characteristic function $\chi$ of a ball. Fig.~\ref{fig:inclusionY_H} shows the non-discretized ball with the regular grid $\Y_h$ in the background. In Fig.~\ref{fig:chiH}, we see the discrete characteristic function $\chara_h$ of this ball as a gray-value image.
 Note that, in general, the input data $\chi_h$ does not allow to recover the phases $\setA$ and $\setB$ exactly as the interface is blurred. Only in the limit as $h\rightarrow 0$ the correct characteristic function and, therefore, the correct sets are recovered. For determining $W^{0,2}_1$ and $S$, in addition the normal directions are needed. {In a weak sense, the unit normal $\n$ of the set $\setB$ at a boundary point is recovered by $\n =- \nabla \chara$, whereas $- \nabla \chara=0$ away from the boundary. This statement may be formalized in terms of functions of bounded variation~\cite{Ambrosio}}. Therefore, we will compute the gradient numerically and establish formulae for $W^{0,2}_1$ and $S$ based on volume averaging, cf.\ Section~\ref{sec:quantities}. To improve the gradient estimation, a smoothing of the characteristic function $\chi_h$ is applied beforehand.
 The algorithm for computing the Minkowski quantities from a given voxel image is summarized in Alg.~\ref{Alg:Minkowski}. First, we apply an image filter $\mathcal{F}_\sigma$ to the characteristic function $\chi_h$. Secondly, we estimate the outward-pointing normal vector by computing the gradient $\g$ from the resulting smoothed image $\smoothIm_h^\sigma$. Finally, the desired quantities $V,S,W^{0,2}_1$ and $\normedW$ are estimated.
 \begin{algorithm}[H]
\caption{Computation of Minkowski quantities}
\label{Alg:Minkowski}
\begin{algorithmic}[1]
\State $\smoothIm_h^\sigma \gets \mathcal{F}_\sigma * \chara_h$\Comment{ Blur image with image filter}
\State $\g(\rr) \gets \nabla_h \smoothIm_h^\sigma(\rr) $\Comment{ Compute gradient}
\State Compute $V$ by \eqref{eq:volumeEstimation}
\State Compute $S$ by \eqref{eq:surfaceEstimation}
\State Compute $W^{0,2}_1$ by \eqref{eq:W5Estimation}
\State Compute $\normedW$ by $W^{0,2}_1/\tr(W^{0,2}_1)$
\State \Return $(V,S,W^{0,2}_1,\normedW)$
\end{algorithmic}
\end{algorithm}

 \subsection{Smoothing by image filters}

\begin{figure}[H]
 \begin{center}
\begin{center}
  \begin{subfigure}{.28\textwidth}
   \includegraphics[width=\textwidth]{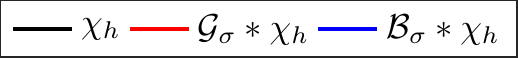}
  \end{subfigure}
  \end{center}
 \begin{subfigure}{0.27\textwidth}
 \includegraphics[width=\textwidth]{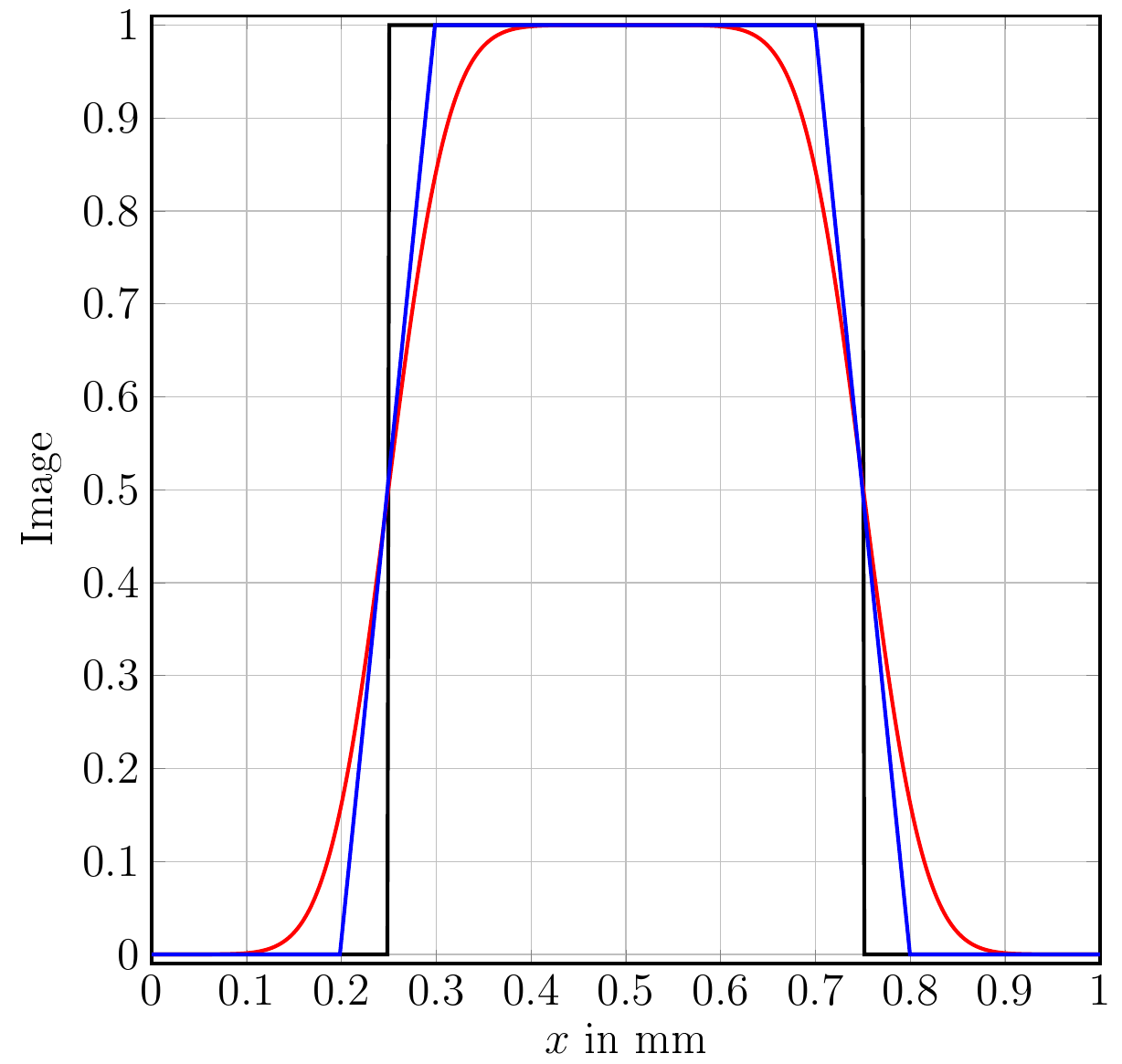}
     \vspace{-0.7cm}
    \caption{{$\sigma = 0.05\frac{\unit{mm}}{h}$}}
    \label{fig:filter005}
    \end{subfigure}
    \hspace{0.2cm}
  \begin{subfigure}{0.27\textwidth}
 \includegraphics[width=\textwidth]{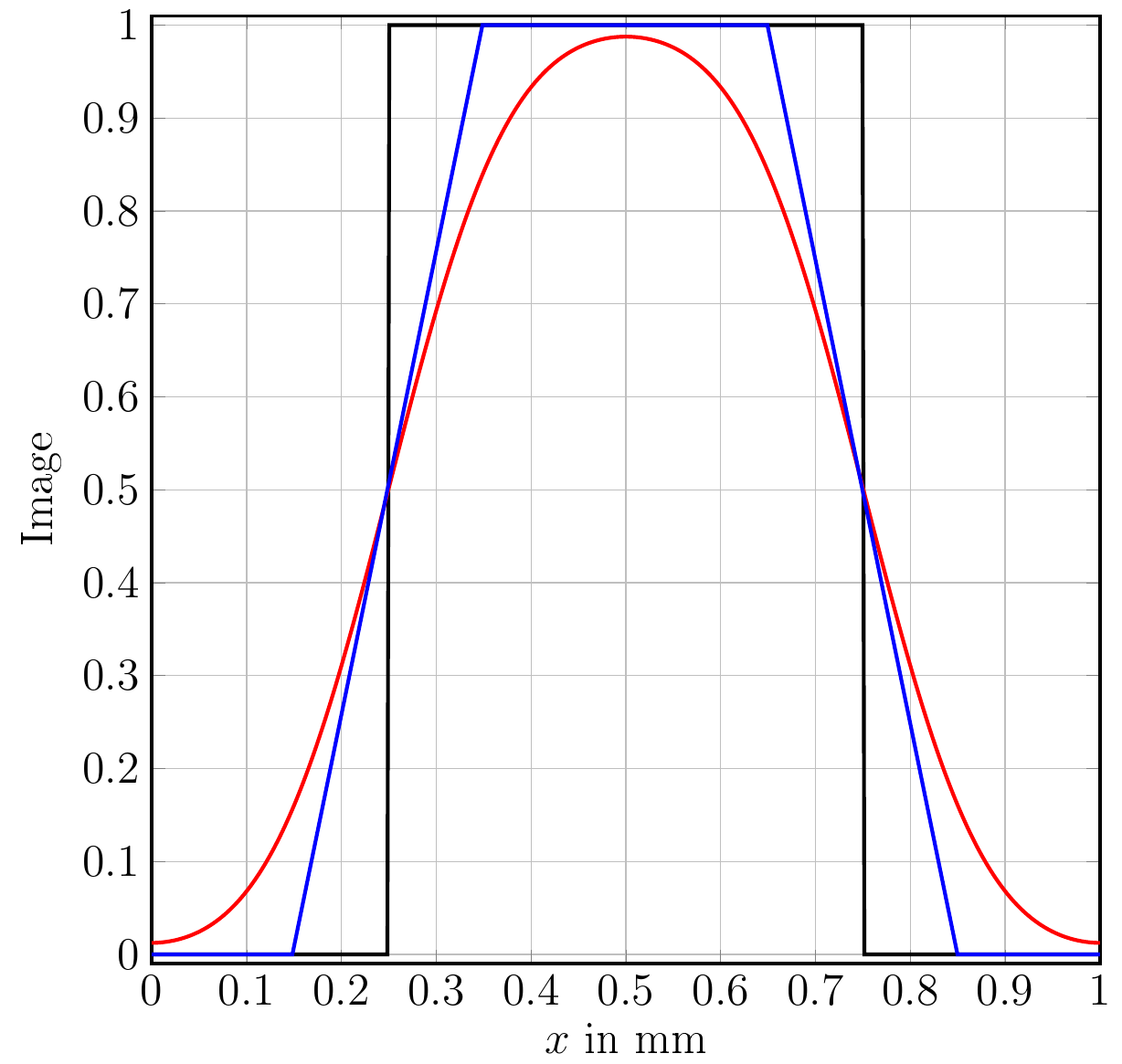}
 \vspace{-0.7cm}
    \caption{{$\sigma = 0.1\frac{\unit{mm}}{h}$}}
    \label{fig:filter01}
 \end{subfigure}
 \hspace{0.2cm}
  \begin{subfigure}{0.27\textwidth}
 \includegraphics[width=\textwidth]{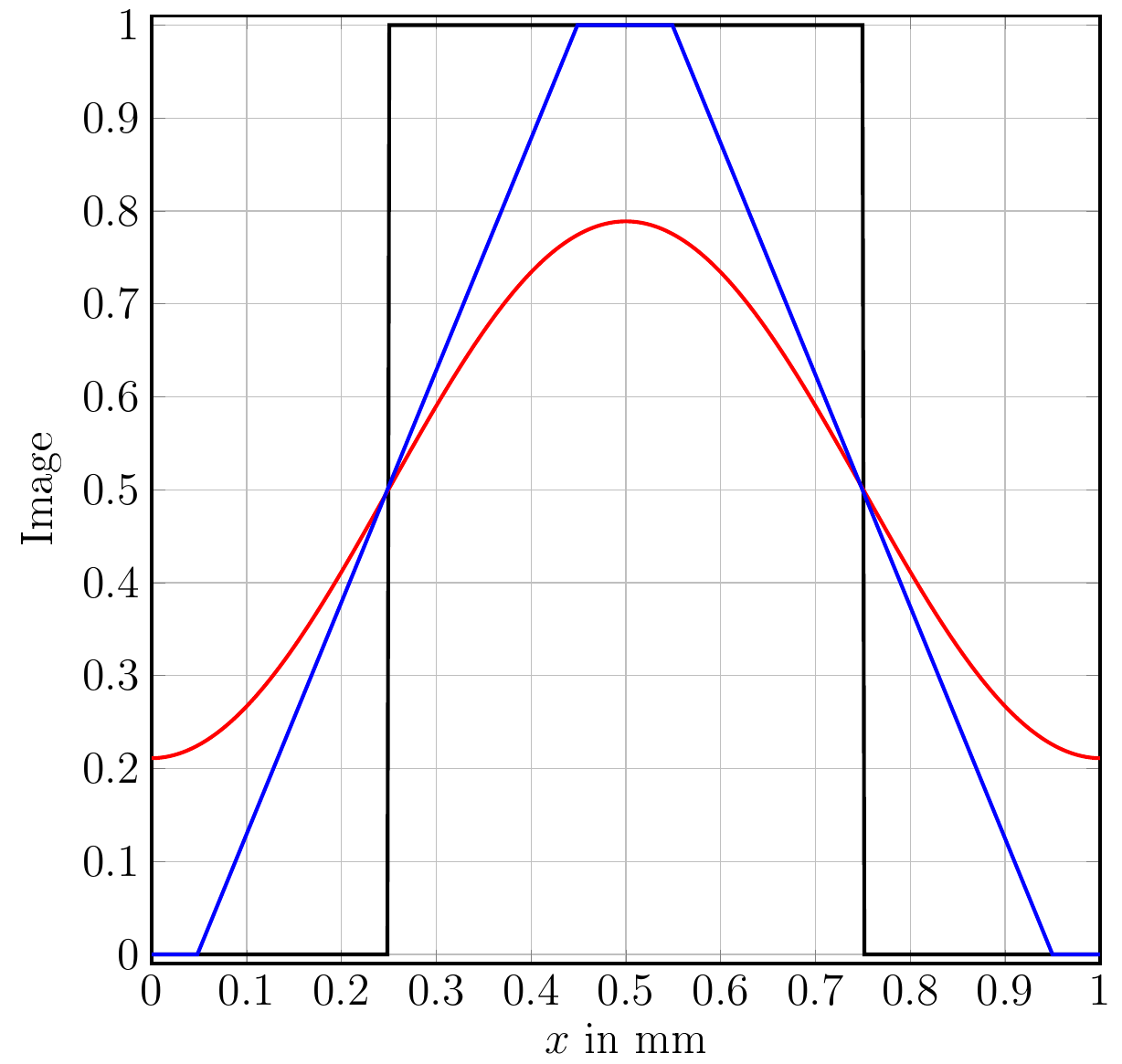}
  \vspace{-0.7cm}
    \caption{{$\sigma = 0.2\frac{\unit{mm}}{h}$}}
    \label{fig:filter02}
 \end{subfigure}
 \caption{Influence of filtering with different kernels and widths $\sigma$.}
 \label{fig:filter}
 \end{center}
 \end{figure}

 Due to the reconstruction procedure, $\mu$-CT scans often exhibit artifacts and impurities. Furthermore, binary voxel-based images do not allow reconstructing interfaces accurately~\cite{MarchingSquare}. To deal with these issues, we apply an image filter to the discrete characteristic function (prior to computing the gradient). As different filters (and different choices of parameters) are available, we will also address choosing an appropriate filter. Applying the filter is realized by convolving the image with a specific filter kernel $\conv_\sigma$. The filter parameter $\sigma$ controls the width of filtering, and the result is the filtered image $\smoothIm_h^\sigma$, {given as the convolution}
 \begin{align*}
     \smoothIm_h^\sigma = \conv_\sigma * \chara_h.
 \end{align*}
 In our implementation, the convolution with $\conv_\sigma$ is implemented via fast Fourier transform (FFT)~\cite{FFT}, see Alg.\ref{Alg:Filter}. Notice that in some cases the Fourier-transformed filter kernel may be computed efficiently without using the FFT.
  \begin{algorithm}[H]
\caption{FFT-based filter application}
\label{Alg:Filter}
\begin{algorithmic}[1]
\State $\widehat{\chara_h} \gets \text{FFT}(\chara_h)$\Comment{Transformation of the characteristic function}
\State $\widehat{\conv_\sigma}\gets\text{FFT}(\conv_\sigma) $\Comment{Transformation of the filter kernel}
\State $\widehat{\smoothIm_h^\sigma}(\xi)\gets \widehat{\chara_h}(\xi)\widehat{\conv_\sigma}(\xi)$\Comment{Multiplication in Fourier space for all frequencies $\xi$}
\State $\smoothIm_h^\sigma\gets \text{IFFT}(\widehat{\smoothIm_h^\sigma})$\Comment{Inverse transformation}
\end{algorithmic}
\end{algorithm}
We shall consider a dimensionless filter parameter $\sigma$ and scale it by the voxel length $h$. As filter kernels, we consider a Gaussian kernel~\cite{Gaussian}
  \begin{align*}
  \gaussFilter(\rr)= \frac{1}{(h\sigma)^3(2\pi)^{\frac{3}{2}}}\exp\bigg(-\frac{\|\rr\|^2}{2(h\sigma)^2}\bigg)
  \end{align*}
  and the characteristic function of the unit ball, scaled to integrate to unity,
  \begin{align*}
  \ballFilter(\rr) = \begin{cases}\frac{3}{4\pi(h\sigma)^3}& \text{if}~ {\|\rr\|\leq h\sigma},\\
  0 & \text{otherwise}.
  \end{cases}
  \end{align*}
 In Fig.~\ref{fig:filter}, the effect of filtering by a Gaussian and a ball kernel, respectively, is shown for a $1$D laminate structure discretized with a voxel length of {\mbox{$h=2\,\unit{\mu m}$}} for three different filter parameters $\sigma$. The red curve illustrates the impact of the Gaussian filter, whereas the blue line represents the ball-filtered image. In the Gaussian case, the resulting image is smooth across the laminate's interface. However, for larger $\sigma$, not only the interface is blurred, but no region of black or white remains. In fact, due to its global support, this even holds for small $\sigma$.\\
 The impact of the ball filter is completely different. The {piecewise constant indicator function} with jumps at the interfaces is transformed into a piecewise linear function with slopes {$\pm\frac{1}{2h\sigma}$}. Therefore, when applying the ball filter to a structure with diameter larger than $2h\sigma$, some region with $\smoothIm_h^\sigma =1$ will remain.

 \subsection{Approximating the surface normal by finite differences} \label{sec:3.3}
 Computing the Minkowski tensor $W^{0,2}_1(\setB)$ requires determining the (unit) normal vector field $\n$ on the surface $\partial \setB$. We approximate the normal field $\n$ by computing the gradient vector field
 \begin{align*}
 \g=\nabla_h \smoothIm_h^\sigma
\end{align*}
of the filtered image $\smoothIm_h^\sigma$ numerically.{ Notice, that $\g$ is dependent on the voxel length $h$. At boundary points, we consider $\n\approx-\g/||\g||$ as the outward pointing unit normal, provided $\g\neq 0$.}
We briefly discuss the choice of the numerical gradient-approximation method.
 Finite-difference approximations are a simple way for approximating the gradient of a function given on a regular voxel grid numerically. Suppose a function $f:\Y\rightarrow \R,~\rr\mapsto f(\rr)$ is given. We consider three finite-difference discretization schemes for the partial derivative in $\e_i$-direction ($i=1,2,3$):
\begin{enumerate}
\item
 first-order approximation by forward differences, i.e.,
 \begin{align*}
 \partial_i^h f(\rr) \approx \frac{f(\rr+h \e_i) - f(\rr)}{h};
 \end{align*}
 \item
 first-order approximation by backward differences, i.e.,
  \begin{align*}
 \partial_i^h f(\rr) \approx \frac{f(\rr) - f(\rr-h \e_i)}{h};
 \end{align*}
 \item
 second-order approximation by central differences, i.e.,
  \begin{align*}
 \partial_i^h f(\rr) \approx \frac{f(\rr+h \e_i) - f(\rr-h \e_i)}{2h}.
 \end{align*}
\end{enumerate}
  \begin{figure}
  \begin{center}
 \begin{subfigure}{0.23\textwidth}
\includegraphics[width=\textwidth,]{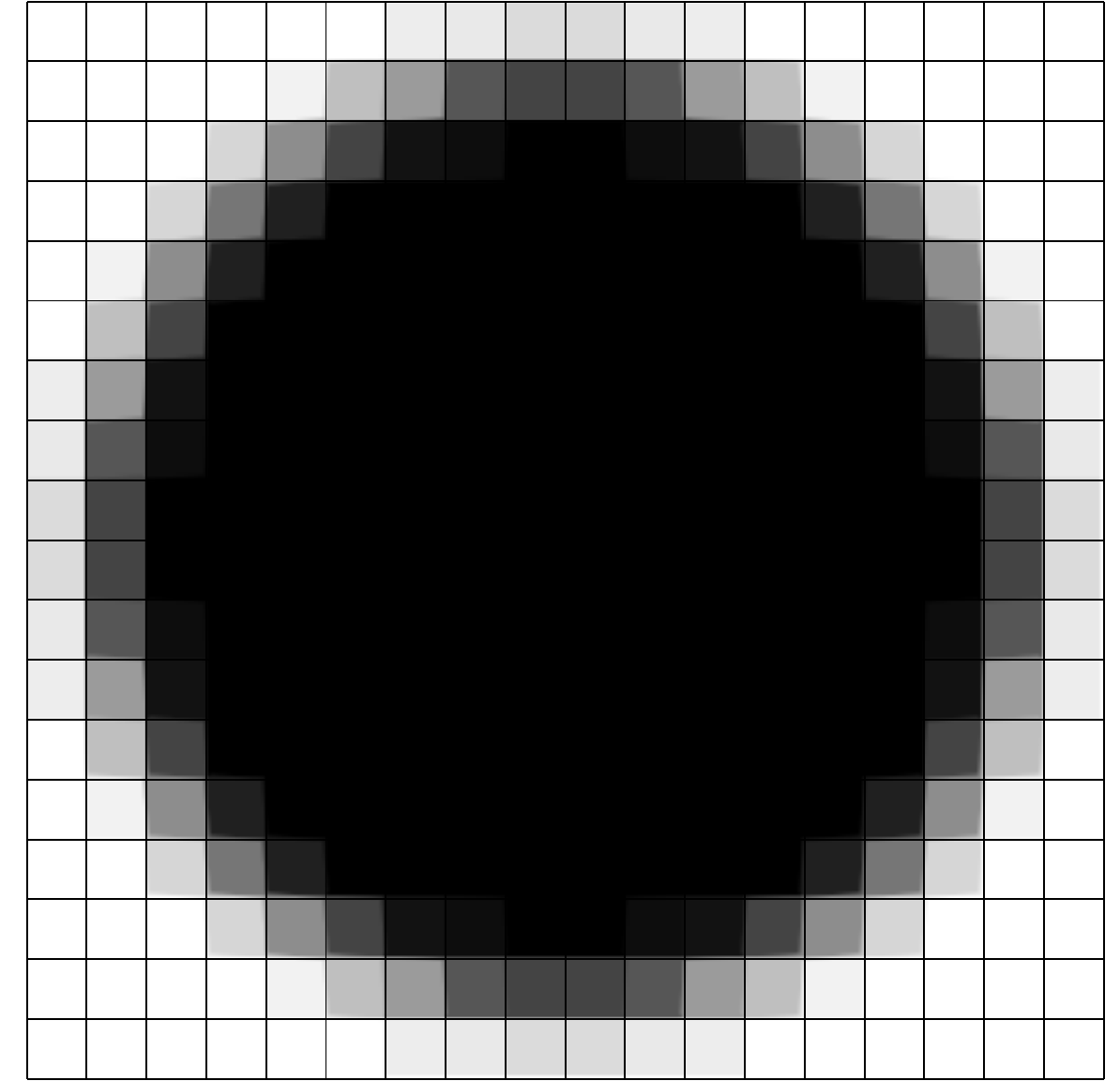}
     \vspace{-.6cm}
    \caption{Filtered image $\smoothIm_h^\sigma$~\\~}
    \label{fig:Gradient_Struct}
    \end{subfigure}
   \begin{subfigure}{0.23\textwidth}
 \includegraphics[width=\textwidth]{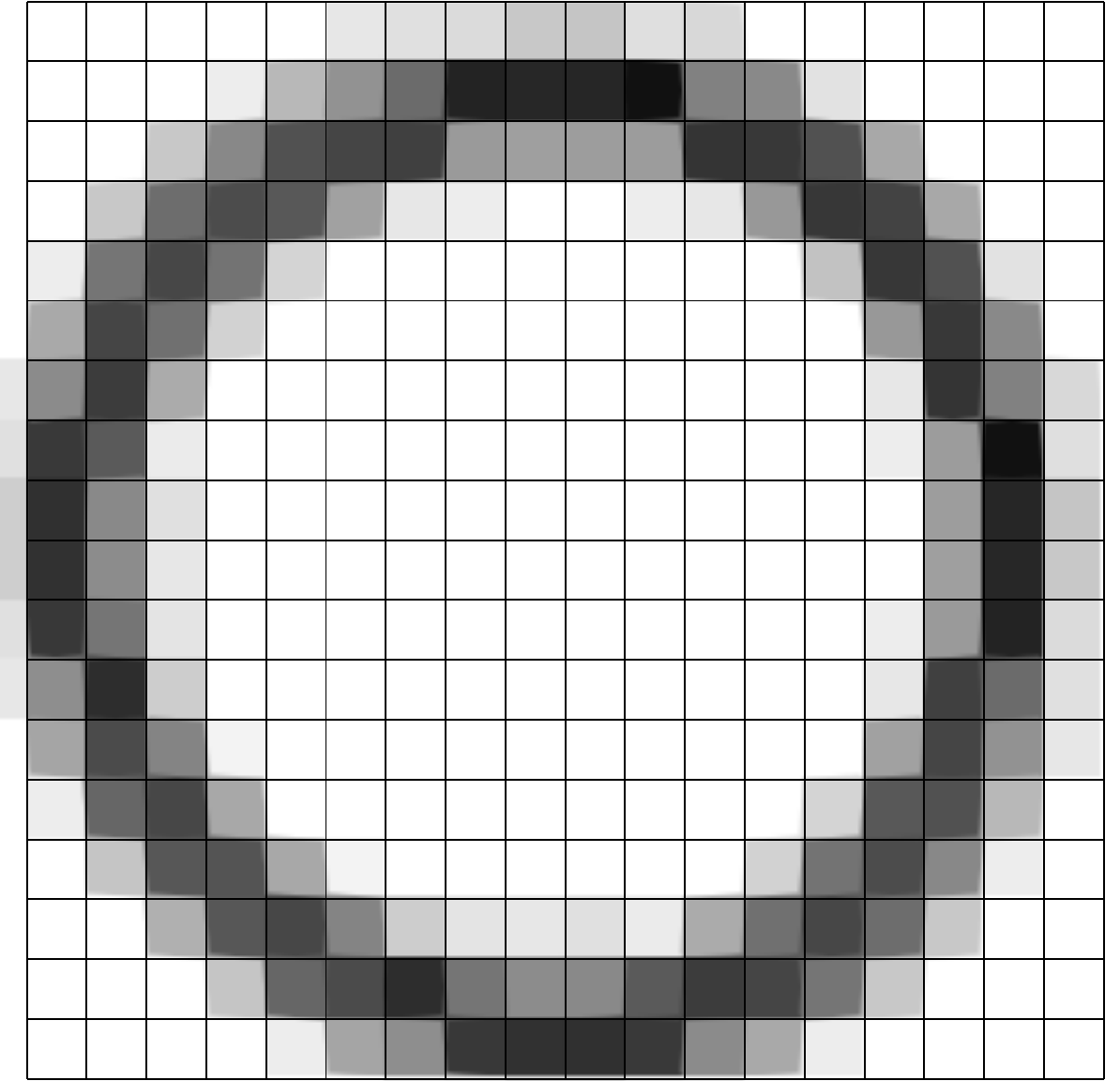}
      \vspace{-.6cm}
    \caption{$\|\nabla_h \smoothIm_h^\sigma\|$, computed via forward differences}
    \label{fig:Gradient_forward}
 \end{subfigure}
  \begin{subfigure}{0.23\textwidth}
 \includegraphics[width=\textwidth]{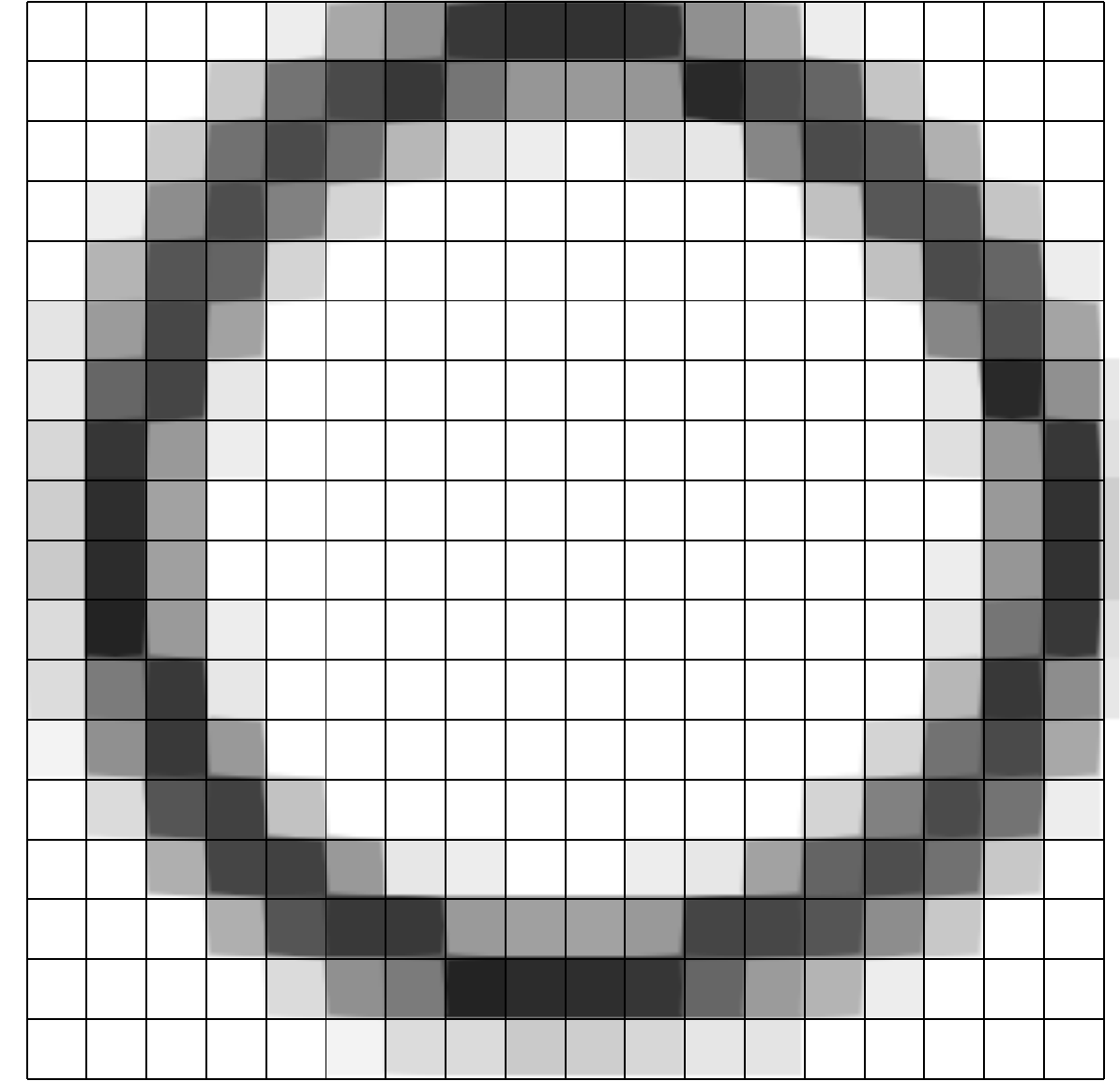}
      \vspace{-.6cm}
    \caption{$\|\nabla_h \smoothIm_h^\sigma\|$, computed via backward differences}
    \label{fig:Gradient_backward}
 \end{subfigure}
 \begin{subfigure}{0.23\textwidth}
 \includegraphics[width=\textwidth]{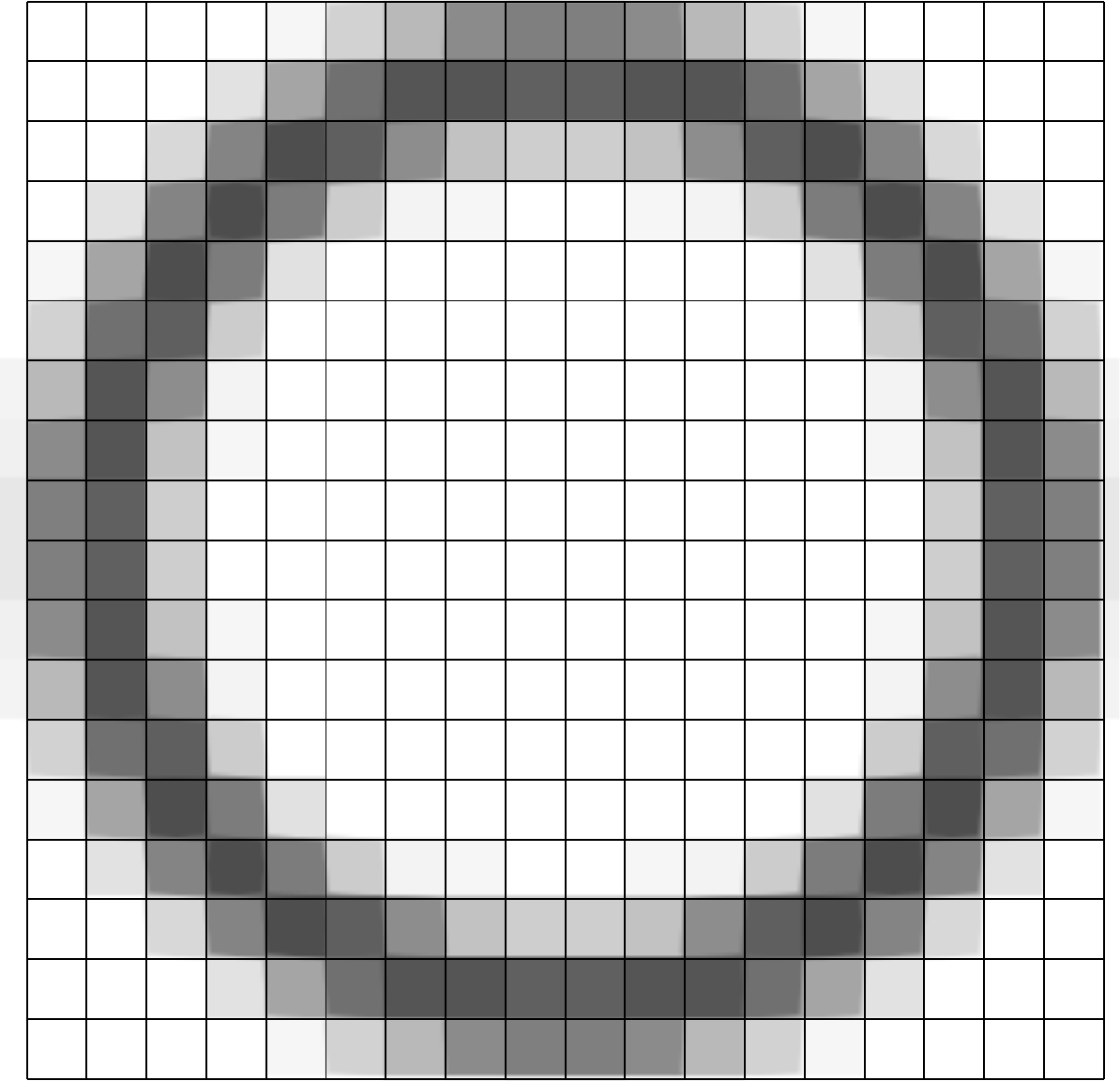}
     \vspace{-.6cm}
    \caption{$\|\nabla_h \smoothIm_h^\sigma\|$, computed via central differences}
    \label{fig:Gradient_central}
 \end{subfigure}
 \caption{Filtered gray-value image of a ball (a) and norm of the gradient computed via the three different  finite-difference approximations (b)-(d).}
 \label{fig:Gradient}
 \end{center}
 \end{figure}
 Since our numerical experiments are performed on periodic structures, we treat the boundary in a periodic fashion.
 Under certain regularity assumptions on the function to differentiate, the first-order approximations converge linearly in $h$ to the exact gradient, whereas the second order approximation converges quadratically as $h\rightarrow 0$, see Olver~\cite{PDE}. However, some further differences arise, which we demonstrate by example, see Fig.~\ref{fig:Gradient}.\\
 Consider the filtered gray-value image of a ball, shown in Fig.~\ref{fig:Gradient_Struct}. When computing the gradient via central differences, the symmetry of the structure is recovered in the symmetry of the gradient field, since $\smoothIm_h^\sigma(\rr)$ and $\n(\rr)$ are evaluated at the same position, cf. Fig¸~\ref{fig:Gradient_central}. However, if we compute the gradient via forward or backward differences, respectively, this will not be the case. The {forward or backward partial derivatives in direction $\e_i$ are not evaluated at $\rr$, but at $\rr \pm h/2 \e_i$, respectively, the faces of the cell. In particular, the partial derivatives in different directions will also be located on different faces}. The resulting gradient fields are shown in Fig.~\ref{fig:Gradient_backward} and Fig.~\ref{fig:Gradient_forward}, respectively. Both appear deformed and uneven compared to the central-differences approach. Furthermore, a diagonal offset is noticeable.
  Numerical tests show that, in our present setting, the gradient approximation based on central differences is more accurate than the other approaches and will therefore be preferred, see also Section \ref{sec:ParamSelect}.

 \subsection{Computing Minkowski tensors}\label{sec:quantities}
 In this section, we propose formulae for computing the volume (fraction) of $\setB$, the surface area of $\partial\setB$ and the Minkowski tensor $W^{0,2}_1(\setB)$. Their accuracy and multigrid convergence will be investigated by numerical means in Section~\ref{sec:ParamSelect}. \\
 The volume is approximated by quadrature, more precisely, by the trapezoidal rule, via
	\begin{align}\label{eq:volumeEstimation}
	V(\setB) \approx \sum_{\rr \in \Y_h}\chara_h(\rr) h^3.
	\end{align}
	{Motivated by results from geometric measure theory, cf. Giusti \cite{Giusti} (Defintion 1.6, Theorem 1.24 and Definition 3.3) and Maggi \cite{Maggi} (Proposition 12.20), we approximate the surface area via}
	\begin{align}\label{eq:surfaceEstimation}
	S(\setB)  \approx \sum_{\rr \in \Y_h}\|\g(\rr)\| h^3,
	\end{align}
	where the gradient $\g(\rr)$ is computed by finite differences and $||\cdot||$ denotes the Euclidean norm.
	{Due to the} relation $S(\setB) = 3\tr(W^{0,2}_1(\setB))$, we approximate the Minkowski tensor $W^{0,2}_1$ by
	\begin{align}\label{eq:W5Estimation}
	W^{0,2}_1(\setB) \approx \frac{1}{3} \sum_{\rr \in \Y_h} \g(\rr)\otimes \g(\rr)  \frac{h^3}{\|\g(\rr)\|+\epsilon},
	\end{align}
	where $\epsilon>0$ is a small constant used to avoid division by zero. The approximation of the {quadratic} normal tensor $\normedW$ is computed from the approximation of $W^{0,2}_1$ by dividing by the trace, as in its definition \eqref{eq:def_DNT}.

\section{Numerical examples}
\label{sec:Numerics}
 \subsection{Setup}
 The algorithms \ref{Alg:Minkowski} and  \ref{Alg:Filter} (as well as  algorithm~\ref{Alg:struct} discussed in Section~\ref{sec:4.3} below) were implemented in Python 3.7 with Cython~\cite{Cython} extensions. Critical operations were parallelized using OpenMP. For the eigenvalue decomposition of the structure tensor, discussed  in Section~\ref{sec:4.3} below, we rely on LAPACK~\cite{LAPACK}. The computations were performed on a desktop computer with a $6$-core Intel i$7$ CPU and $32$GB RAM.

 \subsection{Parameter selection and multigrid convergence}\label{sec:ParamSelect}
\begin{figure}[H]
\begin{center}
\begin{subfigure}{0.19\textwidth}
\centering
\includegraphics[width = \textwidth]{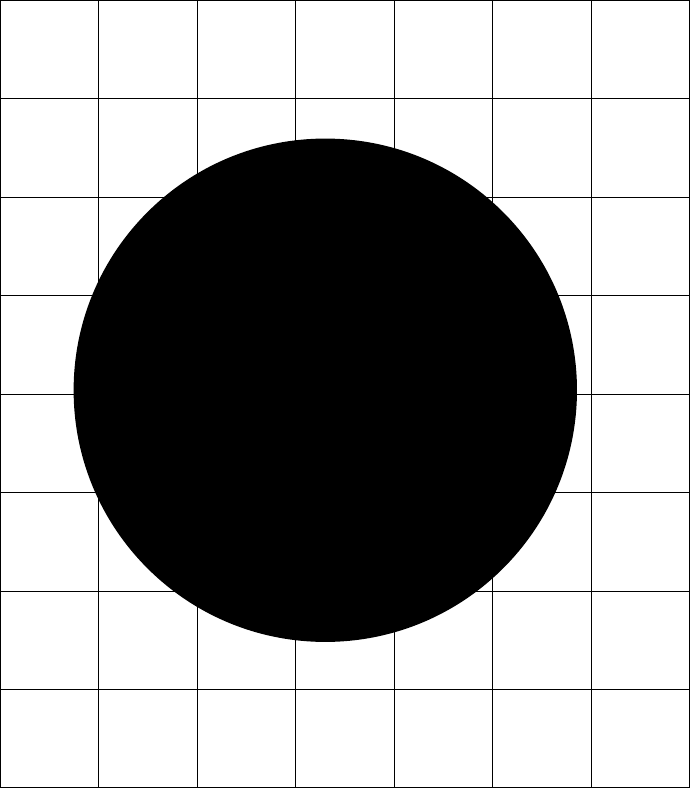}\caption{$\chara$ with grid}\label{fig:charaGrid}
\end{subfigure}
\begin{subfigure}{0.19\textwidth}
\centering
\includegraphics[width = \textwidth]{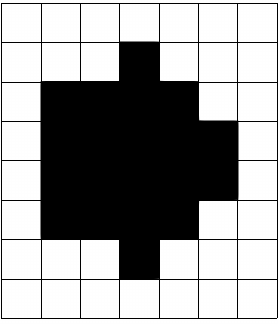}\caption{$\chara_h$ for $p=1$}\label{fig:chara1}
\end{subfigure}
\begin{subfigure}{0.19\textwidth}
\centering
\includegraphics[width = \textwidth]{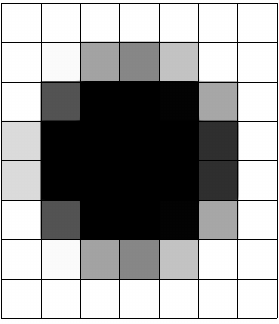}\caption{$\chara_h$ for $p=4$}\label{fig:chara4}
\end{subfigure}
\begin{subfigure}{0.19\textwidth}
\centering
\includegraphics[width = \textwidth]{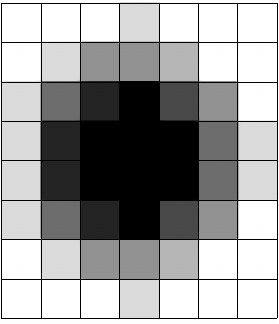}\caption{$\smoothIm_h^{1.2}$ for $p=1$}\label{fig:Filterchara1}
\end{subfigure}
\begin{subfigure}{0.19\textwidth}
\centering
\includegraphics[width = \textwidth]{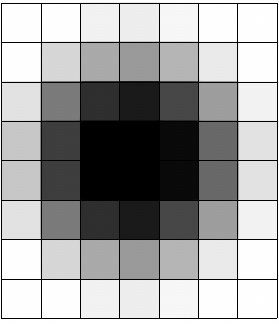}\caption{$\smoothIm_h^{1.2}$ for $p=4$}\label{fig:Filterchara4}
\end{subfigure}
\caption{Characteristic function $\chara$, discrete characteristic function $\chara_h$ and filtered image $\smoothIm_h^{1.2}$ for a single ball using depth $1$ and $4$.}\label{fig:filterDepthComparison}
\end{center}
\end{figure}

The proposed algorithm depends on several basic parameters including the grid size, the gray scale depth, the type and width of the applied filter and the choice of the gradient approximation, which we are free to choose in order to tune the algorithm.
 In this section, we investigate the influence of these parameters and propose suitable choices.\\
 In practical applications, the continuous range $[0,1]$ of gray values is replaced by a discrete set of colors {
 \[ 
\grayColor= 
 \begin{cases}
 \{0,1\}& \text{if}~\depth=1,\\
   \{0,\frac{1}{\depth^3-1},\frac{2}{\depth^3-1},\dots,1\}& \text{if}~\depth\geq 2
  \end{cases}
  \]
  of depth $\depth\geq 1$.} In this context, we consider the discrete characteristic function as the mapping $\chi_h:\Y_h\to \grayColor$.
For $p=1$, we voxelize the object under consideration in a binary manner, by colorizing a voxel if its center lies inside the object. For $p>1$, we compute the binary image on the finer grid $h'=h/p$ and determine the gray-value of a voxel of size $h$ as the mean value of its $p^3$ sub-voxels, resulting in a gray-value image of depth $p$.
We investigate a ball $B_R$ of radius $R>0$ for different $p$. The Minkowski quantities of $B_R$ are known exactly and given by
 \begin{align*}
 V(B_R) = \frac{4\pi R^3}{3}, \quad S(B_R) = 4\pi R^2, \quad W^{0,2}_1(B_R) = \frac{4\pi R^2}{9} \Id, \quad \normedW(B_R) = \frac{1}{3}\Id,
 \end{align*}
 see Appendix \ref{appendix:sphere} {for a derivation of the expressions for $W^{0,2}_1$ and $\normedW$}.
 Hence these quantities can be compared to the corresponding numerically determined quantities $V^{\approx}$, $S^{\approx}$, $W^\approx=W^{0,2,\approx}_1$ and $\normedW^{\approx}$.
 For $W=W^{0,2}_1$ and $\normedW$, we define error measures by
 \begin{align*}
 \tensorError = \frac{\left\|W(B_R) - W^{\approx}(B_R)\right\|}{\|W(B_R)\|} \quad\text{and}\quad \tensorDirError = \frac{\left\|\normedW(B_R) - \normedW^{\approx}(B_R)\right\|}{\left\|\normedW(B_R)\right\|},
 \end{align*}
 where $||\cdot||$ denotes the Frobenius norm.
 Since $W$ is connected to the surface area via $S = 3\tr(W)$, the error $\tensorError$ is directly affected by an error in computing $S$. In contrast, this latter error does not necessarily affect $\tensorDirError$, as both $\normedW$ and $\normedW^{\approx}$ have trace $1$.\\
  We investigate the influence of the different gradient approximations, filter kernels and filter widths $\sigma$, as well as that of the depth $p$ of the initial gray-value image, and we examine multigrid convergence as $h\rightarrow 0$ numerically.
The effect of the filters on the initial gray-value image, depending on the image depth, is exemplified in Fig.~\ref{fig:filterDepthComparison}. In Fig.~\ref{fig:charaGrid}, we see a slice through the characteristic function of a ball with a regular grid in the background.
In Fig.~\ref{fig:chara1} and \ref{fig:chara4}, we see slices of the discrete characteristic functions of the ball for depths $1$ and $4$. The center of the ball does not lie in the center of a voxel, but was chosen with a slight displacement, which results in a more uneven representation of the ball in the discrete images compared to Fig.~\ref{fig:chi}.\\
 Apparently, images with a higher depth give rise to a more accurate representation of a ball than binary images do. Fig.~\ref{fig:Filterchara1} and \ref{fig:Filterchara4} show the image after applying a ball-filter with $\sigma=1.2$. We see that the difference between depth $1$ and $4$ has become smaller, but remains visible. The filtered binary image ($p=1$) seems more uneven than the filtered gray-scale image ($p=4$).

\begin{figure}
\begin{center}
\begin{subfigure}{\linewidth}
\centering
\includegraphics[width = .5\textwidth]{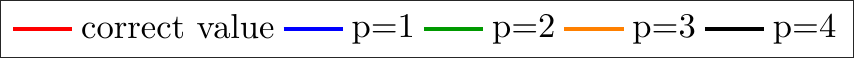}
\end{subfigure}
\begin{subfigure}{0.45\textwidth}
\centering
\includegraphics[width = \textwidth]{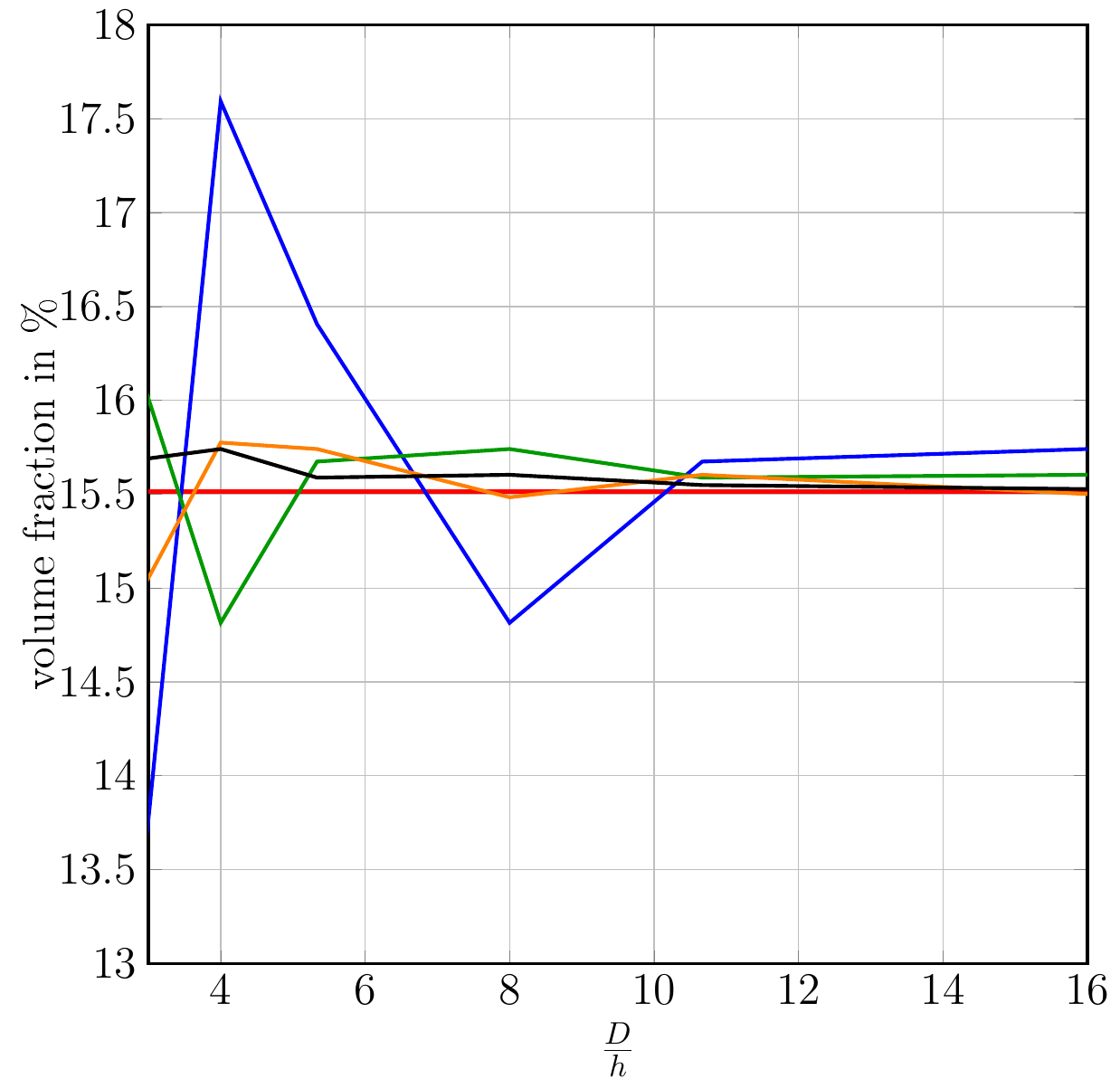}\caption{volume fraction}\label{fig:sphereVol1}
\end{subfigure}
\begin{subfigure}{0.45\textwidth}
\centering
\includegraphics[width = \textwidth]{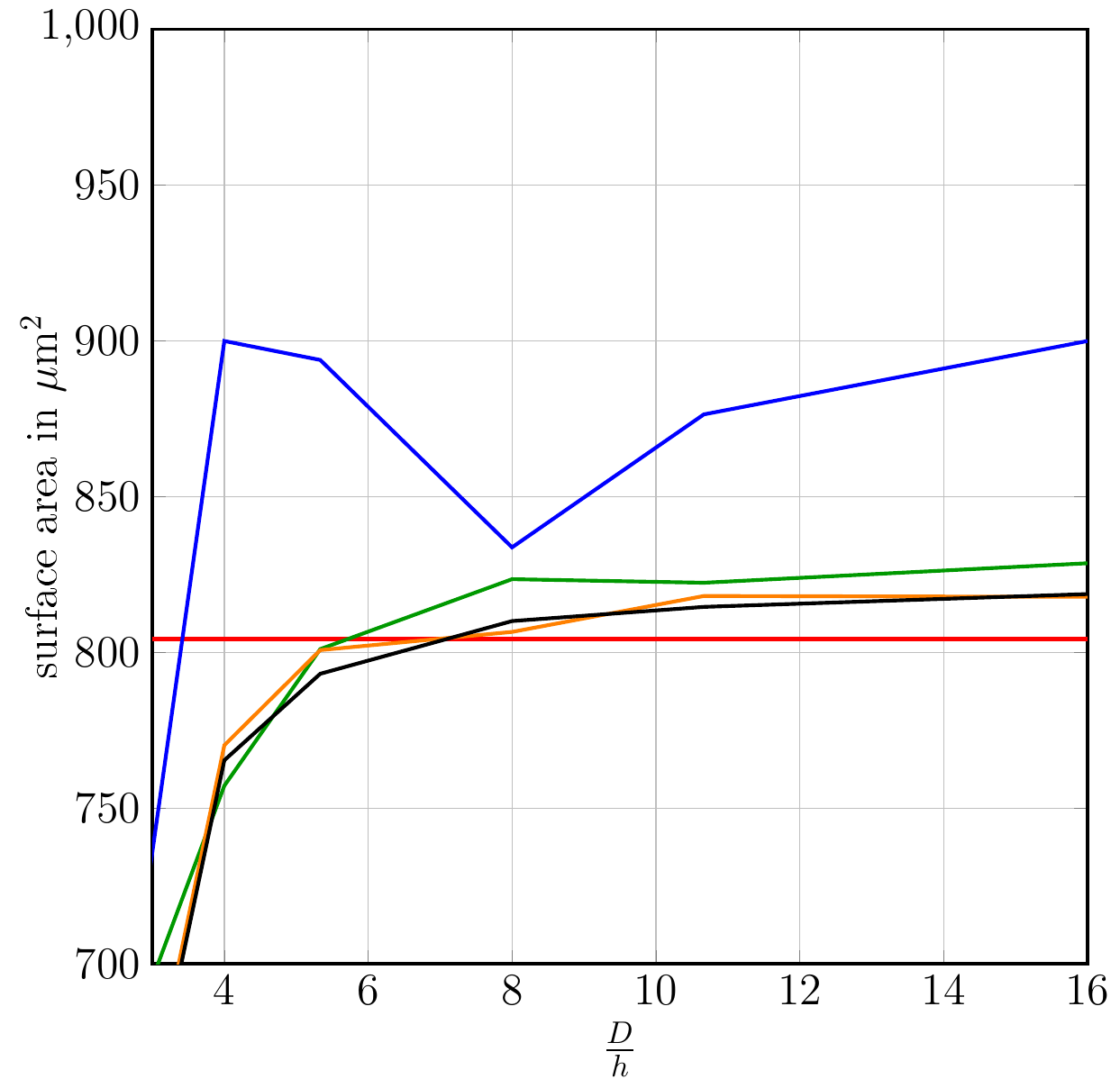}\caption{total surface area}\label{fig:sphereSurf1}
\end{subfigure}
\begin{subfigure}{0.45\textwidth}
\centering
\includegraphics[width = \textwidth]{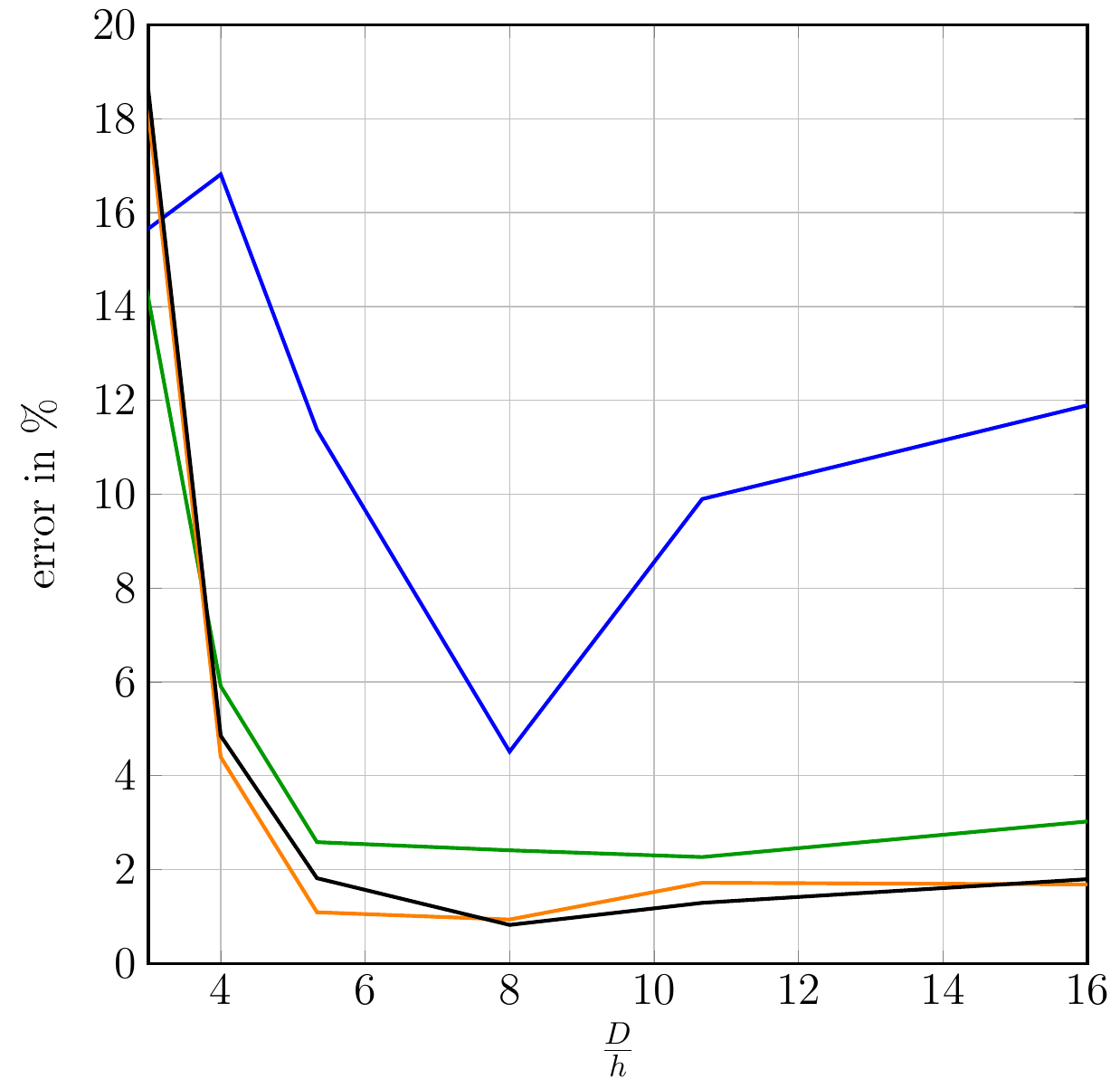}\caption{Minkowski tensor error $\tensorError$}\label{fig:sphereTens1}
\end{subfigure}
\begin{subfigure}{0.45\textwidth}
\centering
\includegraphics[width = \textwidth]{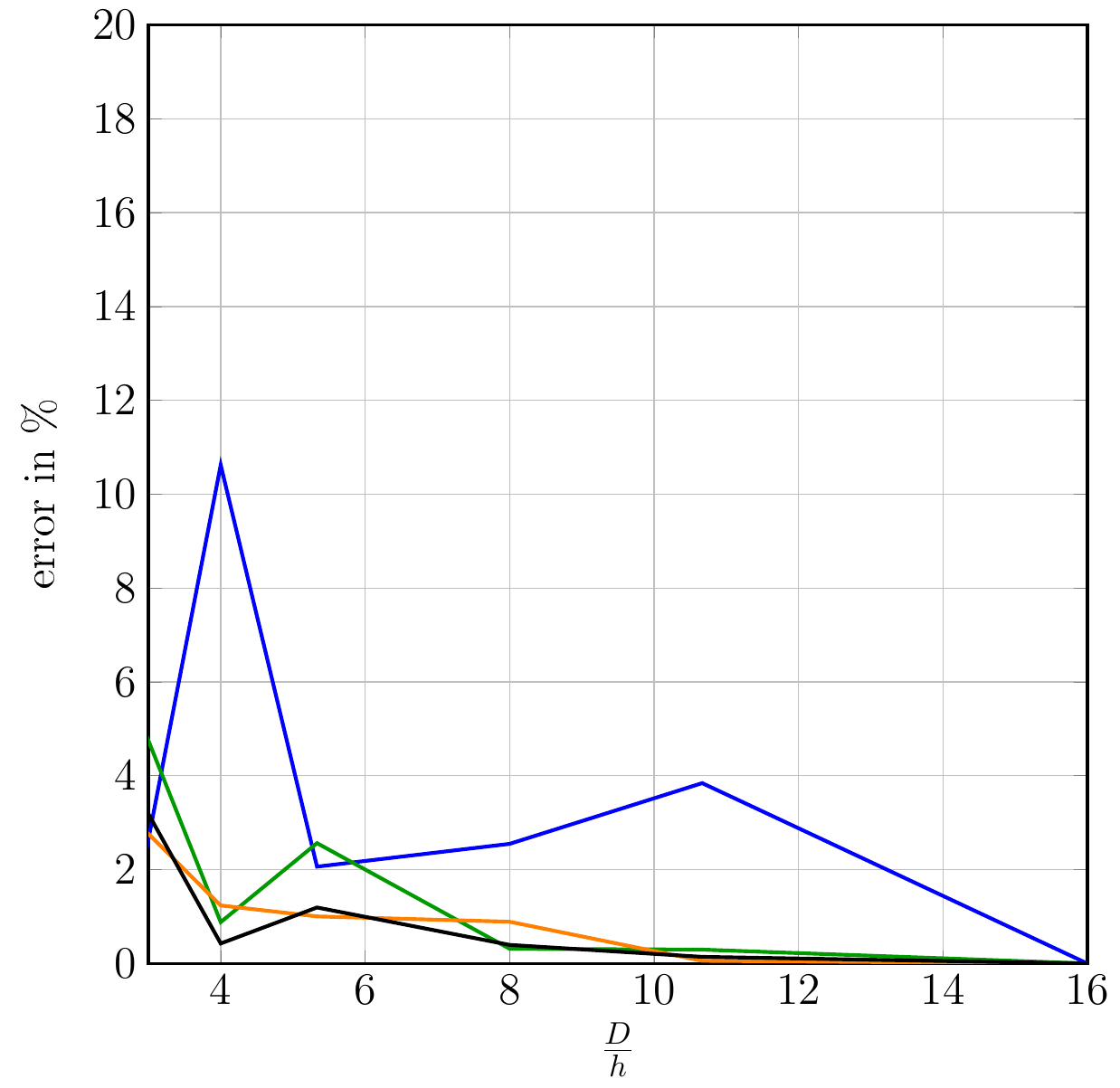}\caption{{QNT} error $\tensorDirError$}\label{fig:sphereTensDir1}
\end{subfigure}
\caption{Volume fraction, total surface area and tensor errors $\tensorError$ and $\tensorDirError$ for the unfiltered image, i.e., $\sigma=0$. The gradient was computed via central differences.}\label{fig:sphere1}
\end{center}
\end{figure}

First, we study the influence of the gray-value depth $p$ of the initial voxel image for different spatial resolutions. The structure under consideration contains a single ball of diameter $16\unit{\mu m}$ in a box of edge length $24\unit{\mu m}$, i.e., the material has a volume fraction of $15.5\unit{\%}$.
For this first study, we omit using a filter and rely on central differences for the gradient estimation.\\ Fig.~\ref{fig:sphereVol1} shows the computed volume fraction vs.\ $D/h$, the diameter of the ball per voxel length, for several gray-value depths $p$. The binary image, i.e., $p=1$, exhibits the largest error and oscillates around the correct value. Only for a high resolution above $D/h=10$, the error is within reasonable bounds. For a higher depth, the volume fraction is accurate even for the lowest resolution.\\
The computed surface area vs. $D/h$ is shown in Fig.~\ref{fig:sphereSurf1}. Using the binary image without any filter overestimates the surface area significantly and does not converge. For $p\geq 2$, we see that the error is reasonable for a resolution of $4$ voxels per diameter and higher. For higher resolution and higher depth, the surface-area computation is rather accurate, but systematically overestimates the correct value by about $2\%$ and does not converge. The error $\tensorError$ of the Minkowski tensor is shown in Fig.~\ref{fig:sphereTens1}. For $p\geq 2$, it is below $6\%$ even for the second-coarsest resolution of $D/h=4$ and stays below $3\%$ at higher resolutions. Finally, the {quadratic} normal tensor $\normedW$ is the one among the computed characteristics which is computed most accurately, see Fig.~\ref{fig:sphereTensDir1}. For $p\geq 2$, the error is below $5\%$ for all spatial resolutions. Additionally, for all image depths, multigrid convergence is visible. This suggests that, to some degree, the error of computing the Minkowski tensor results from the mentioned overestimation of the surface area. Indeed, since $\normedW$ differs from $W^{0,2}_1$ by its trace and $\tr(W^{0,2}_1)=S/3$, the error of computing the surface area present in $W^{0,2}_1$ cancels out to some extent in $\normedW$.

In a second series of numerical experiments we repeated large parts of the above tests using first-order gradient approximations, as described in Section~\ref{sec:3.3}, instead of central differences.  Compared to the latter, both first-order gradient approximations induce much larger errors, exceeding $20\%$. Therefore, we will restrict to central differences for the remainder of the article.

\begin{figure}
\begin{center}
\begin{subfigure}{\textwidth}
\centering
\includegraphics[width = .6\textwidth]{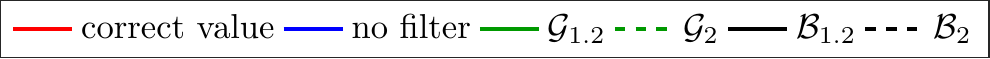}
\end{subfigure}
\begin{subfigure}{0.45\textwidth}
\centering
\includegraphics[width = \textwidth]{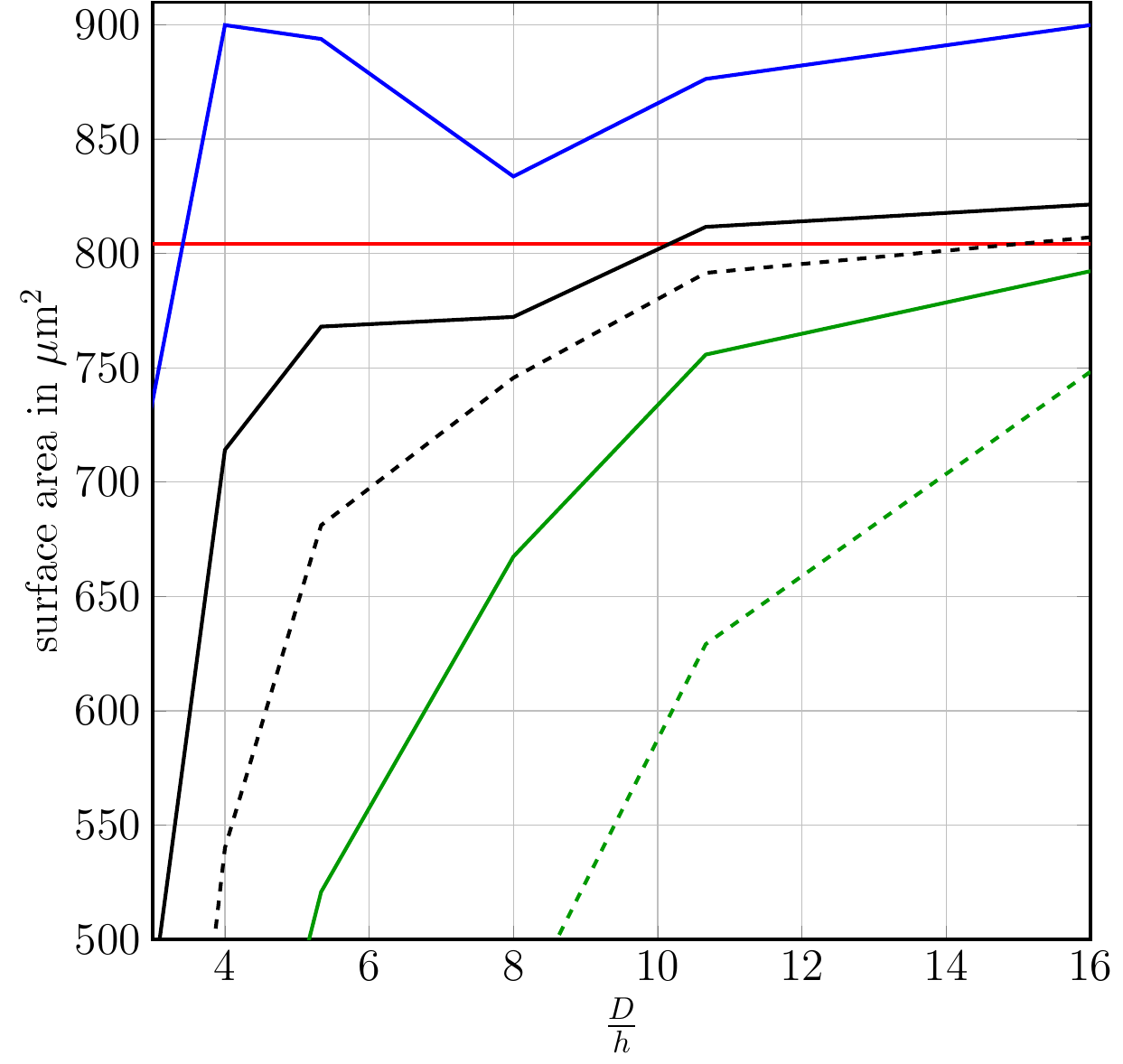}\vspace{-.2cm}\caption{total surface area for $p=1$}\label{fig:sphereSurf2_1}
\end{subfigure}
\begin{subfigure}{0.45\textwidth}
\centering
\includegraphics[width = \textwidth]{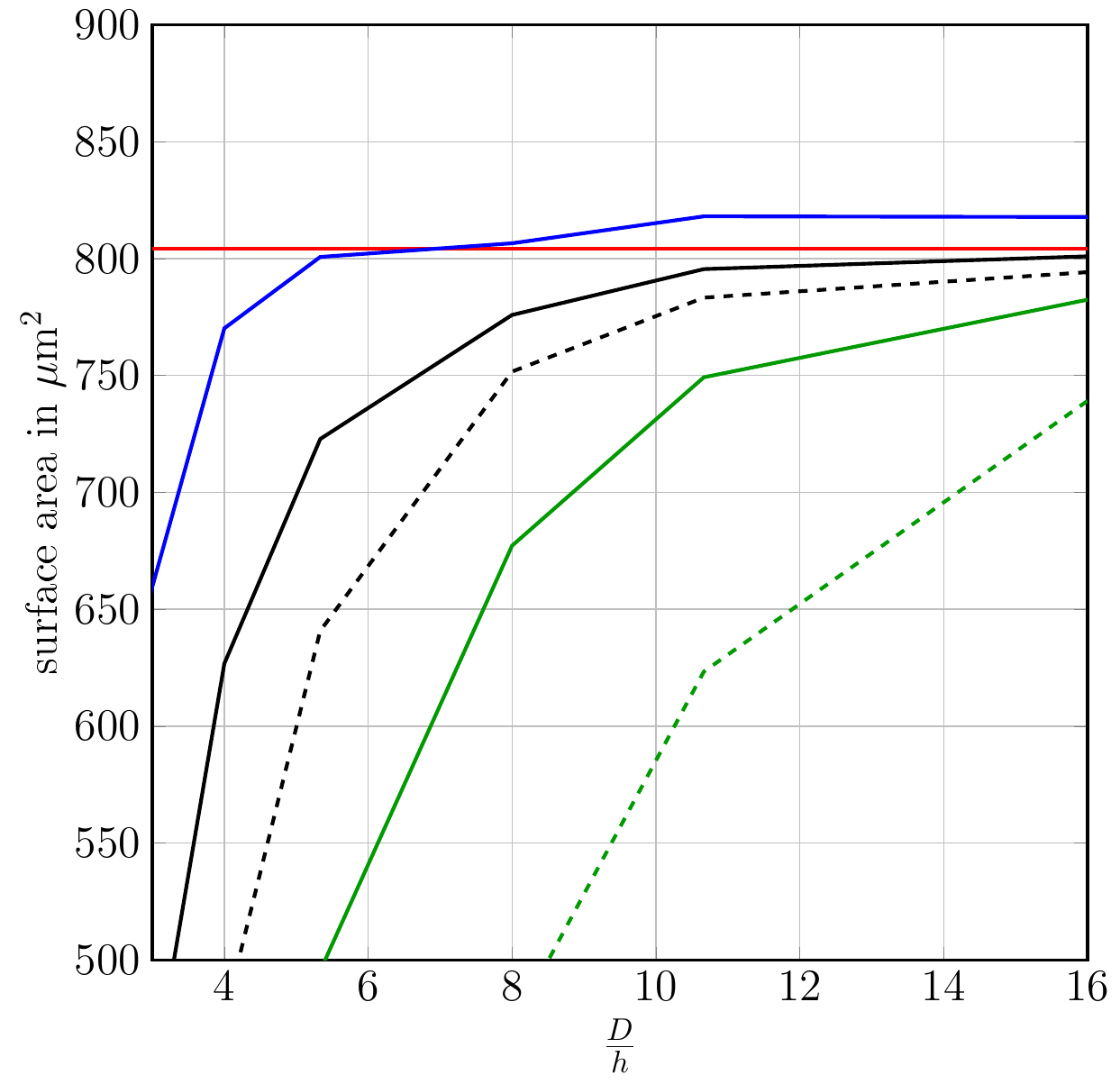}\vspace{-.2cm}\caption{total surface area for $p=3$}\label{fig:sphereSurf2_3}
\end{subfigure}
\begin{subfigure}{0.45\textwidth}
\centering
\includegraphics[width = \textwidth]{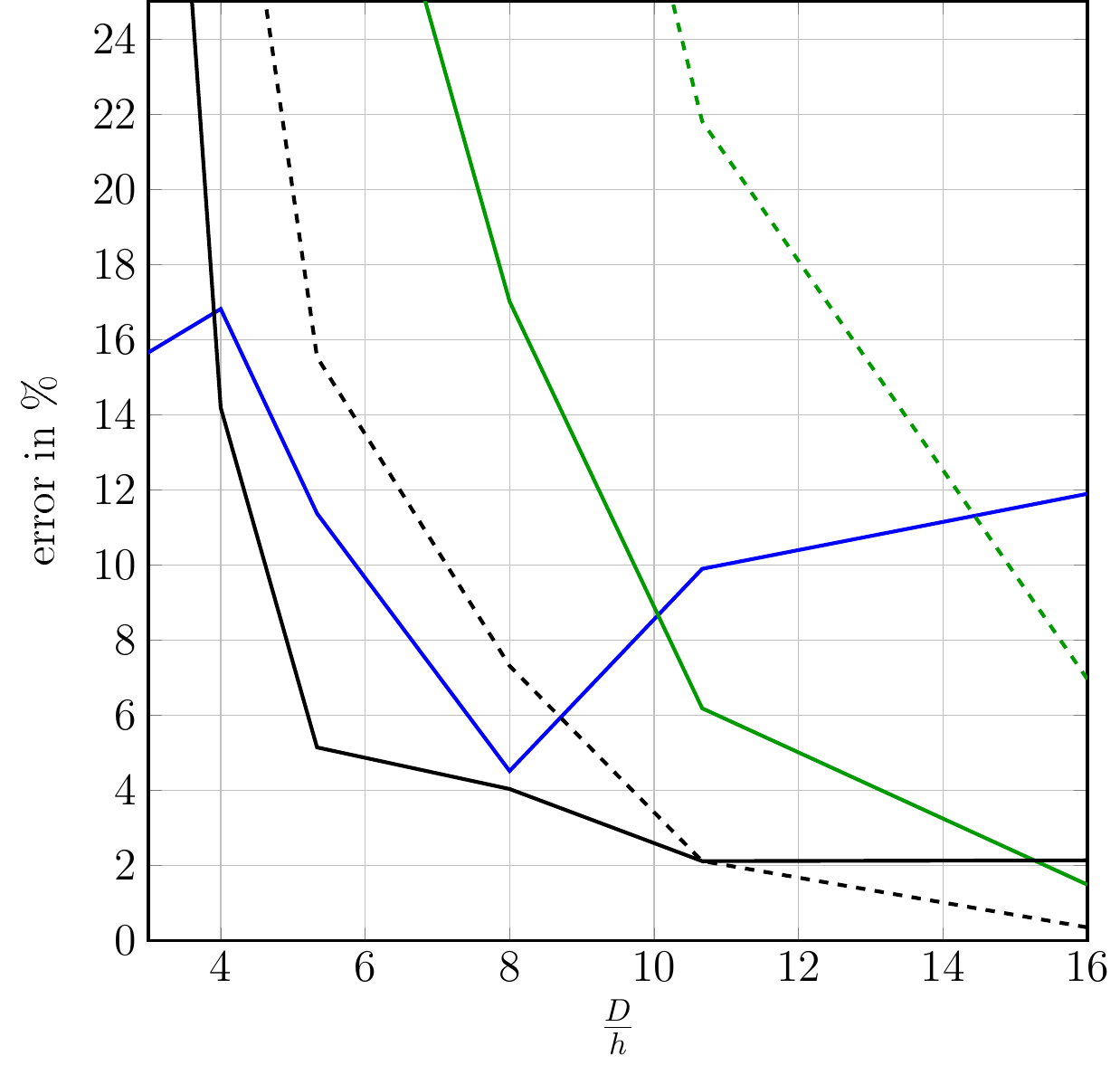}\vspace{-.2cm}\caption{Minkowski tensor error $\tensorError$ for $p=1$}\label{fig:sphereTens2_1}
\end{subfigure}
\begin{subfigure}{0.45\textwidth}
\centering
\includegraphics[width = \textwidth]{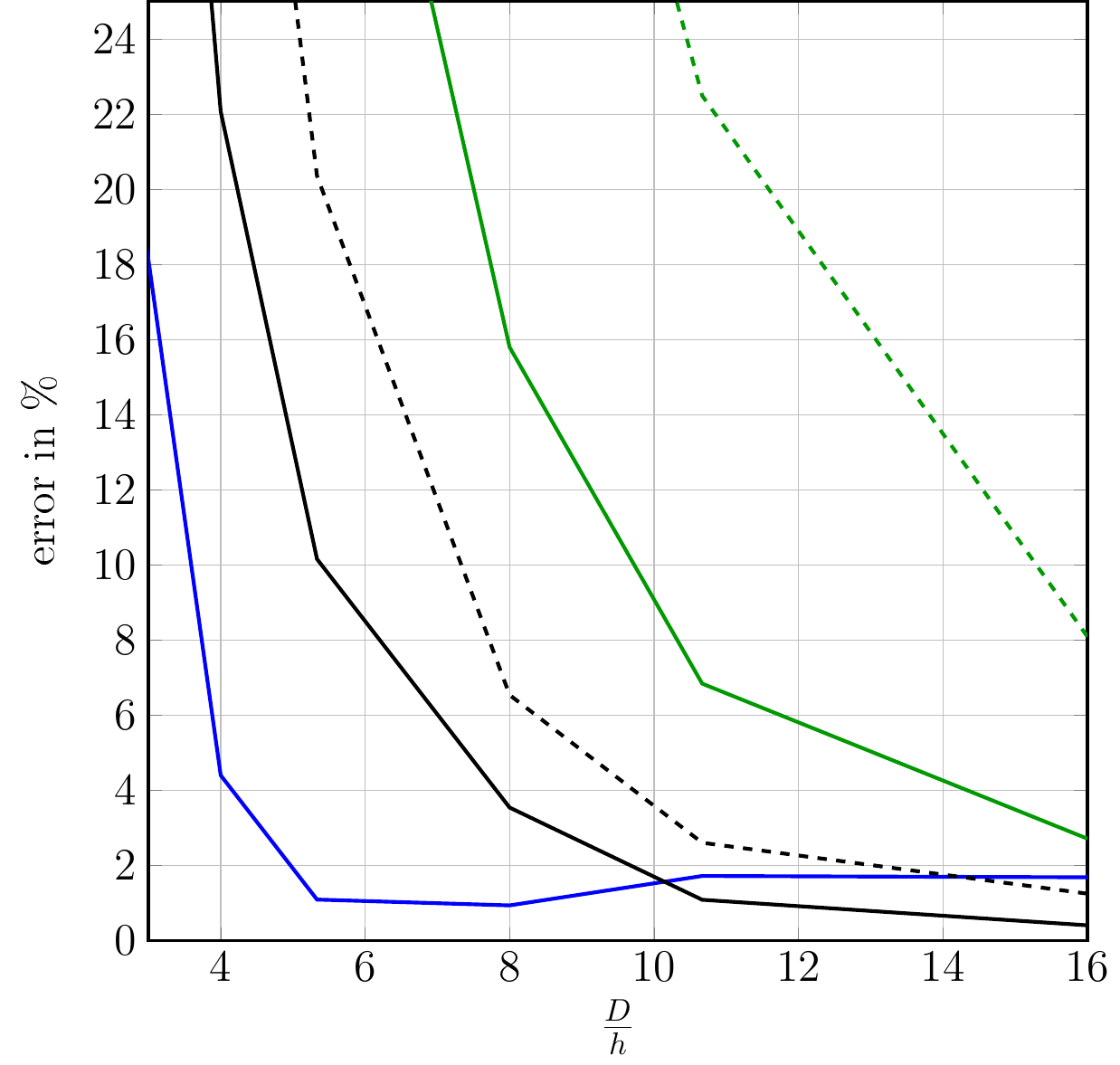}\vspace{-.2cm}\caption{Minkowski tensor error $\tensorError$ for $p=3$}\label{fig:sphereTens2_3}
\end{subfigure}
\begin{subfigure}{0.45\textwidth}
\centering
\includegraphics[width = \textwidth]{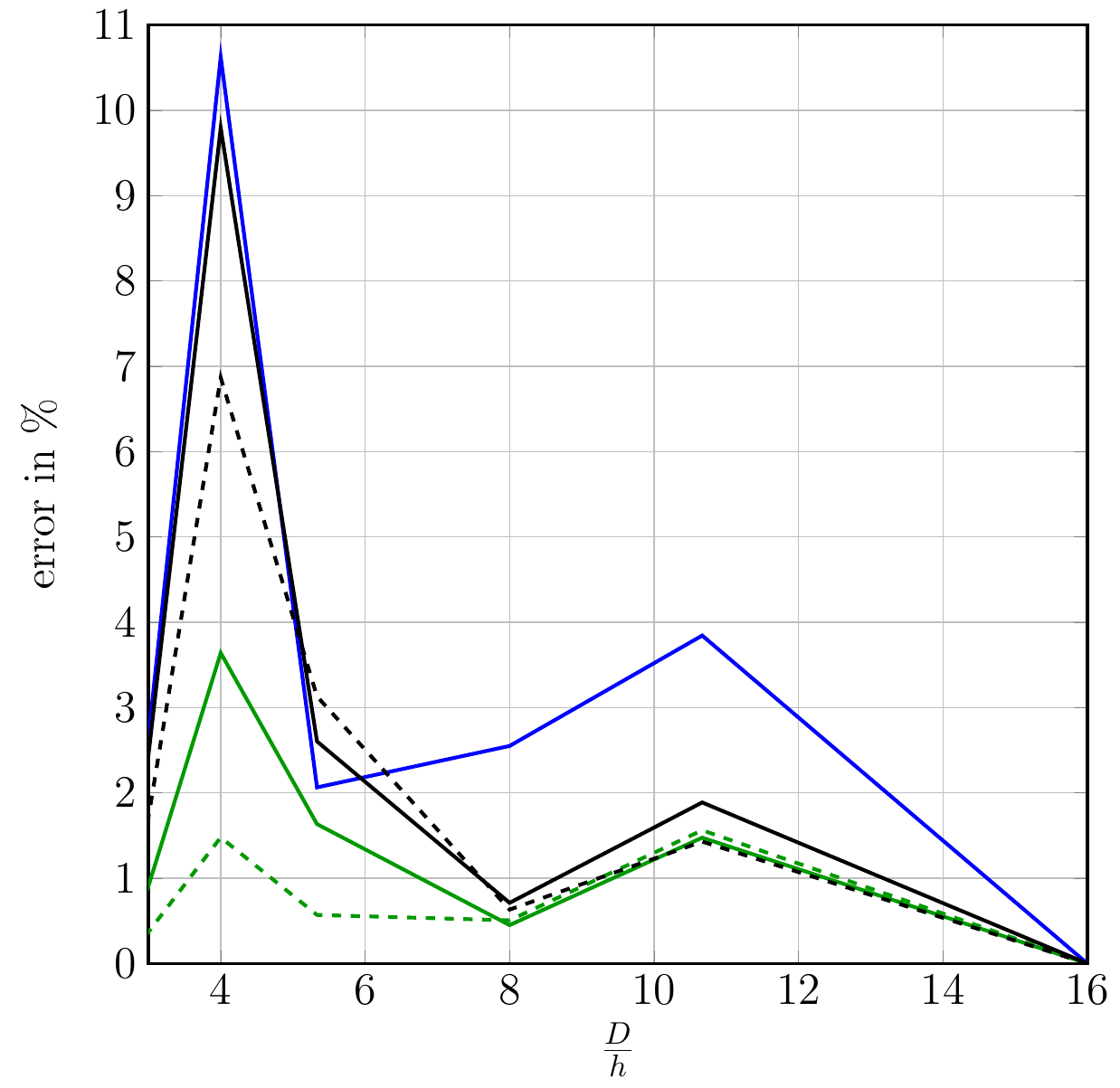}\vspace{-.2cm}\caption{{QNT} error $\tensorDirError$ for $p=1$}\label{fig:sphereTensDir2_1}
\end{subfigure}
\begin{subfigure}{0.45\textwidth}
\centering
\includegraphics[width = \textwidth]{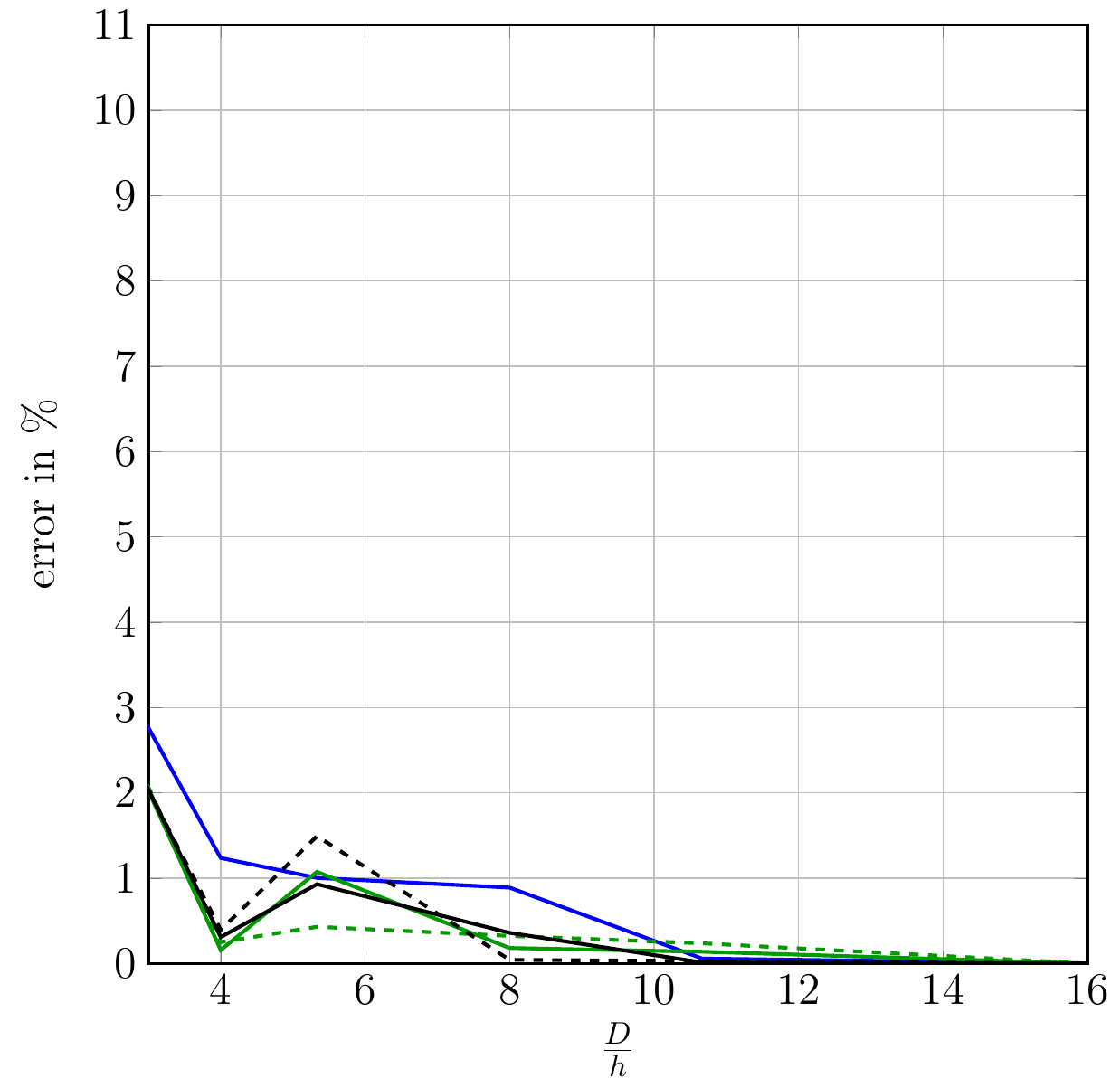}\vspace{-.2cm}\caption{{QNT} error $\tensorDirError$ for $p=3$}\label{fig:sphereTensDir2_3}
\end{subfigure}
\caption{Surface area and tensor errors $\tensorError$ and $\tensorDirError$ for depths $p=1$ and $p=3$ comparing the filter choice.}\label{fig:sphere2}
\end{center}
\end{figure}

Finally, we examine the influence of different filter kernels. Fig.~\ref{fig:sphere2} shows the surface area as well as the two tensor-error measures vs. $D/h$ for gray-value depth $p=1$ (binary) on the left and $p=3$ on the right. We consider the ball filter $\ballFilter$ and the Gaussian filter $\gaussFilter$, both with filter parameters $\sigma=1.2$ and $\sigma=2$, i.e., for a filter width slightly larger than a single voxel and a filter width of $2$ voxels.\\
 In general, the errors for the gray-value image are smaller compared to the binary image. Focusing on the surface-area computation, i.e., Fig.~\ref{fig:sphereSurf2_1} and Fig.~\ref{fig:sphereSurf2_3}, we notice that applying no filter is actually most beneficial for a low spatial resolution. For $p=3$, this even holds up to $D/h=10$. For $p=1$, the surface area is strongly overestimated for higher resolution. Even for $p=3$, no multigrid convergence is achieved, if the filtering step is skipped. To achieve convergence, the ball filter with $\sigma=1.2$ is the most accurate. For $p=1$, the ball filter with $\sigma=1.2$ appears to be the best choice for resolutions up to $D/h=10$. Above that threshold, the choice $\sigma=2$ exhibits the smallest error. Nevertheless, the ball filter with $\sigma = 1.2$ serves as a good compromise. {For both gray-image depths}, applying the ball filter leads to better results than applying the Gaussian filter for computing the surface area of the structure.\\
Investigating the error $\tensorError$, see Fig.~\ref{fig:sphereTens2_1} and Fig.~\ref{fig:sphereTens2_3}, permits us to draw similar conclusions. Fig.~\ref{fig:sphereTensDir2_1} and Fig.~\ref{fig:sphereTensDir2_3} show that the filter choice plays a subordinate role compared to the image depth for the error $\tensorDirError$. For the binary image, the Gaussian filter with $\sigma=2$ exhibits the lowest error, staying below the threshold of $2\%$ for all resolutions. For gray-value images, however, the  error of computing the {quadratic} normal tensor is below $3\%$ for all resolutions and filters, which is accurate enough for most applications.

 \subsection{A short-fiber reinforced composite} \label{sec:4.3}
 \subsubsection{Characterization of fiber-reinforced composites}
 Short-fiber reinforced composites enjoy great popularity owing to their high (mass-)specific stiffness~\cite{Jones}. The local fiber alignment is strongly dependent on the manufacturing process~\cite{ChungKwon}. The effective material behavior of short-fiber reinforced composites is anisotropic, in general, and strongly dependent on the local fiber orientation. Each fiber is interpreted as a straight {spherical} cylinder of length $L$ and diameter $D$, axis-aligned with unit vector {$\p$}.
 Frequently used microstructure characteristics for fiber-reinforced composite materials are the volume fraction, the aspect ratio $L/D$ and the fiber-orientation tensors of second or fourth order~\cite{Kanatani1984,AdvaniTucker}. For fibers of equal length and equal diameter, the resulting fiber-orientation tensors {(of order 2 and 4)} of a structure with $N$ fibers and orientation vectors {$\p_1,\dots,\p_N$} are defined by
 \begin{align*}
 {A = \frac{1}{N}\sum_{i=1}^N \p_i \otimes \p_i \quad \text{and} \quad \mathbb{A} = \frac{1}{N}\sum_{i=1}^N \p_i \otimes \p_i \otimes \p_i \otimes \p_i.}
 \end{align*}
For varying fiber length and diameter, similar expressions have been proposed in Bay-Tucker~\cite{BayTucker} based on length- or volume-weighted averaging.\\
 For a gray-value $\mu$-CT image, the fiber-orientation tensors of second and fourth order may be computed by a variety of methods, see Pinter et al.~\cite{Pinter2018}. A popular approach uses the structure tensor~\cite{StructureTensor}, cf.\ Alg.~\ref{Alg:struct}.
Alternatively, fibers may be segmented individually, see Hessmann et al.~\cite{Hessmann2019} for recent work.\\
To gain insight into the relation between the fiber-orientation tensor $A$ and the Minkowski tensor $W^{0,2}_1$, we compare their expressions for a single fiber of length $L$ and diameter $D$, oriented in direction {$\p$}, see Appendix \ref{appendix:cylinder} for the detailed computation:
\begin{align}\label{eq:fiberTensor1}
A &= \p\otimes \p, \\ \label{eq:fiberTensor2}
W^{0,2}_1 &=  \frac{\pi D^2}{6} \bigg[ \p\otimes \p + \frac{L}{D} \bigg( \Id - \p\otimes \p\bigg)\bigg] \quad \text{and}\\ \label{eq:fiberTensor3}
\normedW &= \frac{1}{1 + 2\frac{L}{D}}\bigg[\p\otimes \p + \frac{L}{D} \bigg( \Id - \p\otimes \p\bigg)\bigg].
\end{align}
For microstructures containing $N$ fibers, $A$ is computed by averaging the single-fiber expression \eqref{eq:fiberTensor1}. $W^{0,2}_1$ is computed  by summing \eqref{eq:fiberTensor2} over all fibers. The resulting quadratic normal tensor $\normedW$ may be computed as a surface-area weighted average of expression \eqref{eq:fiberTensor3}.\\
{The fiber-orientation tensor and the {quadratic} normal tensor} need to be interpreted differently:
\begin{itemize}
\item For a single fiber $K$, the fiber-orientation tensor of second order is a singular matrix (of rank $1$) describing the projection onto the fiber axis. In contrast, the Minkowski tensor $W^{0,2}_1(K)$ of a single fiber $K$ is a full rank matrix, which arises as a weighted sum {of the orthogonal projection onto the fiber axis and the complementary projection onto the plane perpendicular to this axis}.
\item For high aspect ratios, i.e., for $L\gg D$, {for the} {QNT}, the {prefactor in front of the complementary projection} is much {larger} than the {other prefactor}.
\item Using the fiber-orientation tensor as {a} descriptor of a microstructure {rests upon} specific assumptions that are often not satisfied for real structures. Typically, fibers are not of equal length, because they break during to the manufacturing process~\cite{FiberBreakage}. Furthermore, the assumption that fibers are straight cylinders is not met in most of the cases, as longer fibers bend during manufacturing and therefore exhibit curvature~\cite{FiberBending}. {In such situations, the structure-tensor based computation of the fiber-orientation tensor} still gives \emph{some} tensorial quantity as output. However, interpreting this result as {a} fiber-orientation tensor may not be justified.\\
The Minkowski tensors, on the other hand, are not restricted to specific geometric assumptions such as particular shapes. Therefore, for structures containing curved fibers of different lengths or mixtures of fibers with other objects etc., $W^{0,2}_1$ is still a geometrically well-defined quantity. {As Minkowski tensors are integrals of locally computable quantities, cf. \eqref{eq:MinkowskiTensors}, they are even well-defined locally on any piece of a {complex geometric structure}. In contrast, the fiber orientation tensor is a non-local quantity intrinsically tied to cylindrical shapes.} 
\end{itemize}

\begin{table}
\resizebox{\textwidth}{!}{
\begin{tabular}{|c|c|c|c|c|}
\hline
&$A$&$\normedW ~\text{for}~\frac{L}{D}=10$&$\normedW ~\text{for}~\frac{L}{D}=25$&$\normedW ~\text{for}~\frac{L}{D}=50$\\
\hline
$\# 1$&$\sv 1&0&0\\0&0&0\\0&0&0\ev$& $\sv0.048 &0 &0 \\0 &0.476 &0 \\0 &0 &0.476 \ev$& $\sv0.025 &0 &0 \\0 &0.4875 &0 \\0 &0 &0.4875 \ev$&$\sv0.012 &0 &0 \\0 &0.494 &0 \\0 &0 &0.494 \ev$\\
\hline
$\# 2$&$\sv 0.79&0&0\\0&0.19&0\\0&0&0.02\ev$&$\sv0.1379 &0 &0 \\0 &0.3946 &0 \\0 &0 &0.4675 \ev$&$\sv0.1219 &0 &0 \\0 &0.3996 &0 \\0 &0 &0.4785 \ev$&$\sv0.1131 &0 &0 \\0 &0.4024 &0 \\0 &0 &0.4845 \ev$\\
\hline
$\# 3$&$\sv 0.49&0&0\\0&0.49&0\\0&0&0.02\ev$&$\sv0.266 &0 &0 \\0 &0.266 &0 \\0 &0 &0.468 \ev$&$\sv0.2607 &0 &0 \\0 &0.2607 &0 \\0 &0 &0.4785 \ev$&$\sv0.258 &0 &0 \\0 &0.258 &0 \\0 &0 &0.484 \ev$\\
\hline
$\# 4$&$\sv 0.6&0&0\\0&0.3&0\\0&0&0.1\ev$&$\sv0.219 &0 &0 \\0 &0.348 &0 \\0 &0 &0.433 \ev$&$\sv0.21 &0 &0 \\0 &0.349 &0 \\0 &0 &0.441 \ev$&$\sv0.205 &0 &0 \\0 &0.349 &0 \\0 &0 &0.446 \ev$\\
\hline
$\# 5$&$\sv 0.33&0&0\\0&0.33&0\\0&0&0.33\ev$&$\sv0.333 &0 &0 \\0 &0.335 &0 \\0 &0 &0.332 \ev$&$\sv0.333 &0 &0 \\0 &0.335 &0 \\0 &0 &0.332 \ev$&$\sv0.333 &0 &0 \\0 &0.335 &0 \\0 &0 &0.332 \ev$\\
\hline
\end{tabular}}
\caption{Comparison of fiber-orientation tensor $A$ and {quadratic} normal tensor {QNT} for microstructures of different orientation and aspect ratio.}
\label{tab:orientationComparison}
\end{table}
An overview of how the fiber-orientation tensor {compares with} the {quadratic} normal tensor for varying aspect ratios is given in Tab.~\ref{tab:orientationComparison}. For this study, we generated $5\times3$ different microstructures, each containing $20\%$ fibers of equal length and diameter, using the sequential addition and migration algorithm~\cite{SAM}. This algorithm draws fibers from an angular central Gaussian {distributions} on the two-dimensional sphere~\cite{ACG}. Indeed, the set of possible angular central Gaussian {distributions} may be parameterized by the second-order fiber-orientation tensors, see Montgomery-Smith et al.~\cite{closure}.\\
Across the microstructures we varied the orientation distribution (5 different ones $\#1-\#5$) and the aspect ratio (3 different choices: $L/D=10,25$ and $50$). For convenience, all matrices are chosen to be diagonal w.r.t.\ the {standard basis $\{\e_1, \e_2, \e_3\}$}.\\
Microstructure $\#1$ {is composed of aligned fibers} in $e_1$-direction. The second microstructure {lies} almost {entirely within} the {$\e_1-\e_2$}-plane, with preferred direction {$\e_1$}. The almost planar-isotropic case in the {$\e_1-\e_2$}-plane is realized via microstructure $\#3$. A general anisotropic case with preferred direction {$\e_1$} and least preferred direction {$\e_3$} is given in case $\#4$. And, finally, microstructure $\#5$ {shows} the isotropic case.
For {the} isotropic orientation ($\#5$), all tensors are nearly equal. {The} $\normedW$ for the almost planar orientation ($\#3$) results in one larger ({corresponding} to the normal {vector} of the plane) and two equal smaller eigenvalues, indicating no preferred direction within the plane. {The} $\normedW$ of {structure} $\#2$ exhibits three different eigenvalues. The largest is equal to the largest eigenvalue of $\#3$. Of the two smaller eigenvalues, the smallest indicates a preferred direction. The same interpretation holds for {structure} $\#4$. For the uni-directional case ($\#1$), the largest eigenvalue appears twice, which indicates a planar symmetry in both planes normal to the corresponding eigenvectors. By the smallest eigenvalue, again a preferred direction is indicated.\\
In contrast to the fiber orientation tensor $A$, the {quadratic} normal tensor varies also with the {aspect} ratio $L/D$ of the fibers. This may also be seen from the eigenvalue ratio $\beta$ of $\normedW$, cf.~\eqref{eq:eigenval_ratio}, listed in Tab.\ref{tab:degreeOfAnisotropy}. This scalar measure of anisotropy is smallest in case of a unidirectional orientation distribution ($\#1$) and almost $1$ in the isotropic case $\#5$. The degree of anisotropy is amplified for higher aspect ratios, which results in a lower $\beta$.
\begin{table}[H]
\centering
\begin{tabular}{|c|c|c|c|}
\hline
&$\beta$ for $\frac{L}{D}=10$&$\beta$ for $\frac{L}{D}=25$&$\beta$ for $\frac{L}{D}=50$\\
\hline
$\# 1$&$0.1003$&$ 0.0503$&$ 0.0250$\\
\hline
$\# 2$&$0.2943$&$ 0.2544 $&$0.2343$\\
\hline
$\# 3$&$0.5690$&$  0.5448 $&$0.5327$\\
\hline
$\# 4$&$0.5055 $&$0.4751$&$ 0.4598$\\
\hline
$\# 5$&$0.991$&$  0.9903$&$ 0.9899$\\
\hline
\end{tabular}
\caption{The degree of anisotropy of the different structures considered in Tab.~\ref{tab:orientationComparison} measured by means of the eigenvalue ratio $\beta(\normedW)$ of the {quadratic} normal tensor, cf. equation~\eqref{eq:eigenval_ratio}. {Apparently,} the anisotropy does not only {depend on the fiber-orientation distribution,} but is also sensitive to the {aspect} ratio $L/D$ of the fibers}.
\label{tab:degreeOfAnisotropy}
\end{table}
\subsubsection{{Sensitivity w.r.t.\ inter-fiber spacing}}

A well-known challenge when computing fiber{-}orientation measures on $\mu$-CT scans is the sensitivity w.r.t.\ spatial resolution, as well as overlapping or touching fibers~\cite{touchingFibers}. In the following study, we investigate the influence of the inter-fiber distance.
   \begin{figure}\begin{center}
 \begin{subfigure}{0.45\textwidth}
 \includegraphics[width=\textwidth,trim = 140 0 130 100, clip]{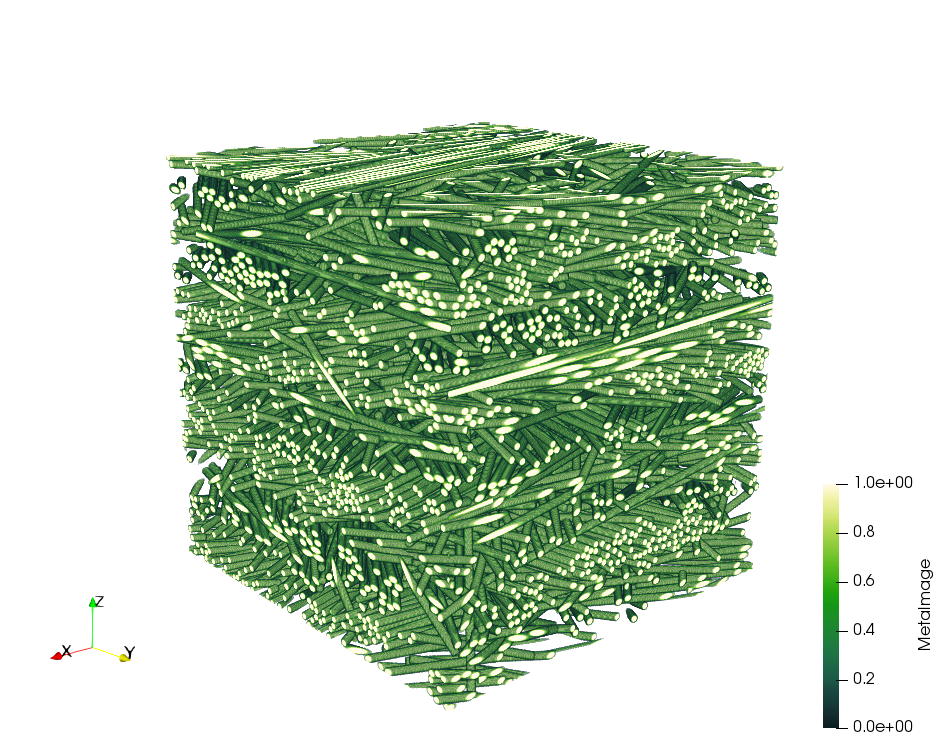}
     \vspace{-0.7cm}
    \caption{$1\%$ relative distance }
    \label{fig:phi1}
    \end{subfigure}
    \hspace{0.2cm}
    \begin{subfigure}{0.45\textwidth}
 \includegraphics[width=\textwidth,trim = 160 60 130 100, clip]{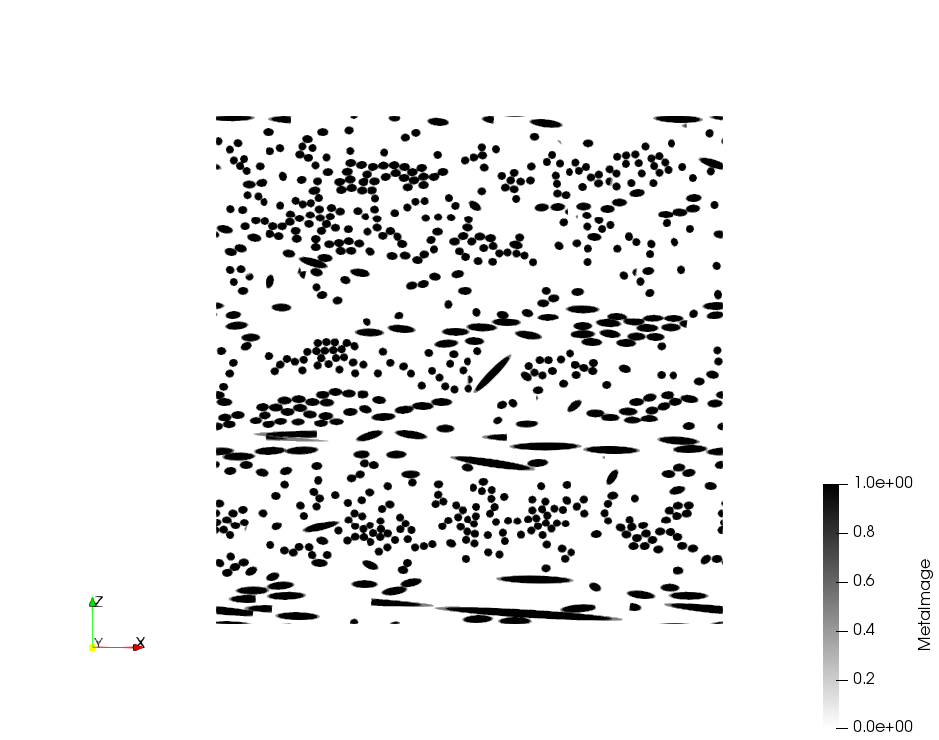}
     \vspace{-0.7cm}
    \caption{$1\%$ relative distance - slice }
    \label{fig:slicePhi1}
    \end{subfigure}
    \hspace{0.2cm}
  \begin{subfigure}{0.45\textwidth}
 \includegraphics[width=\textwidth,trim = 140 0 130 100, clip]{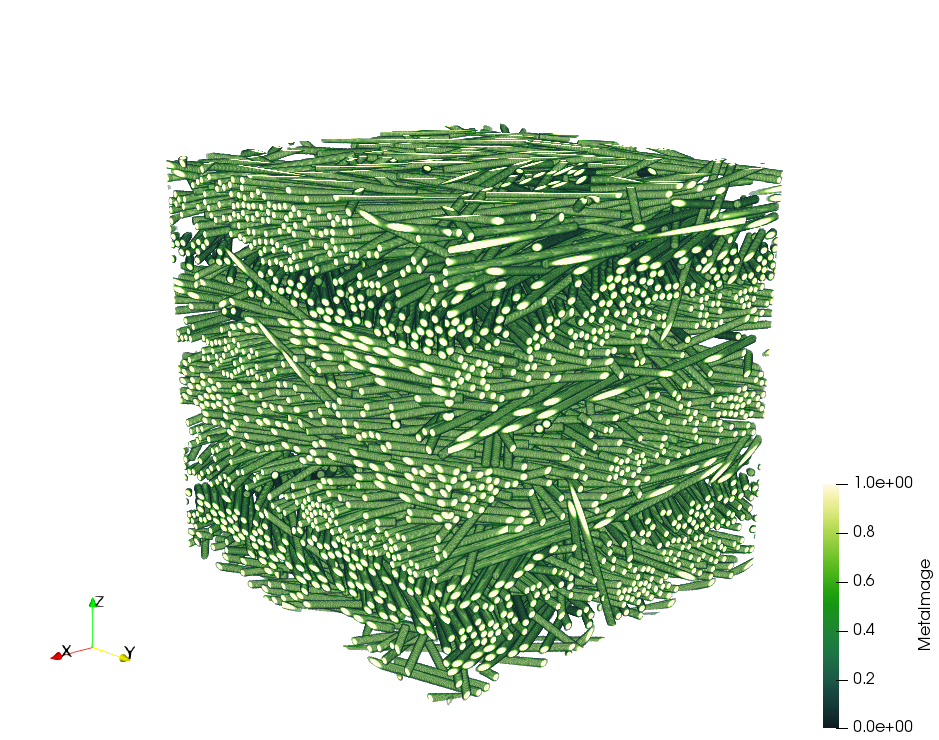}
 \vspace{-0.7cm}
    \caption{$10\%$ relative distance}
    \label{fig:phi10}
 \end{subfigure}
 \hspace{0.2cm}
 \begin{subfigure}{0.45\textwidth}
 \includegraphics[width=\textwidth,trim = 160 60 130 100, clip]{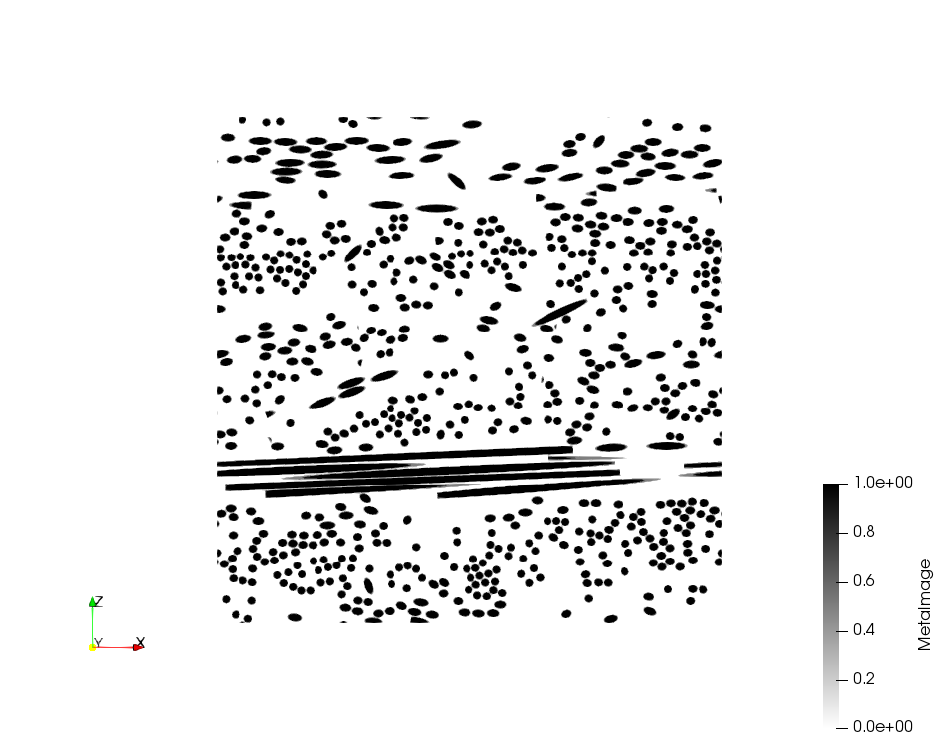}
 \vspace{-0.7cm}
    \caption{$10\%$ relative distance - slice}
    \label{fig:slicePhi10}
 \end{subfigure}
 \hspace{0.2cm}
  \begin{subfigure}{0.45\textwidth}
 \includegraphics[width=\textwidth,trim = 140 0 130 100, clip]{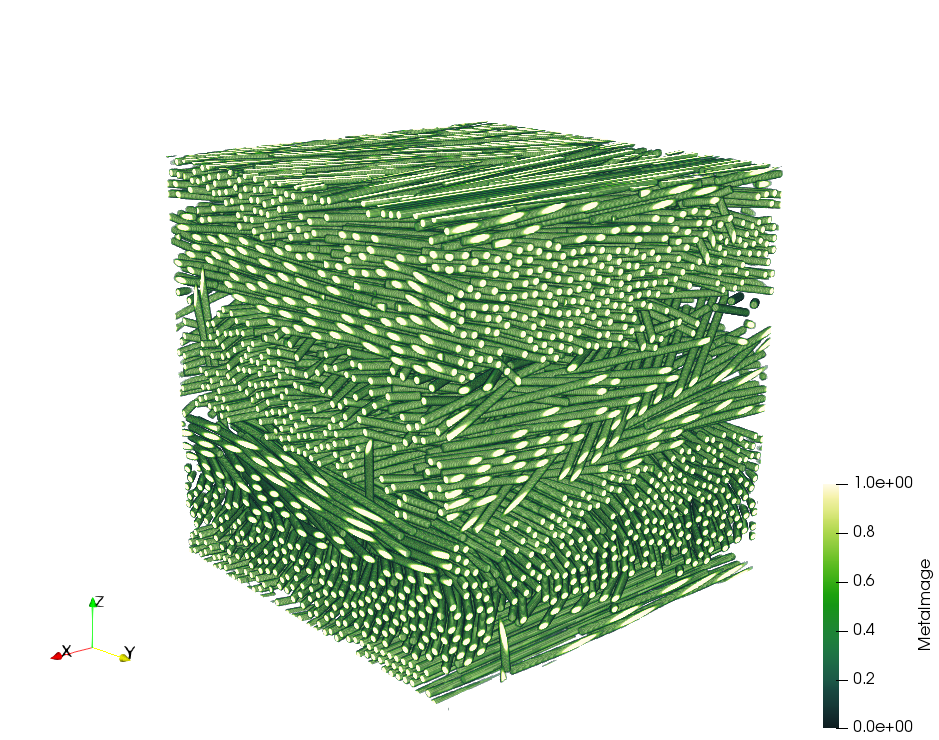}
  \vspace{-0.7cm}
    \caption{$50\%$ relative distance}
    \label{fig:phi50}
 \end{subfigure}
  \begin{subfigure}{0.45\textwidth}
 \includegraphics[width=\textwidth,trim = 160 60 130 100, clip]{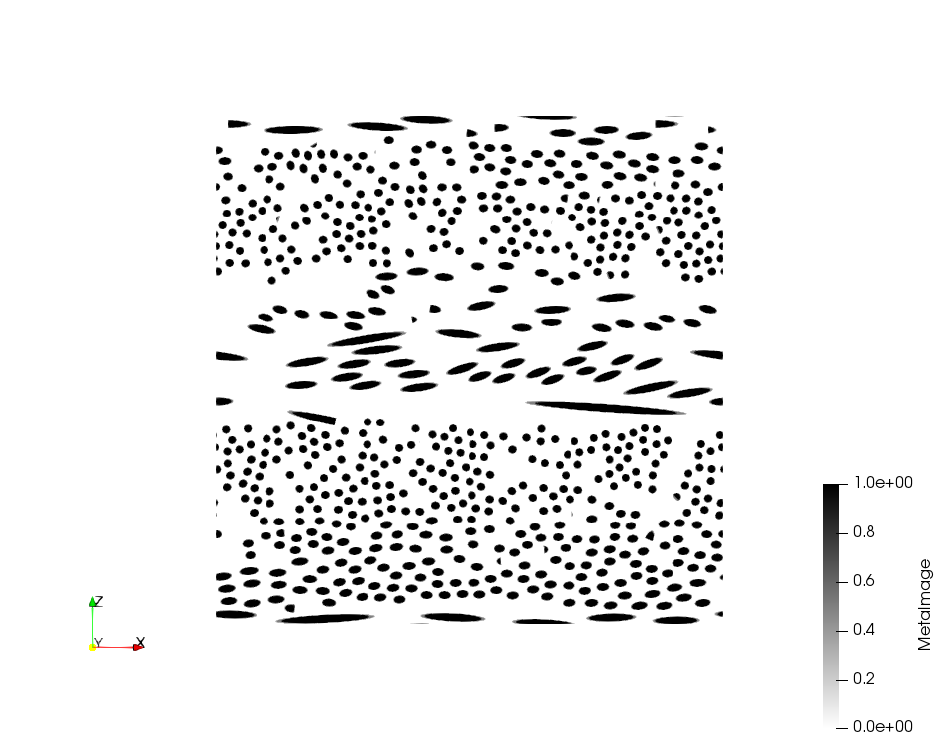}
  \vspace{-0.7cm}
    \caption{$50\%$ relative distance - slice}
    \label{fig:slicePhi50}
 \end{subfigure}
 \caption{Fiber-reinforced composite containing $1336$ fibers of equal length and varying inter-fiber spacing. The structures were generated synthetically using the sequential addition and migration algorithm~\cite{SAM}.}
 \label{fig:Fibers}
 \end{center}
 \end{figure}
 Using the sequential addition and migration algorithm~\cite{SAM}, we generate{d} structures with $20\%$ fibers of {aspect} ratio $25$, containing a total of $1336$ inclusions. The fiber-orientation tensor {was} chosen almost planar isotropic with $A=\text{diag}(0.49,0.49,0.02)$. The minim{um} distance {between} the fibers compared to their diameter {can be chosen} as an input for the microstructure generator. We generated $6$ microstructures with minim{um} relative distance varying from  $1\%$ to $50\%$ . Volumetric views and {transverse} slice{s} of {three} of these structures are shown in Fig.~\ref{fig:Fibers}. For $1\%$ relative distance, several bundles of touching or almost touching fibers are visible, whereas, for $50\%$, each fiber is {comfortably} surrounded by matrix material. All structures were voxelized with gray-value depth $\depth=2$ and for three spatial resolutions of $D/h=4, D/h=8$ and $D/h=12$, resulting in {volume images} with $256^3, 512^3$ and $768^3$ voxels{, respectively}.\\
For this data set, we compare the surface{-}area computation and the errors for the tensors $W^{0,2}_1$ and $\normedW$. As processing options, we compare no filter and the ball filter $\ballFilter$ with filter parameter $\sigma = 1.2$. The central{-}difference approximation is used for the gradient.
 \begin{figure}
\begin{center}
\begin{subfigure}{\textwidth}
\centering
\includegraphics[width = .4\textwidth]{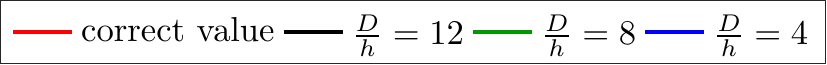}
\end{subfigure}
\begin{subfigure}{0.43\textwidth}
\centering
\includegraphics[width = \textwidth]{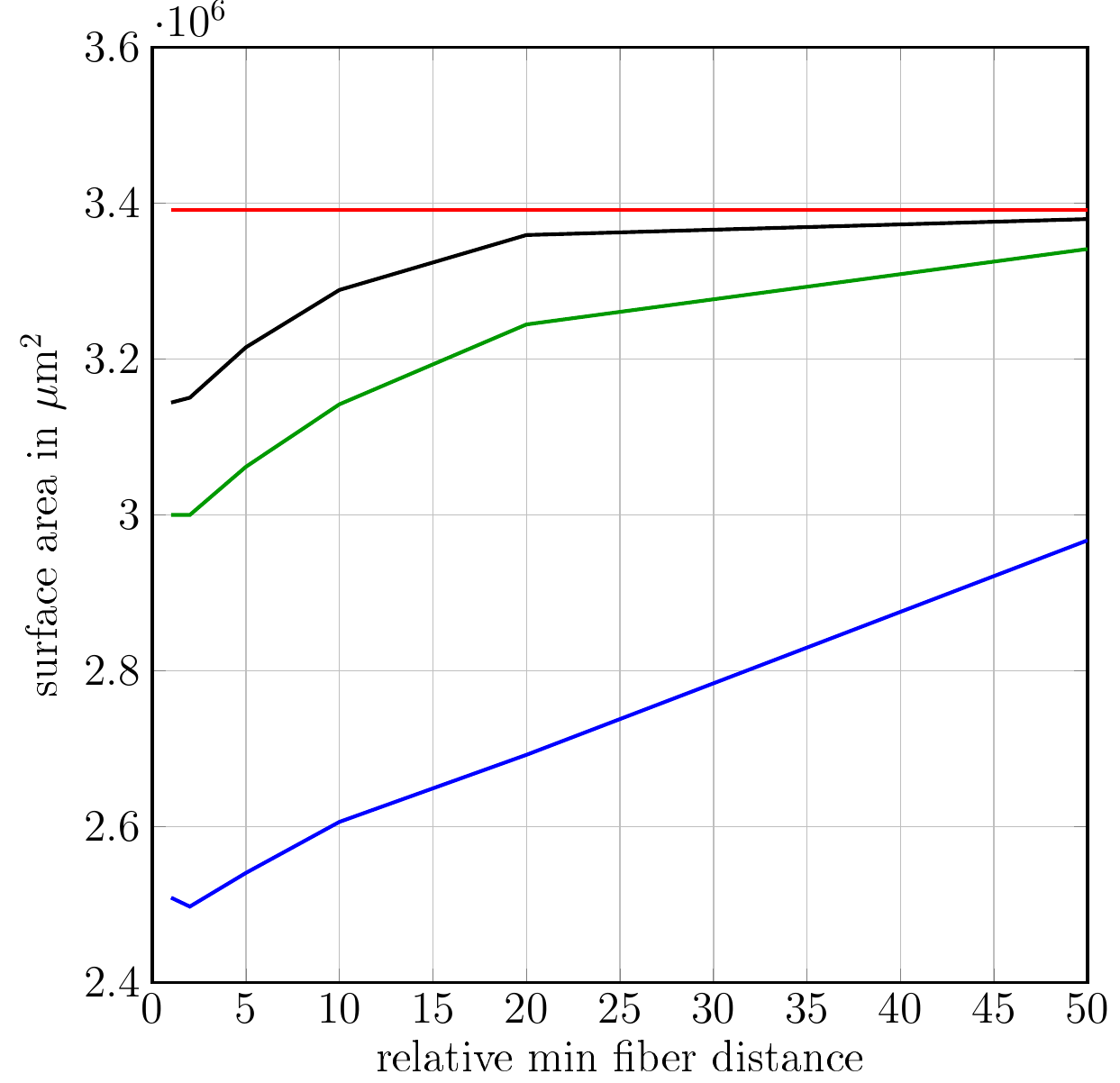}\caption{total surface area, $\sigma = 1.2$}\label{fig:FiberSurf_12}
\end{subfigure}
\begin{subfigure}{0.43\textwidth}
\centering
\includegraphics[width = \textwidth]{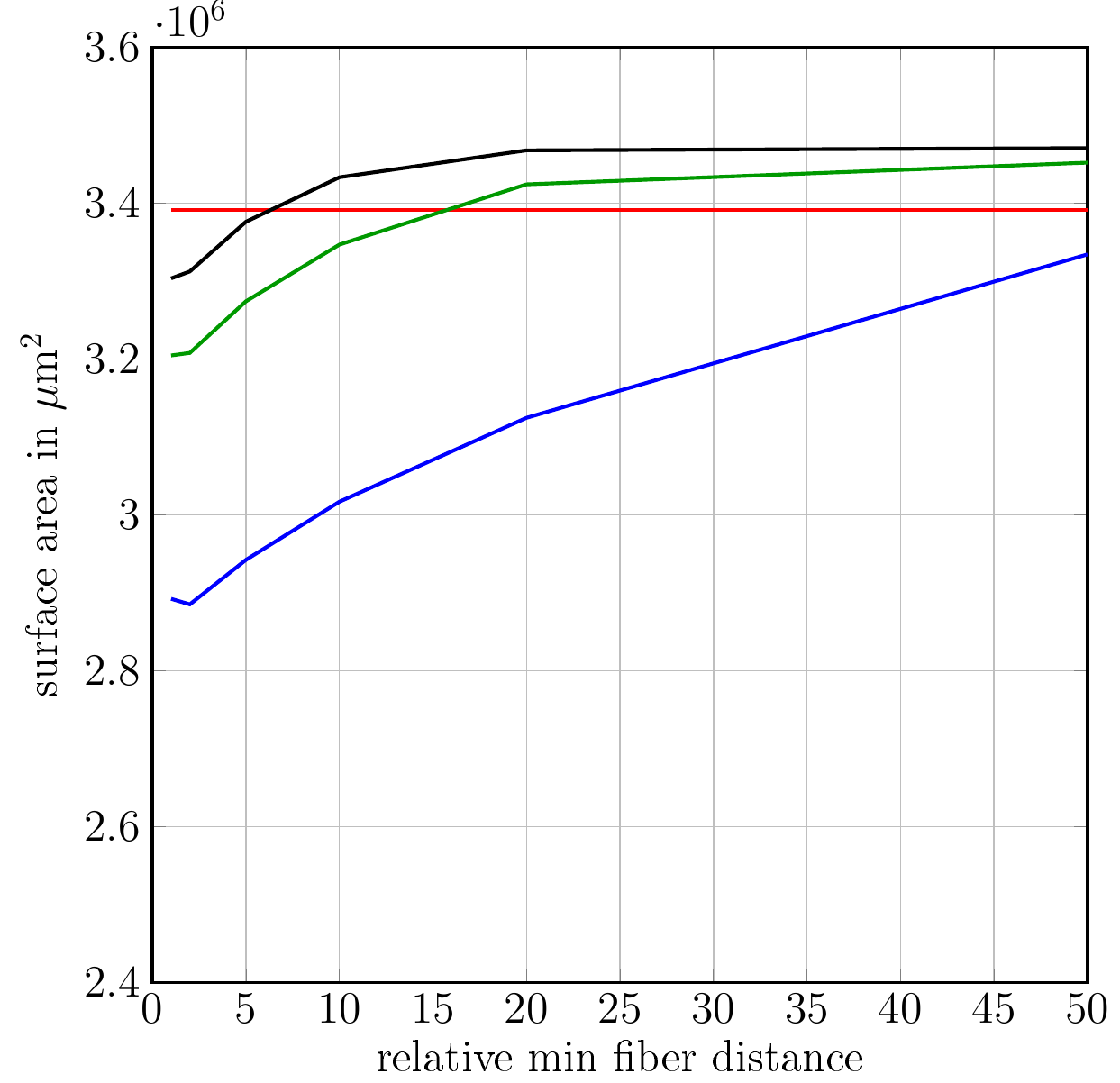}\caption{total surface area, no filter}\label{fig:FiberSurf_0}
\end{subfigure}
\begin{subfigure}{0.43\textwidth}
\centering
\includegraphics[width = \textwidth]{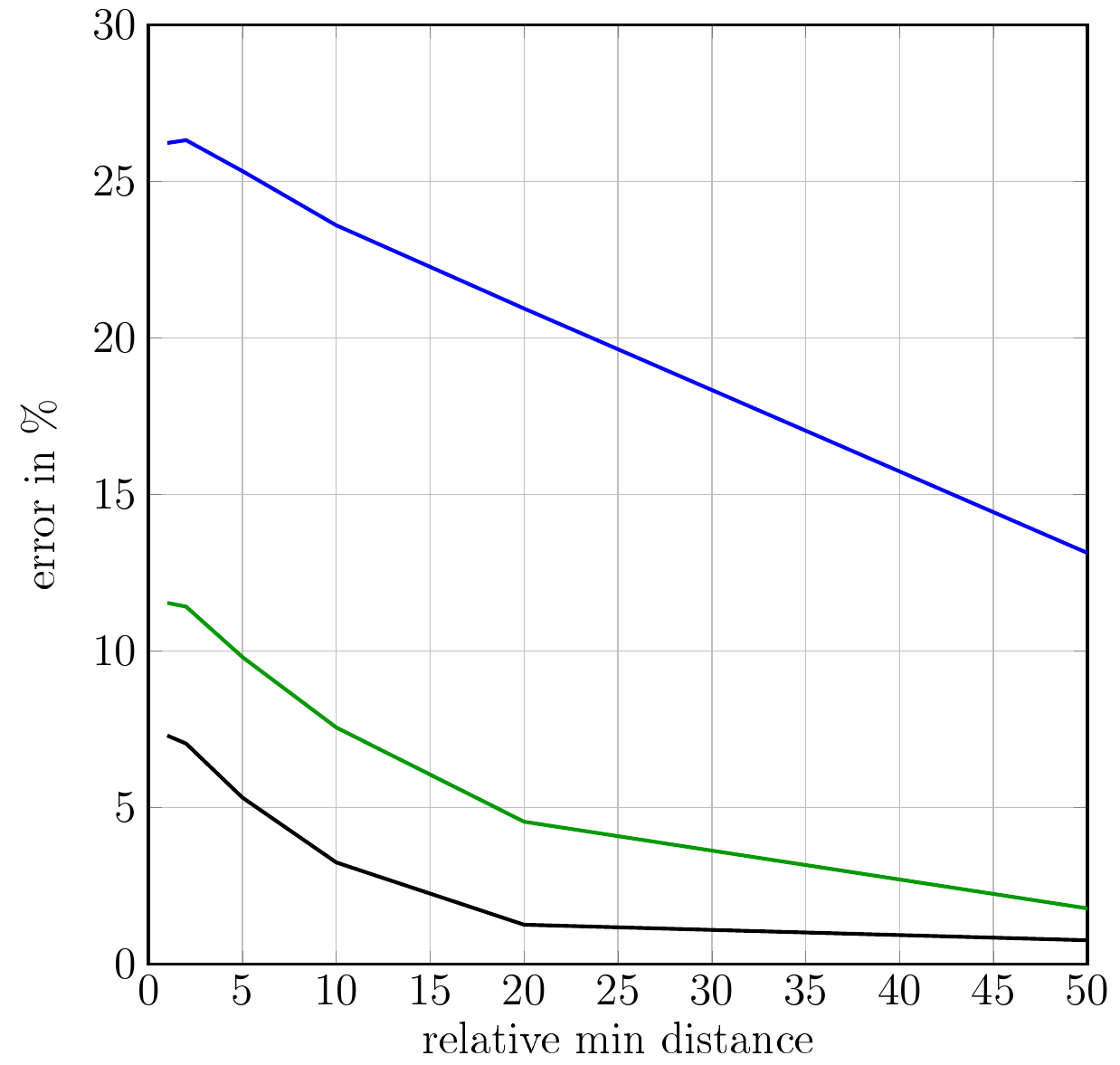}\caption{Minkowski tensor error $\tensorError$, $\sigma = 1.2$}\label{fig:FiberTensor_12}
\end{subfigure}
\begin{subfigure}{0.43\textwidth}
\centering
\includegraphics[width = \textwidth]{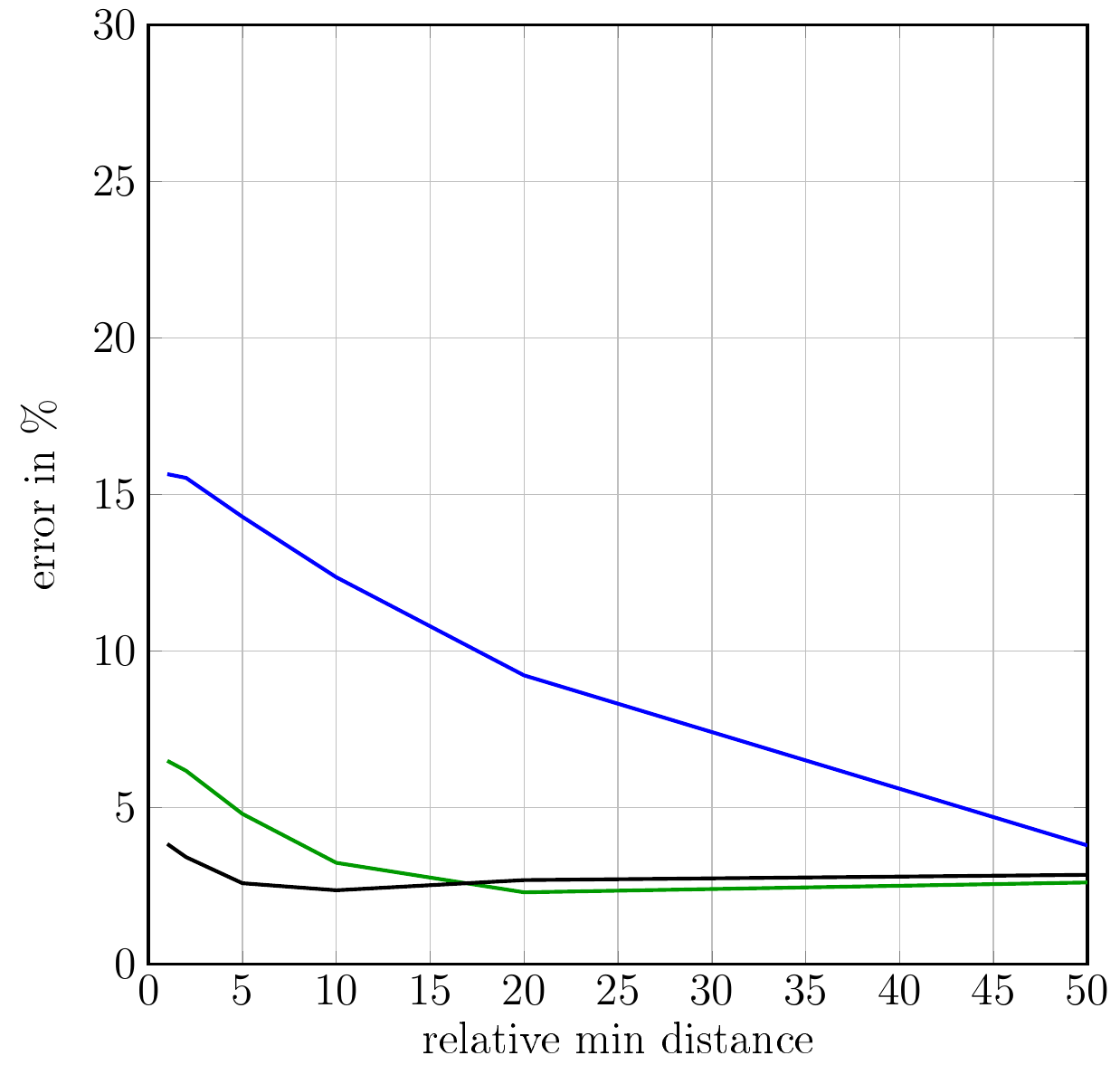}\caption{Minkowski tensor error $\tensorError$, no filter}\label{fig:FiberTensor_0}
\end{subfigure}
\begin{subfigure}{0.43\textwidth}
\centering
\includegraphics[width = \textwidth]{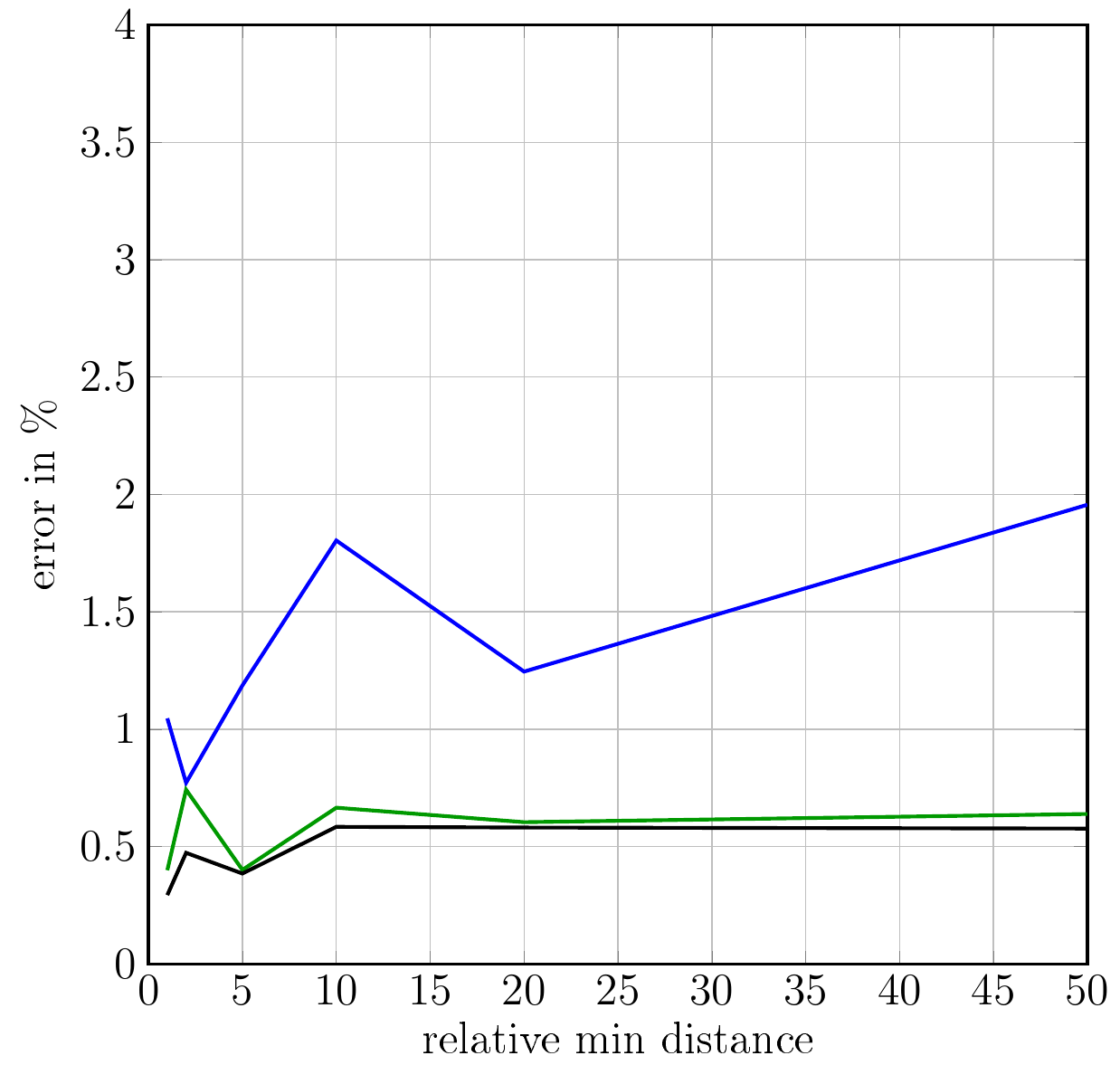}\caption{{QNT} error $\tensorDirError$, {$\sigma = 1.2$}}\label{fig:FiberTensorDir_12}
\end{subfigure}
\begin{subfigure}{0.43\textwidth}
\centering
\includegraphics[width = \textwidth]{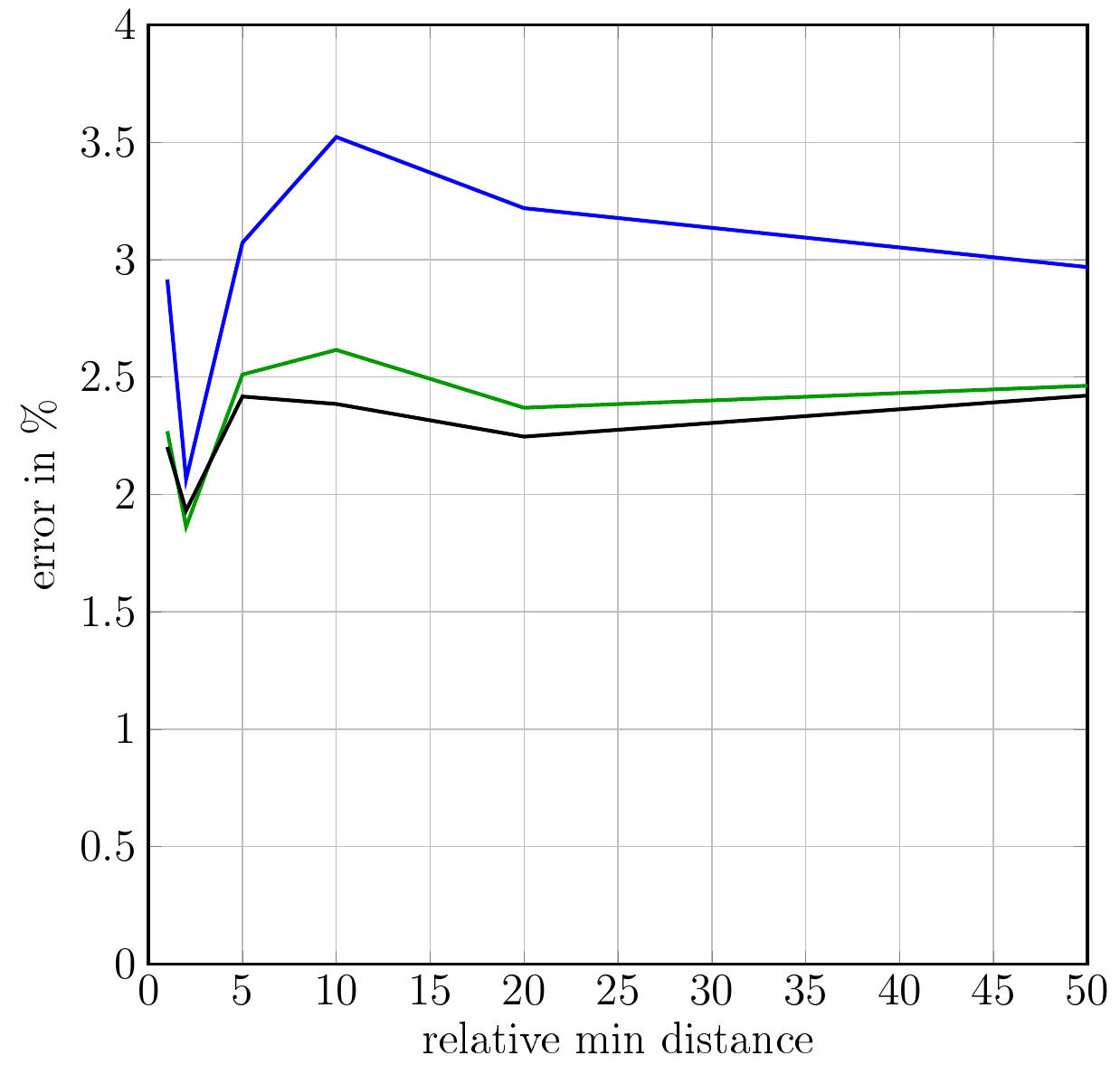}\caption{{QNT} error $\tensorDirError$, no filter}\label{fig:FiberTensorDir_0}
\end{subfigure}
\caption{total surface area and the errors $\tensorError$ and $\tensorDirError$ plotted vs.\ the minimum relative distance between fibers.}\label{fig:FiberStudy}
\end{center}
\end{figure}
Fig.~\ref{fig:FiberSurf_12} shows the computed total surface area vs.\ the minimum fiber distance relative to the diameter for the three spatial resolutions under consideration, using the ball filter with $\sigma=1.2$. We observe two trends{. Firstly,} the surface area is generally underestimated for all spatial resolutions. However, we clearly see multigrid convergence. Furthermore, the error is smaller for the larger minimum distance. This {observation conforms to} our expectations, as the surface area of touching or almost touching fibers is not computed accurately enough by a gradient-based approximation. In Fig.~\ref{fig:FiberSurf_0}, we see the results of the surface area computation without applying any filter. The errors are, in general, lower than in the case of $\sigma=1.2$. However, neither multigrid convergence, nor a convergence as the minimum fiber distance increases is observed. Fig.~\ref{fig:FiberTensor_12} and Fig.~\ref{fig:FiberTensor_0} show the error $\tensorError$ for both filter choices. Again, the results reflect the relative error of the surface area estimation.
Fig.~\ref{fig:FiberTensorDir_12} and Fig.~\ref{fig:FiberTensorDir_0} contain the error{s} of the {quadratic} normal tensor for both filter choices.
For no filter application, the error is below $4\%$, and for $\sigma=1.2$, it is even below $2\%$ for all spatial resolutions and minimum inter-fiber distances. No clear trend w.r.t. {the} inter-fiber spacing is visible.
Hence, the {quadratic} normal tensor $\normedW$ may serve as a microstructure descriptor that is robust w.r.t.\ small inter-fiber spacing.\\
  \begin{figure}
\begin{center}
\begin{subfigure}{\textwidth}
\centering
\includegraphics[width = .3\textwidth]{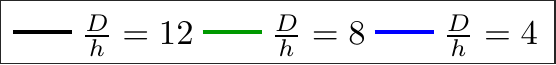}
\end{subfigure}
\begin{subfigure}{0.45\textwidth}
\centering
\includegraphics[width = \textwidth]{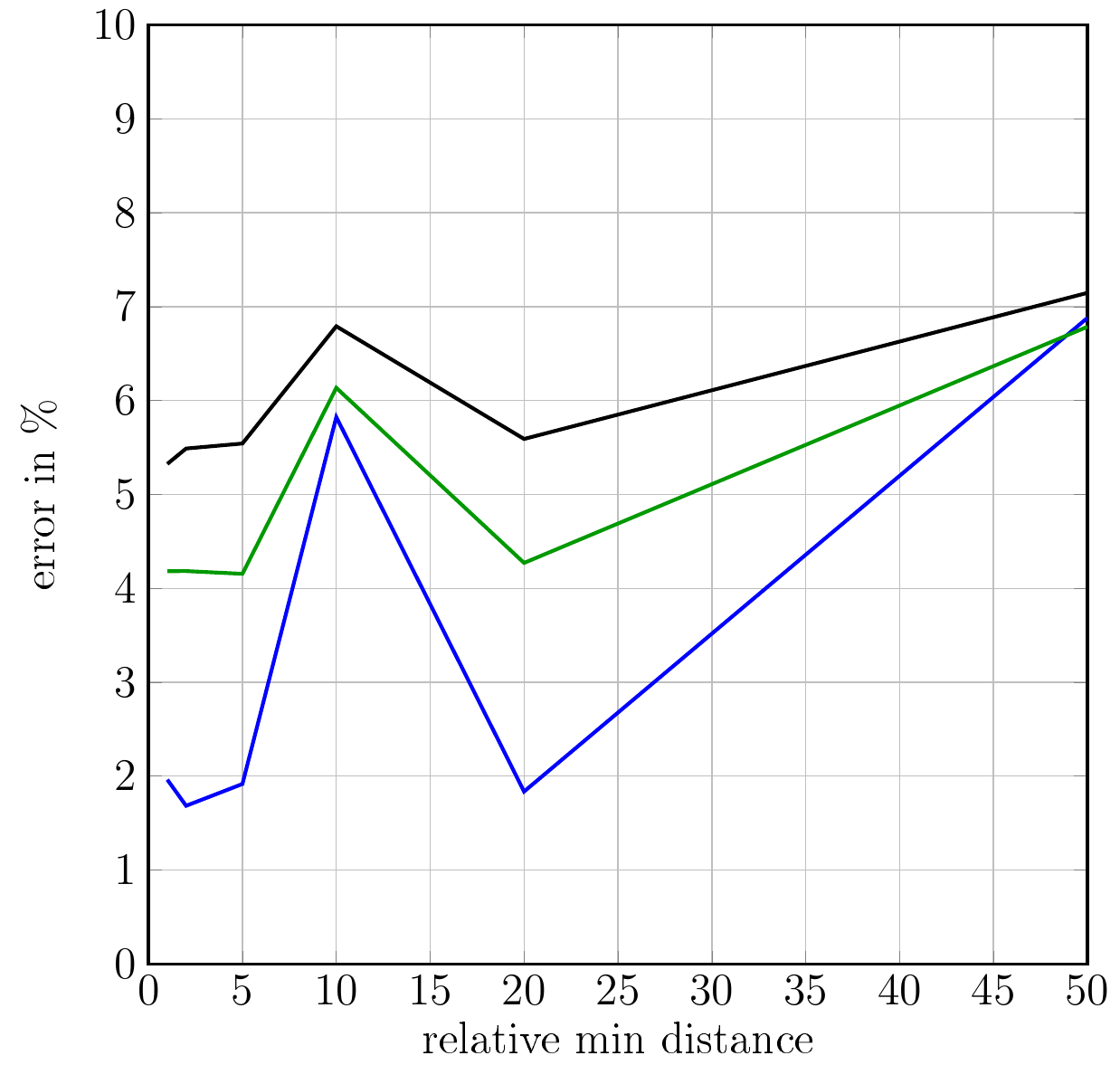}\caption{{Fiber-orientation tensor} error $E_A$, $\sigma = 0,~\mu = 3$}\label{fig:sig0mu3}
\end{subfigure}
\begin{subfigure}{0.45\textwidth}
\centering
\includegraphics[width = \textwidth]{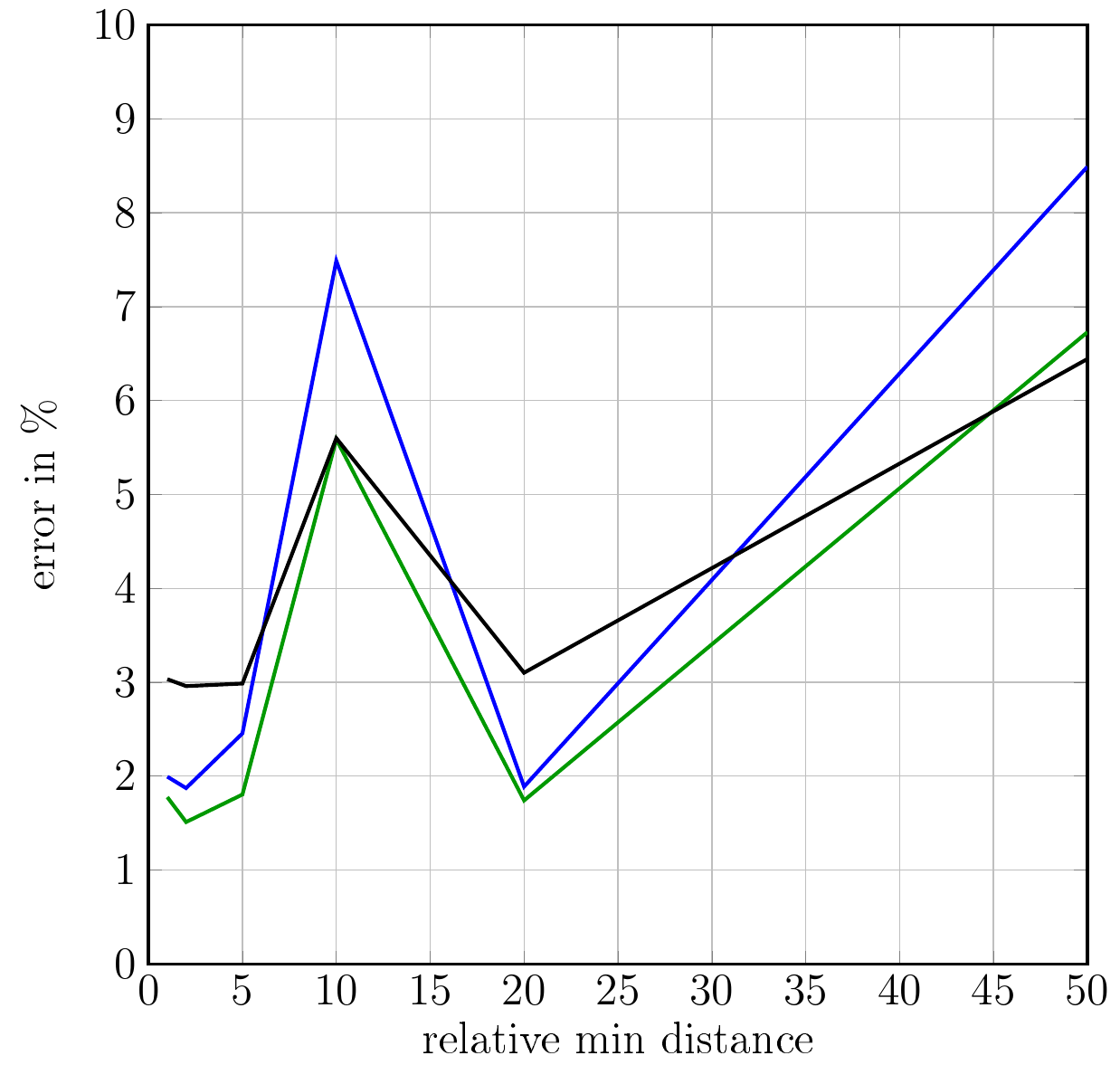}\caption{{Fiber-orientation tensor} error $E_A$, $\sigma = 0,~\mu = 6$}\label{fig:sig0mu6}
\end{subfigure}
\begin{subfigure}{0.45\textwidth}
\centering
\includegraphics[width = \textwidth]{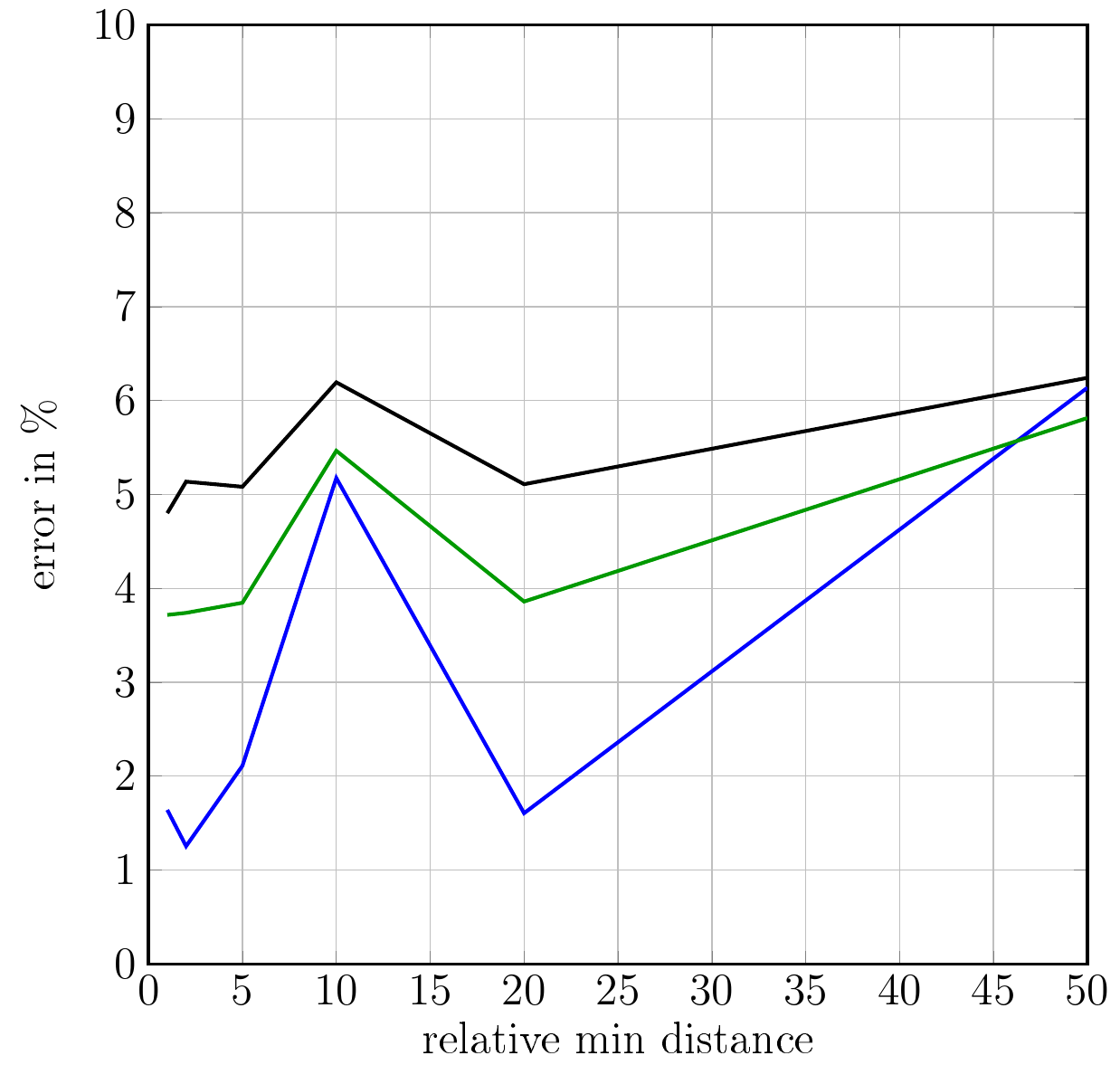}\caption{{Fiber-orientation tensor} error $E_A$, $\sigma = 1.2,~\mu = 3$}\label{fig:sig1mu3}
\end{subfigure}
\begin{subfigure}{0.45\textwidth}
\centering
\includegraphics[width = \textwidth]{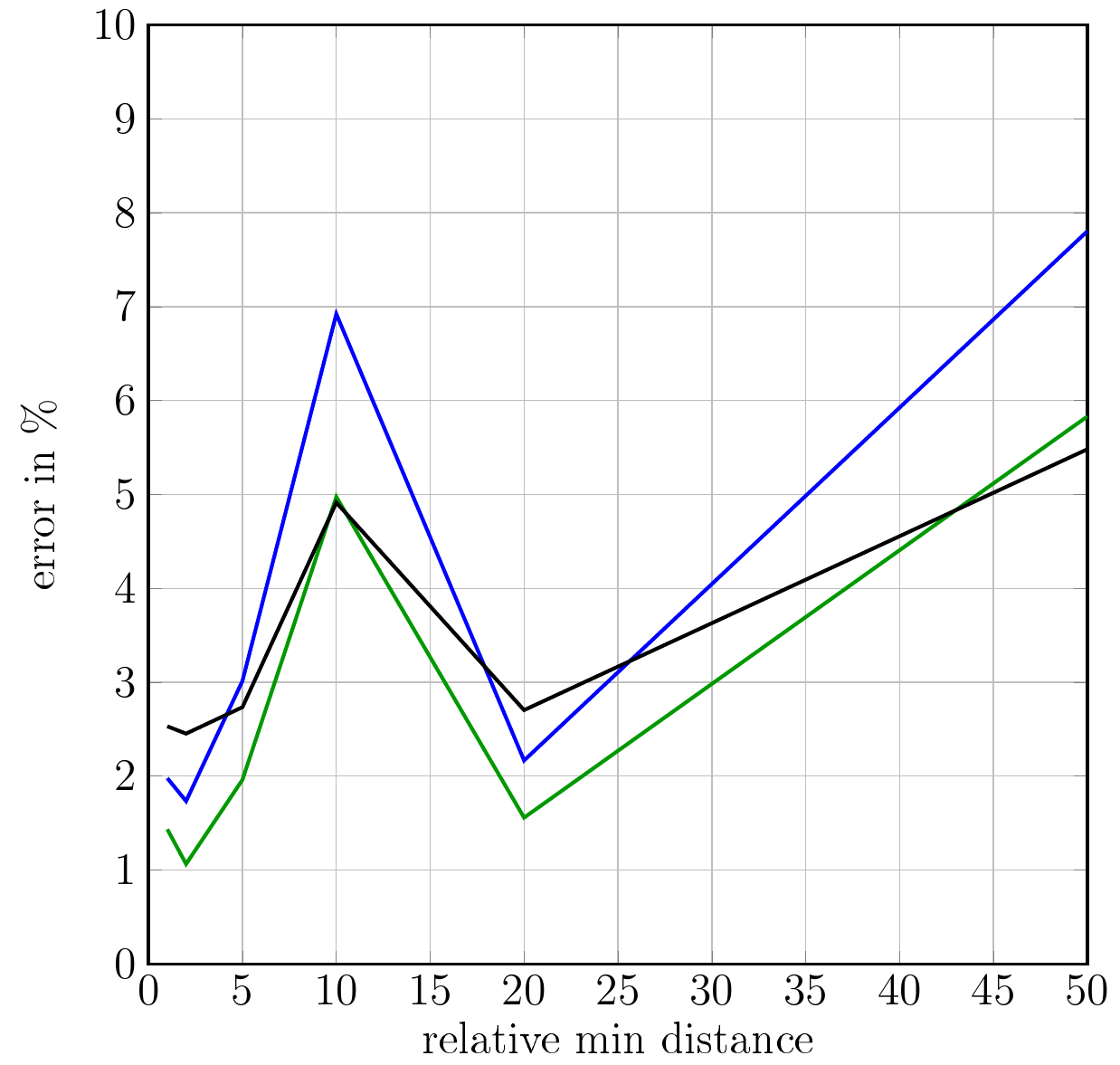}\caption{{Fiber-orientation tensor} error $E_A$, $\sigma = 1.2,~\mu = 6$}\label{fig:sig1mu6}
\end{subfigure}
\caption{Error of the {structure-tensor based fiber orientation tensor computation} using different filter parameters for the first and second filter.}\label{fig:structureTensor}
\end{center}
\end{figure}
 We compare our approach with the well{-}established structure{-}tensor {method}, see Algorithm \ref{Alg:struct}, which we implemented into {our} code.
  \begin{algorithm}
\caption{{Computing the fiber-orientation tensor via the structure-tensor method~\cite{StructureTensor}}}
\label{Alg:struct}
\begin{algorithmic}[1]
\State $\smoothIm_h^\sigma \gets \mathcal{F}_\sigma * \chara_h$\Comment{ Blur image with image filter}
\State $\g(\rr)\gets \nabla_h \smoothIm_h^\sigma $\Comment{ Apply discrete gradient}
\State $I(\rr) \gets \g(\rr)\otimes \g(\rr)$\Comment{ Compute local tensor}
\State $I_\mu \gets \mathcal{F}_{\mu}*I $\Comment{ Blur local tensor with second filter}
\State $\{\lambda_{i}(\rr), \mathbf{v}_{i}(\rr)\}\gets\text{Eig}(I_\mu(\rr))$\Comment{ Local eigenvalue decomposition (sorted, smallest first) }
\State $A^\approx = \sum_{\rr \in \Y_h}\mathbf{v}_1(\rr)\otimes \mathbf{v}_1(\rr)$\Comment{Extract local orientation tensor}
\State \Return $A^\approx/\tr(A^\approx)$
\end{algorithmic}
\end{algorithm}
 Computing the fiber-orientation tensor numerically via the structure-tensor algorithm requires applying a second filter $\conv_\mu$ with filter parameter $\mu$ to the tensor field $\n(\rr) \otimes \n(\rr)$ (component-wise, this tensor field denotes said `structure tensor'). Pinter et al.~\cite{Pinter2018} recommend that for {the filter parameter for} the second filter should be {larger than} for the first filter, which should be rather small. In our case, this is best recovered by choosing the ball filter with small filter parameter (i.e., $\sigma=1.2$) or no filter (i.e., $\sigma=0$) as the first filter. For the second filter, we choose a Gaussian kernel $\mathcal{G}_\mu$ with $\mu = 3$ and $\mu =6$.
{To evaluate the accuracy of the method, we introduce the fiber-orientation tensor error measure (similar to $\tensorDirError$)}
\begin{align*}
E_A = \frac{\|A - A^\approx\|}{\|A\|},
\end{align*}
where $A^\approx$ is the approximated fiber-orientation tensor computed by the structure tensor approach.\\
Fig.~\ref{fig:structureTensor} shows the error of {this method} for the four filter combinations $\sigma = 0, 1.2;~\mu=3, 6$. The error is below $9\%$ for all structures and resolutions. The first filter width $\sigma=1.2$ results in a lower error than {for} $\sigma = 0$. This holds for all spatial resolutions and fiber-distance thresholds. For the second filter, however, the optimal choice depends on the spatial resolution. The two finer resolutions benefit from a larger second filter and even exhibit a larger error for the smaller $\mu$ than {for} the coarse resolution. With respect to the relative minimum distance of fibers, no clear trend is visible. The error fluctuates between $1\%$ and $8\%$ for the different microstructures. The error of the {quadratic} normal tensor $\normedW$, on the other hand, was below $2\%$ for all spatial resolutions and hence provides a reliable option for characterizing fiber-reinforced composites.\\
All computations were performed in {a} matter of minutes.

 \subsection{{Sand grains and sand-binder composites}} \label{sec:4.4}
\begin{figure}[H]
 \begin{center}
 \begin{subfigure}{0.3\textwidth}
 \includegraphics[width=\textwidth,trim = 120 0 160 110, clip]{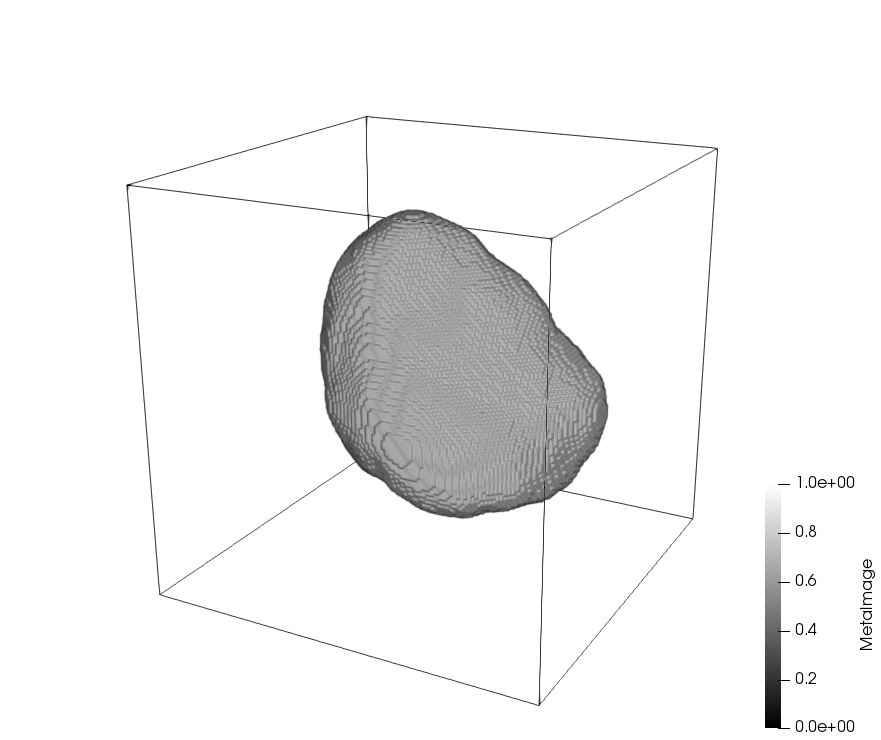}
     \vspace{-0.7cm}
    \caption{Grain \#1 }
    \label{fig:grain1}
    \end{subfigure}
    \begin{subfigure}{0.3\textwidth}
 \includegraphics[width=\textwidth,trim = 120 0 160 110, clip]{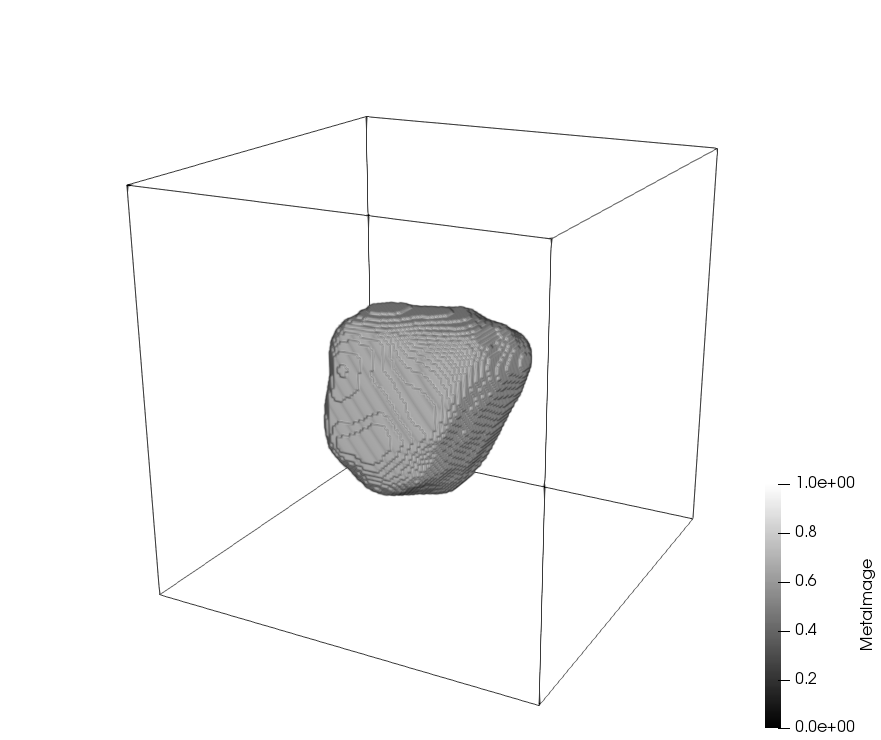}
     \vspace{-0.7cm}
    \caption{Grain \#2 }
    \label{fig:grain2}
    \end{subfigure}
  \begin{subfigure}{0.3\textwidth}
 \includegraphics[width=\textwidth,trim = 120 0 160 110, clip]{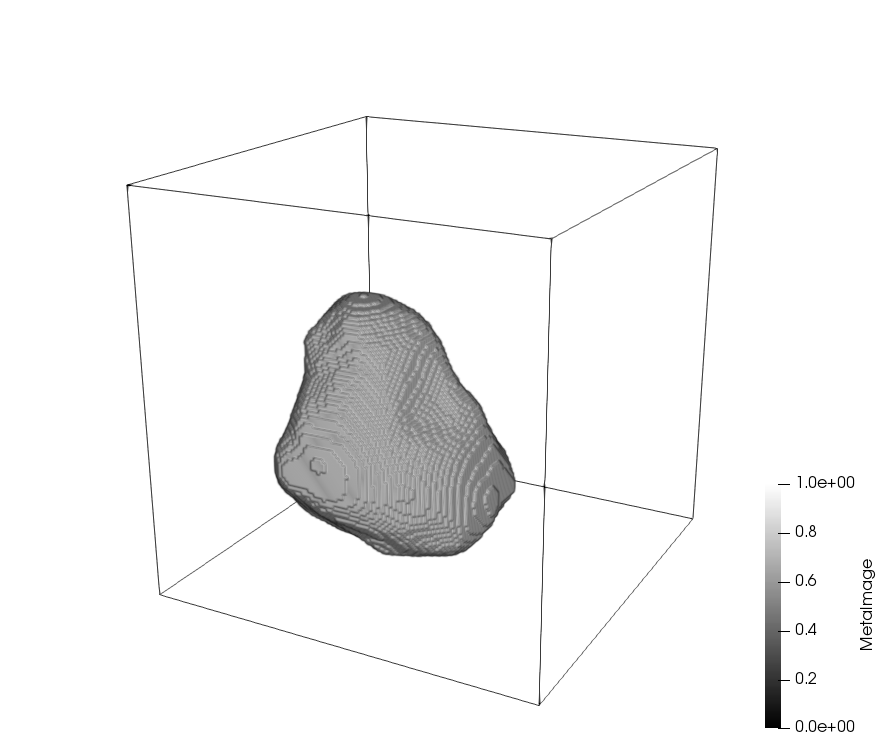}
 \vspace{-0.7cm}
    \caption{Grain \#3}
    \label{fig:grain3}
 \end{subfigure}
 \begin{subfigure}{0.3\textwidth}
 \includegraphics[width=\textwidth,trim = 120 0 160 110, clip]{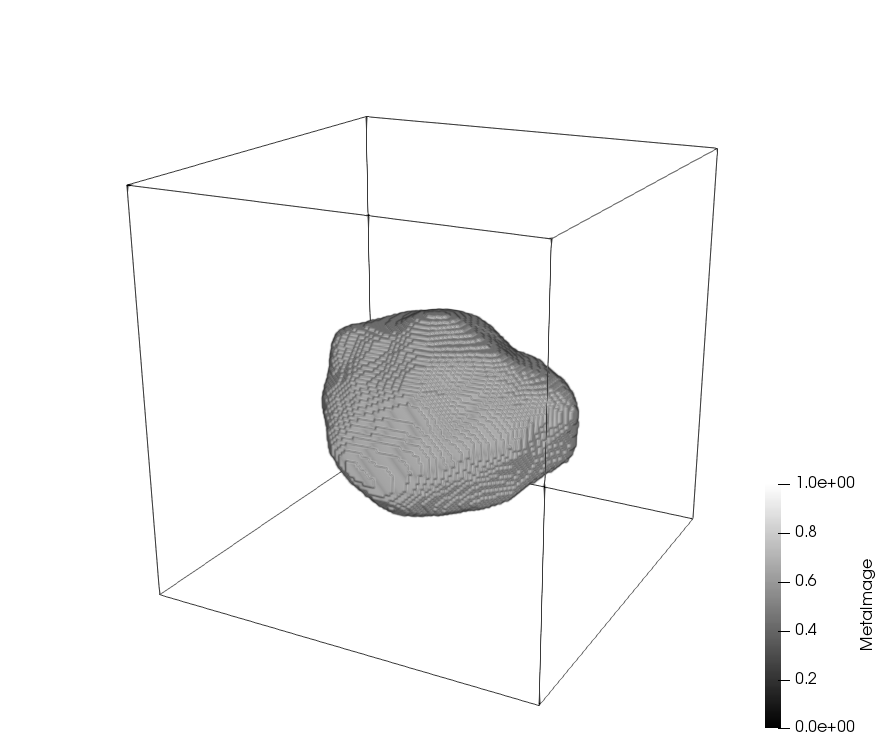}
 \vspace{-0.7cm}
    \caption{Grain \#4}
    \label{fig:grain4}
 \end{subfigure}
  \begin{subfigure}{0.3\textwidth}
 \includegraphics[width=\textwidth,trim = 120 0 160 110, clip]{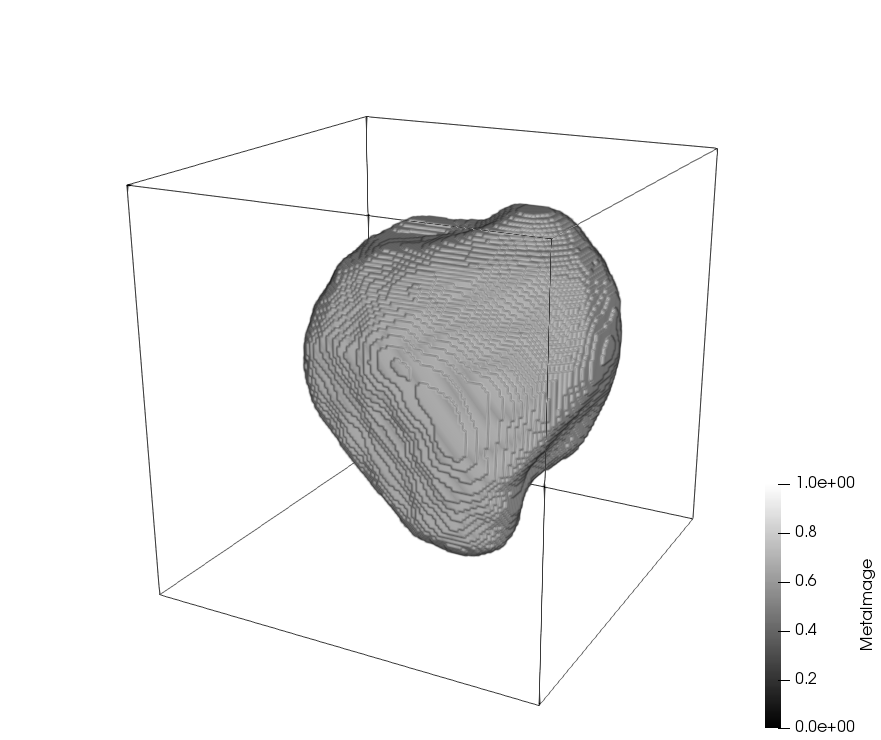}
  \vspace{-0.7cm}
    \caption{Grain \#5}
    \label{fig:grain5}
 \end{subfigure}
  \begin{subfigure}{0.3\textwidth}
 \includegraphics[width=\textwidth,trim = 120 0 160 110, clip]{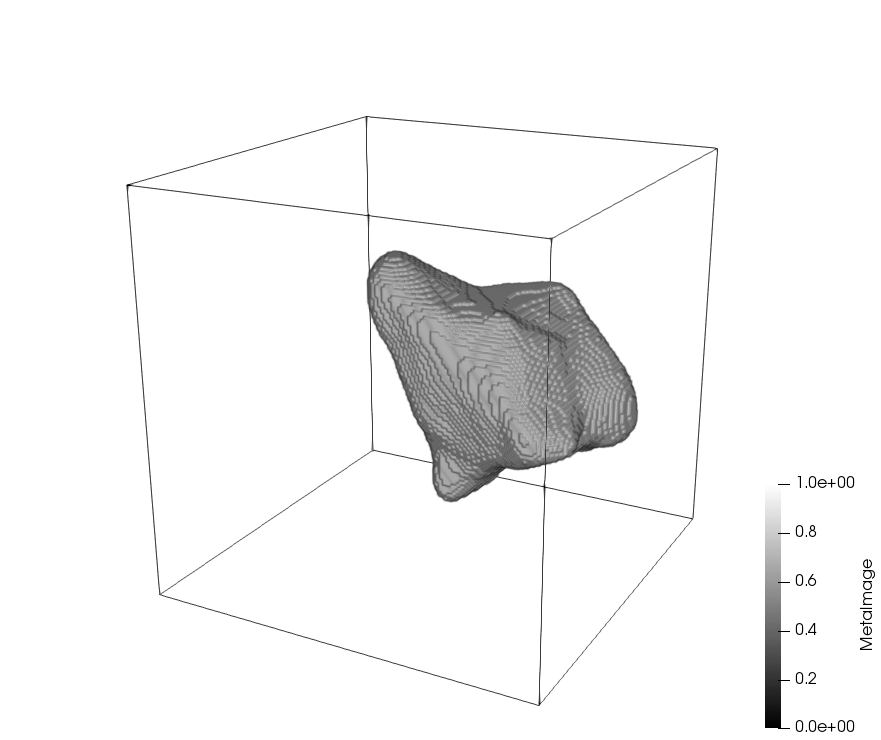}
  \vspace{-0.7cm}
    \caption{Grain \#6}
    \label{fig:grain6}
 \end{subfigure}
 \caption{Six different sand grains whose shapes are analyzed using $\normedW$, see Tab.~\ref{tab:sand}.}
 \label{fig:grains}
 \end{center}
 \end{figure}

For manufacturing parts with complex geometry, casting is often the {preferred} choice~\cite{Rao2003}. For casting, the mold enters a cavity of the specified shape. This cavity, in turn, is realized as a sand core, which has to be destroyed after the casting process. Such sand cores are composed of sand grains which are held together by an organic or inorganic binder. These constituents, their proportion and shape, strongly influence the overall material behavior of the sand-binder aggregate~\cite{Sand}. Loosely speaking, if the strength of the aggregate is too low, the part will not survive the casting process. On the other hand, excessive strength may prevent the part to be extracted unscathed from the sand core.\\
 In this section, we compute the {quadratic} normal tensors of sand cores to study their anisotropy and to demonstrate the wide range of applicability of {{quadratic} normal} tensors. We consider six different sand{-}grain shapes which were obtained from fitting cleaned up and binarized $\mu$-CT scans~\cite{Sand}. The individual grains are shown in Fig.~\ref{fig:grains}.
 These sand grains are non-convex and anisotropic. The computed {quadratic} normal tensors $\normedW$ are listed in Tab.~\ref{tab:sand}. For the computation, we chose the ball filter $\ballFilter$ with $\sigma = 1.2$ voxels and used central differences for the gradient-approximation.
 \begin{table}
  \begin{center}
 \resizebox{\textwidth}{!}{
 \begin{tabular}{|c|c|c||c|c|c|}
 \hline
 Grain & $\normedW$& $\beta$&Grain &$\normedW$&$\beta$\\
 \hline\hline
  \#1& $\sv0.2743&0.0518&0.0722\\0.0518&0.2486&0.0179\\0.0722&0.0179&0.4772\ev$&$0.4044$& \#2&$\sv0.3669&-0.0428&0.0405\\-0.0428&0.2949&-0.0204\\0.0405&-0.0204&0.3382\ev$ &$0.6643$\\
 \hline\hline
  \#3 &$\sv0.382&0.0405&0.0055\\0.0405&0.2945&-0.0462\\0.0055&-0.0462&0.3235\ev$ & $0.626$&  \#4&$\sv0.2195&0.0407&0.0319\\0.0407&0.3979&0.0936\\0.0319&0.0936&0.3826\ev$&$0.4244$ \\
 \hline \hline
  \#5&$\sv0.28&0.0092&0.0013\\0.0092&0.2875&0.07\\0.0013&0.07&0.4325\ev$ &$0.5564$& \#6&$\sv0.3123&0.07&0.0428\\0.07&0.3202&-0.023\\0.0428&-0.023&0.3674\ev$ &$0.5795$\\
 \hline
 \end{tabular}}
 \caption{{Quadratic} normal tensor $\normedW$ and eigenvalue ratio $\beta$ of the six grains in Fig.~\ref{fig:grains}.}
 \label{tab:sand}
 \end{center}
 \end{table}
 In addition to the {quadratic} normal tensor, we quantify the degree of anisotropy by listing the eigenvalue ratios \eqref{eq:eigenval_ratio} of $\normedW$. We observe that all sand grains have a distinct degree of anisotropy, varying between $\beta = 0.4$ and $\beta = 0.63$. To gain further insight into the anisotropy of the grains, we compute the eigenvalue decomposition of $\normedW$ for grain \#1. The eigensystem reads
 \begin{align*}
 \lambda_1 = 0.5046, ~\mathbf{v}_1 = \sv -0.3233\\ -0.1308\\ -0.9372 \ev;\quad
 \lambda_2 = 0.2914,~ \mathbf{v}_2 = \sv-0.6666 \\-0.6715\\  0.3236\ev; \quad \lambda_3 = 0.204,~ \mathbf{v}_3 = \sv-0.6717 \\ 0.7294\\  0.1299\ev.
 \end{align*}
The largest eigenvalue indicates a somewhat disc-like shape within the plane normal to $\mathbf{v}_1$. The vectors $\mathbf{v}_2$ and $\mathbf{v}_3$ lie in that plane, the lower eigenvalue $\lambda_3$ indicates a slight extension in direction $\mathbf{v}_3$. For a better understanding, {we point at} Tab.~\ref{tab:orientationComparison}, where fiber-reinforced composites are analyzed and a `translation' to well{-}known orientation tensors is {provided}.
\begin{figure}[H]
\begin{center}
 \begin{subfigure}{0.45\textwidth}
 \includegraphics[width=\textwidth,trim = 120 0 120 70, clip]{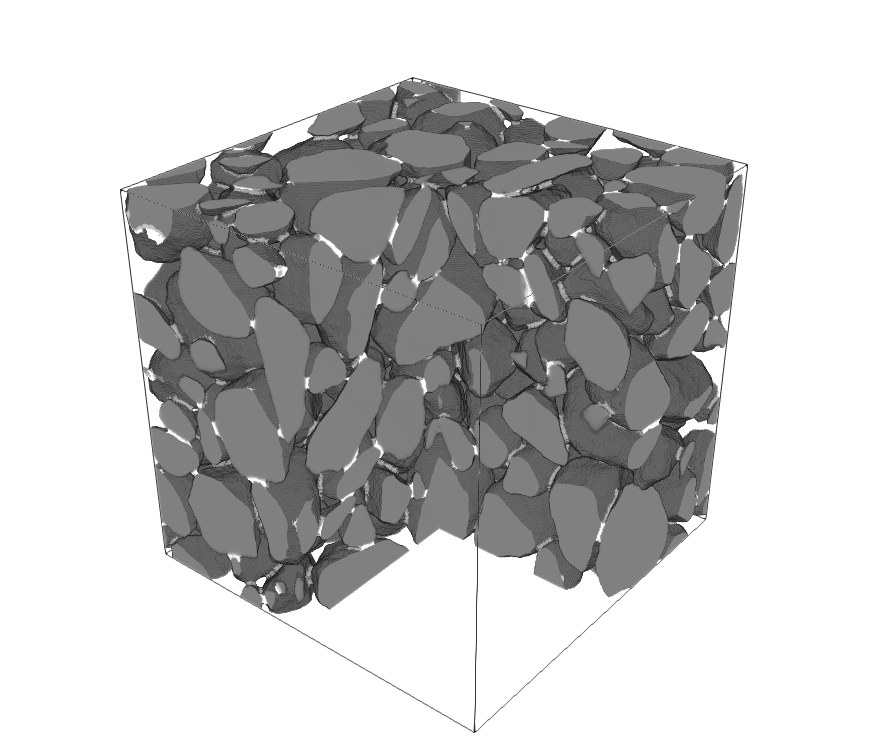}
     \vspace{-0.7cm}
    \caption{Structure $\#1$, containing $216$ sand grains}
    \label{fig:E}
    \end{subfigure}
     \begin{subfigure}{0.45\textwidth}
 \includegraphics[width=\textwidth,trim = 120 0 120 70, clip]{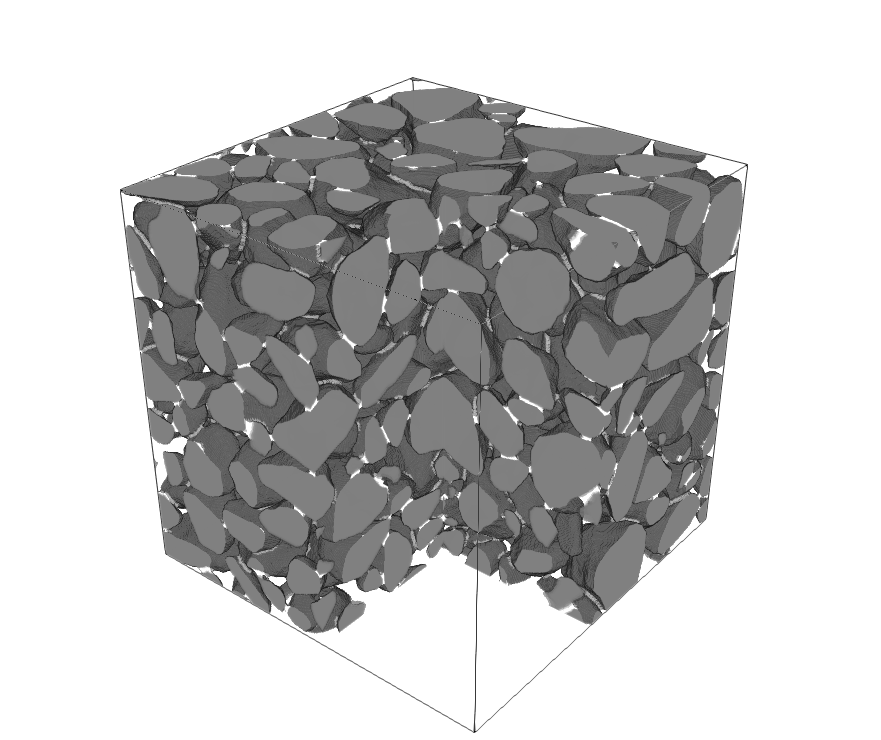}
     \vspace{-0.7cm}
    \caption{Structure $\#2$, containing $343$ sand grains}
    \label{fig:F}
    \end{subfigure}
    \caption{Sand core structures, containing $58.58\%$ sand and $1.28\%$ {i}norganic binder. The structures were generated {by the} mechanical contraction method~\cite{Sand}.}\label{fig:SandCoreStruct}
\end{center}
\end{figure}
These six sand grains of Fig.~\ref{fig:grains} were used for generating sand-binder composite microstructures, characteristic for casting applications, using {the} mechanical contraction method~\cite{Sand}. Two realizations, containing $216$ and $343$ sand grains, are shown in Fig.~\ref{fig:SandCoreStruct}. Both structures consist of $58.58\%$ sand {grains} and $1.28\%$ inorganic binder. In contrast to particle-filled composites, these microstructures involve an interpenetrating porous phase.\\
On $\mu$-CT images of sand-binder composites, the binder phase cannot be distinguished from the sand phase, cf.\ Schneider et al.~\cite{Sand}. Therefore, we investigate how the presence of the binder phase affects the Minkowski tensors. We compare the {quadratic} normal tensor $\normedW$ of the sand-binder composite to the one with only sand {grains} for both structures in Fig.~\ref{fig:SandCoreStruct}.
We chose the ball filter $\ballFilter$ with $\sigma = 1.2$ voxels and central differences for the gradient approximation. For all structures, the resulting tensor $\normedW$, the degree of anisotropy $\beta$ and the total surface area are listed in Tab.~\ref{tab:sandStruct}.
 \begin{table}
  \begin{center}
 \resizebox{\textwidth}{!}{
 \begin{tabular}{|c|c|c|c|c|c|c|}
 \hline
 &\multicolumn{3}{c|}{Sand {grains} alone}&\multicolumn{3}{c|}{Sand-binder composite}\\
 \hline
 & $\normedW$& $\beta$&$S[\unit{mm}^2]$&$\normedW$&$\beta$&$S[\unit{mm}^2]$\\
 \hline\hline
 Structure $\#1$& $\sv 0.3294& -0.0002 &   -0.0025\\ -0.0002 &    0.3412& -0.0094\\ -0.0025 &-0.0094 & 0.3293 \ev$&$0.9329$&$51.74$&$\sv 0.3292& -0.0015 &-0.0006 \\-0.0015 & 0.3396& -0.0074\\ -0.0006 &-0.0074&  0.3312\ev$ &$0.9485$&$47.26$\\
 \hline\hline
 Structure $\#2$&$\sv 0.3218& -0.0004 &-0.0011\\ -0.0004&  0.3366& -0.0028\\ -0.0011& -0.0028 & 0.3415\ev$ & $0.9385$&$80.72$&$\sv 0.3233& -0.001&  -0.0014\\ -0.001&   0.3362& -0.0026\\ -0.0014& -0.0026&  0.3405 \ev$&$0.9451$ &$73.96$\\
 \hline
 \end{tabular}}
 \caption{{Quadratic} normal tensor, degree of anisotropy and total surface area for grain structures $\#1$ and $\#2$ with and without binder.}\label{tab:sandStruct}
 \end{center}
 \end{table}
 The {quadratic} normal tensor is almost isotropic in all four cases. Removing the binder phase 
  leads to slightly more anisotropic {quadratic} normal tensors compared to the sand-binder composite. However, the change is marginal. Without the binder, the surface area of every grain is fully exposed, which results in a $9.5\%$ larger total surface area in case of structure $\#1$ and $9.1\%$ larger surface area in case of structure $\#2$. In general, we see that, although the grains within both structures are highly anisotropic, the resulting microstructure as a whole is almost isotropic. Hence, mechanical contraction of anisotropic shapes results in an overall isotropic microstructure.
 This conforms to the results of Schneider et al.~\cite{Sand}, where elastic homogenization studies on similar structures were performed. An isotropic approximation of the effective stiffness {tensor} was {shown to be accurate}.

\section{Conclusion and Outlook}
\label{sec:Conclusion}
In this study, we {proposed using} Minkowski tensors, a tensor{-valued} generalization of {the} scalar-valued Minkowski functionals, for the analysis of microstructures given {implicitly on} voxel images. Due to their tensorial nature, Minkowski tensors  naturally contain information about the anisotropy of geometric structures {and can be incorporated into continuum mechanical or other physical modeling approaches}.\\
We provide an efficient and compact algorithm for computing the Minkowski tensor $W^{0,2}_1$ and the resulting {quadratic} normal tensor ({QNT}) from 3D gray-value image data. This algorithm is based on image filtering and a numerical gradient computation. We demonstrated the multigrid convergence of our algorithm on a {single-ball structure}. Central differences and a ball filter with low filter parameter turned out to be the most accurate for binary images. For gray-value images of low resolution, {skipping the filtering step may be beneficial}. Furthermore, we demonstrated that the {quadratic} normal tensor is rather insensitive to errors in the surface area computation, {thus} providing a robust measure of microstructure anisotropy.\\
For fiber-reinforced composites, we compared characterizations based on the {QNT} to the well-established fiber-orientation tensors. We compared our approach to the common structure-tensor approach and demonstrated the accuracy and robustness of the {quadratic} normal tensor.\\
Finally, we studied the {QNT} of sand-core {micro}structures. Its applicability to complex grain geometries demonstrates the versatility of the Minkowski{-}tensor approach.\\
In future applications, further Minkowski tensors may be used for describing and characterizing a variety of microstructures, including curved fibers, fibers of different length and diameter, mixtures of several different shapes within a matrix, or polycrystalline structures.
For a robust curvature-approximation technique based on voxel{-}image data, for instance, the curvature-dependent Minkowski tensor $W_2^{0,2}$ may be computed, providing additional information on the microstructure.\\
{The Minkowski tensors of the second rank may reflect only three types of material symmetries: isotropy, transverse isotropy and orthotropy. To detect finer material symmetries, working with higher-order Minkowski tensors is necessary.} {Mickel et al.~\cite{mickel13} suggested using irreducible Minkowski tensors for anisotropy characterization, a decomposition of the surface{-}normal density into those of some basic shapes in the spirit of Fourier analysis. This approach may also be beneficial {for} fiber{-}orientation analysis. Moreover, the concept of Minkowski maps~\cite{Klatt12, Klatt13} may allow study{ing} the local differences of the fiber orientation across an inhomogeneous medium.}\\
Last but not least, Minkowski tensors may serve as input for further studies. {Similar} to fiber-orientation tensor based mean-field models~\cite{Benveniste,KehrerWicht}, models based on Minkowski tensors {may be developed}. The {quadratic} normal tensor is able to provide insights for structures containing curved fibers and may serve as a tool for investigating their mechanical behavior.

\section*{Acknowledgements}
We thank S. Gajek, J. G\"orthofer, D. Wicht and T.-A. Langhoff for support during the preparation of this manuscript.
The authors acknowledge financial funding by the KIT center \emph{MathSEE} (Mathematics in Sciences, Engineering, and Economics). M. Schneider and T. B\"ohlke acknowledge
partial financial support by the German Research Foundation (DFG) within the International Research Training Group ``Integrated engineering of continuous-discontinuous long fiber reinforced polymer structures'' (GRK 2078). Support from the DFG for the project SCHN 1595/2-1 is gratefully acknowledged by M. Schneider and F. Ernesti.
\appendix

\section{{Minkowski tensors for specific shapes}}
\subsection{Minkowski tensor of a {ball}}\label{appendix:sphere}
Consider the {ball} $B_R(0)$, parameterized by spherical coordinates $(r,\varphi,\theta)$, with $r\in [0,R),~\varphi \in [0,2\pi],$ and $\theta \in [0,\pi]$. The transformation to Cartesian coordinates reads
\begin{align*}
\rr(r,\varphi,\theta)  = \sv r\sin(\theta)\cos(\varphi)\\ r\sin(\theta)\sin(\varphi)\\ r \cos(\theta)\ev.
\end{align*}
The outward-pointing unit normal on $\partial B_R(0)$ is given by $\n(r,\varphi,\theta) = \rr(1,\varphi,\theta)$ and is thus independent of $r$. With this parameterization at hand, the Minkowski tensor $W^{0,2}_1$ computes as
\begin{align*}
W^{0,2}_1&(B_R(0)) = \frac{R^2}{3}\int_0^{2\pi}\int_0^\pi \n(\varphi,\theta)\otimes \n(\varphi,\theta) \sin(\theta)\dd \theta \dd \varphi\\
&= \frac{R^2}3 \int_0^{2\pi}\int_0^\pi \sv
\sin^2(\theta)\cos^2(\varphi) & \sin^2(\theta)\sin(\varphi)\cos(\varphi) &\sin(\theta)\cos(\theta)\cos(\varphi)\\
\sin^2(\theta)\sin(\varphi)\cos(\varphi)&\sin^2(\theta)\sin^2(\varphi) & \sin(\theta)\cos(\theta)\sin(\varphi) \\ \sin(\theta)\cos(\theta)\cos(\varphi)&\sin(\theta)\cos(\theta)\sin(\varphi)&\cos^2(\theta)\ev\sin(\theta) \dd \theta \dd \varphi\\
&=\frac{4\pi R^2}{9} \Id.
\end{align*}

\subsection{{Minkowski tensor of a cylinder}}\label{appendix:cylinder}
We consider a cylinder $K$ in $\R^3$, oriented in $z$-direction. We parameterize it by cylindrical coordinates $(r,\varphi,z)$ with $r\in [0,R),\varphi \in [0,2\pi]$ and $z\in(0,L)$.
The transformation to Cartesian coordinates reads
\begin{align*}
\rr(r,\varphi,z)= \sv r \cos(\varphi)\\ r \sin(\varphi)\\ z\ev.
\end{align*}
We divide the boundary into three subsets, describing the side, top and bottom of the cylinder $\partial K =\partial K_s\cup \partial K_t \cup \partial K_b$. The side $\partial K_s$ is parameterized by $r=R$, $\varphi\in (0,2\pi]$, $z\in[0,L]$, the bottom $\partial K_b$ by $r\in[ 0,R]$, $\varphi\in (0,2\pi]$, $z=0$ and the top $\partial K_t$ by $r\in[0,R]$, $\varphi\in (0,2\pi]$, $z=L$.
The outward-pointing unit normals for the side, bottom and top boundary, respectively, read
\begin{align*}
\mathbf{n}_s=
\sv\cos(\varphi)\\ \sin(\varphi)\\ 0\\\ev,
\quad \mathbf{n}_t=\sv 0\\0\\ 1 \ev \quad \text{and}
\quad \mathbf{n}_b=
\sv
 0\\
0\\
- 1\\
\ev.
\end{align*}
With this parametrization at hand, we compute the Minkowski tensor $W^{0,2}_1$ of $K$ by
\begin{align*}
W^{0,2}_1(K) &= \frac 13\int_0^{2\pi}\int_0^L \sv \cos^2(\varphi) & \cos(\varphi)\sin(\varphi) & 0\\\cos(\varphi)\sin(\varphi)& \sin^2(\varphi) & 0\\0&0&0 \ev R\dd \varphi \dd z + \frac{2}{3} \int_0^R\int_0^{2\pi} \e_z \otimes \e_z r\dd r \dd \varphi\\
&= \frac{\pi}{3}LR (\e_x \otimes \e_x + \e_y \otimes \e_y)+ \frac{2 \pi}{3} R^2 \e_z \otimes \e_z \\&=  \frac{2\pi}3 R^2\bigg[\e_z \otimes \e_z + \frac{L}{2R}\bigg(\Id - \e_z \otimes \e_z \bigg)\bigg].
\end{align*}

Dividing $W^{0,2}_1(K)$ by its trace gives the quadratic normal tensor of $K$:
\[
\normedW(K) = \frac{R}{R+L} \e_z\otimes \e_z + \frac{L}{2(R+L)}\bigg(\Id - \e_z \otimes \e_z\bigg).
\]
If $R\ll L$ holds, then $R/(R+L)$ is the smallest eigenvalue, which indicates an extension in $\e_z$-direction. The larger eigenvalue $L/(2(R+L))$ has multiplicity 2, indicating some symmetry within the $\e_x-\e_y$-plane.\\
{If $R\gg L$ holds, the smaller eigenvalue has multiplicity 2, indicating a disc-like shape within the $\e_x-\e_y$-plane.
}
\bibliographystyle{unsrt}
\bibliography{literature}
\end{document}